\documentclass[times,final]{elsarticle}

\usepackage[english]{babel}

\usepackage[letterpaper,top=2cm,bottom=2cm,left=3cm,right=3cm,marginparwidth=1.75cm]{geometry}

\usepackage{standalone} 
\usepackage{amsmath}
\usepackage{graphicx}
\usepackage{epstopdf}
\usepackage{tikz-cd}
\usepackage{tikz}
\DeclareGraphicsExtensions{.eps}

\usepackage{comment}

\usepackage{placeins}

\usepackage[colorlinks=true, allcolors=blue]{hyperref}
\usepackage{subcaption}
\usepackage[ruled,vlined]{algorithm2e}

\usepackage{array}
\newcolumntype{C}[1]{>{\centering\arraybackslash}p{#1}}

\usepackage{lineno}
\usepackage{setspace}

\DeclareMathAccent{\svec}{\mathord}{letters}{126}
\newcommand\stvec[1]{\mathbf #1}				
				
\newcommand\ssvec[1]{\svec{\stvec{#1}}}

\definecolor{red}{RGB}{255, 0, 0}
\definecolor{blue}{RGB}{0, 0, 255}
\definecolor{martin_color}{RGB}{0, 200, 0}

\begin{document}

\begin{frontmatter}
\title{
Reinforcement learning for anisotropic p-adaptation and error estimation in high-order solvers
}
\author[1]{David Huergo\corref{cor1}}
\cortext[cor1]{Corresponding author}
\ead{david.huergo.perea@upm.es}
\author[1]{Mart\'in de Frutos}
\author[1]{Eduardo Jané}
\author[1]{Oscar A. Marino}
\author[1,2]{Gonzalo Rubio}
\author[1,2]{Esteban Ferrer}

\address[1]{ETSIAE-UPM-School of Aeronautics, Universidad Politécnica de Madrid, Plaza Cardenal Cisneros 3, E-28040 Madrid, Spain}
\address[2]{Center for Computational Simulation, Universidad Politécnica de Madrid, Campus de Montegancedo, Boadilla del Monte, 28660 Madrid, Spain}


\begin{abstract}
We present a novel approach to automate and optimize anisotropic p-adaptation in high-order h/p solvers using Reinforcement Learning (RL). 
The dynamic RL adaptation uses the evolving solution to adjust the high-order polynomials. We develop an offline training approach, decoupled from the main solver, which shows minimal overcost when performing simulations. In addition, we derive an inexpensive RL-based error estimation approach that enables the quantification of local discretization errors. The proposed methodology is agnostic to both the computational mesh and the partial differential equation to be solved.

The application of RL to mesh adaptation offers several benefits. It enables automated and adaptive mesh refinement, reducing the need for manual intervention. It optimizes computational resources by dynamically allocating high-order polynomials where necessary and minimizing refinement in stable regions. This leads to computational cost savings while maintaining the accuracy of the solution. Furthermore, RL allows for the exploration of unconventional mesh adaptations, potentially enhancing the accuracy and robustness of simulations. 
%
This work extends our original research in \cite{huergo2024reinforcement}, offering a more robust, reproducible, and generalizable approach applicable to complex three-dimensional problems. We provide validation for laminar and turbulent cases: circular cylinders, Taylor Green Vortex and a 10MW wind turbine to illustrate the flexibility of the proposed approach.  
\end{abstract}

\end{frontmatter}



\onehalfspacing

\section{Introduction}\label{sec:Introduction}
The integration of machine learning techniques with scientific computing is transforming the field of computational fluid dynamics (CFD) \citep{LECLAINCHE2023108354,Vinuesa_2022}. Machine learning, and particularly reinforcement learning (RL), has undergone remarkable advances in recent times. The application of RL to a variety of fields, including robotics, gaming, finance, and healthcare, has been notable for its ability to manage sequential decision-making problems and, more generally, to provide the autonomy of systems in complex tasks \cite{Sutton1998}.

In fluid dynamics, RL has been proposed to enable flow control of time-varying flows \citep{GARNIER2021104973,vignon2023recent, font2024active}. Indeed, RL employs algorithms that allow the agent to learn optimal control strategies through the interaction with an environment and receiving feedback in the form of rewards or penalties. Agents learn to make optimal decisions considering the current state of the environment, selecting actions, and observing the results. This learning process enables RL agents to navigate complex environments, optimize behavior, and achieve set goals.

%
When considering flow simulations, where unsteady and turbulent flows evolve in time and space, RL shows great potential, since it offers the possibility of automatically selecting numerical parameters and mesh resolutions as flows evolve. An example is the selection of constants in turbulence models \cite{kurz2023deep,beck2023toward}, optimal parameters in high-order schemes \cite{feng2023deep,huergo2024multigrid} or mesh adaptation. 

 Mesh adaptation or adaptive mesh refinement (AMR) is essential in numerical methods, allowing the refinement or coarsening of the computational mesh based on solution smoothness and computational cost. Traditionally, this process has depended on heuristics, manual intervention or error estimators, which can be time-consuming and limited in capturing complex flow phenomena. A review of the current state of the art is provided in Section \ref{sec:state_of_the_art}.
 The integration of RL for mesh adaptation offers a novel approach to automate the process, minimizing user intervention while providing a flexible and general framework applicable to a wide range of problems. However, it is still in its early stages of development.
RL has been successfully used for mesh adaptation (h-refinement) in recent works, where agents were trained to increase the number of elements in the mesh, increasing the accuracy in critical regions. In particular, Foucart et al. \cite{foucart2023deep} compared deep reinforcement learning with heuristic approaches to perform an h-adaptive mesh refinement that is generalizable for different partial differential equations (PDE). In addition, in the research conducted by Yang et al.~\cite{yang2023reinforcement} new policy architectures are considered, which are independent of the mesh size and are trained directly from simulations. In the field of multi-agent RL, Dzanic et al. \cite{dzanic2024dynamo} showed a novel approach based on independent PPO that can be used for h- and p-refinement, with the limitation that only periodic meshes are used and a single level of refinement is allowed.
These approaches lead to accurate solutions with reduced computational cost compared to traditional mesh refinement.

%

This work builds on our foundational research presented in \cite{huergo2024reinforcement}, where we established the basis for applying reinforcement learning to p-adaptation in 1D partial differential equations. Based on this original concept, we have developed a novel and improved methodology for RL-based p-adaptation, resulting in a more robust and reproducible approach that can tackle complex three-dimensional problems. Here, we focus on the Navier-Stokes equations, but the methodology is flexible and could be used to solve other PDEs.
We explore the uncharted domain of anisotropic high-order p-adaptation using RL, with the aim of automating and optimizing the mesh adaptation process. By incorporating RL algorithms into simulations, we can train agents to adjust the polynomial order in high-order solvers, according to evolving solutions. 
The RL agent interacts with the numerical solver, observes the flow state, and makes decisions on mesh modifications. During offline training, the agent receives rewards based on the accuracy and efficiency of the numerical solution. This training is decoupled from the real simulation (the solver is not required during the training), enhancing the efficiency of the proposed approach.
Furthermore, this novel methodology allows us to train highly generalizable RL agents that are mesh-independent and can be used to potentially solve any PDE with a high-order solver.

Finally, to contextualize the novelty of our work, we summarize in Table \ref{tab:AMR_comparison} the main characteristics of our RL-based p-adaptation strategy and other state-of-the-art RL-based adaptive mesh refinement strategies proposed in \cite{foucart2023deep} and \cite{dzanic2024dynamo}.  It can be seen that our work considers p-adaptation and is trained in 1D problems but can be used for 3D flows, resulting in computational cost savings during the training. We do not need to retrain the RL agent for new cases; once trained, it works for any (tested) case. In addition, we propose using a \emph{value iteration} algorithm that is fully reproducible and robust (since it does not rely on neural networks). Finally, we provide an inexpensive RL-based error estimation and apply the p-adaptation strategy to 3D turbulent cases, which has not been done previously.

The rest of the paper is organized as follows: first, in Section \ref{sec:state_of_the_art} we comment the state-of-the-art in p-adaptation for high-order solvers. In Sections \ref{subsec:DGSEM} and \ref{subsec:RL_framework} we provide some background on discontinuous Galerkin solvers and the reinforcement learning framework. Then, in Section \ref{subsec:Value_iteration} we describe the main RL algorithm that is the core of the entire methodology, and in Section \ref{subsec:RL_strategy} we define all the required elements that are necessary to compute the previous algorithm. In Section \ref{subsec:Implementation_p_adaptation}, all the underlying information in relation to the implementation of the previous methodology in a real solver is provided. Finally, in Section \ref{sec:Results} we show different results to validate the proposed strategy and in Section \ref{sec:Conclusions} we provide some final conclusions.
\newpage

\begin{table}[h!]
\centering
\begin{tabular}{| >{\raggedright}p{3.75cm} | >{\raggedright}p{3.33cm} | >{\raggedright}p{3.33cm} | >{\raggedright\arraybackslash}p{3.33cm}| }
\hline
 & \textbf{Present Method} & \textbf{Foucart et al. \cite{foucart2023deep}} & \textbf{Dzanic et al. \cite{dzanic2024dynamo}} \\
\hline\hline
\textbf{Adaptation Type} & p-adaptation & h-adaptation & h- and p-adaptation (not h/p-)\\
\hline
\textbf{Mesh Applicability} & Valid for any mesh & Valid for any mesh & Cartesian periodic meshes\\
\hline
\textbf{PDE Generalization} & Valid for any PDE without retraining the agent & Valid for any PDE, but requires retraining in each case & Valid for hyperbolic conservation laws, and requires retraining in each case\\
\hline
\textbf{Dimensional Generalization} & The agent trained in 1D cases generalizes to 2D and 3D cases & The agent can be trained for 1D and 2D cases separately & 2D cases only\\
\hline
\textbf{Anisotropic Adaptation} & Yes & No & No\\
\hline
\textbf{Dynamic Adaptation} & Yes & Yes & Yes and anticipatory\\
\hline
\textbf{Adaptation Levels} & Between $p_{\min}$ and $p_{\max}$ & Based on defined budget & One level\\
\hline\hline
\textbf{RL Method} & Value iteration:  reproducible and robust & Deep RL (DQN, A2C, and PPO): Supports continuum state definitions & Independent PPO\\
\hline
\textbf{State Definition} & Complete with local contribution & Complete with local/global contribution & Complex and lacks detailed information of the solution\\
\hline
\textbf{Reward: Trade-off} & Between accuracy and computational cost & Between accuracy and computational cost & None\\
\hline
\textbf{Reward: Reliability} & High: Accuracy measured against a reference solution &  Intermediate: Accuracy measured based on the change of the approximation &  Intermediate: Accuracy measured based on the change of the approximation \\
\hline
\textbf{Reward: Accuracy Contribution} & Can be chosen to achieve certain accuracy before the cost matters &  Both cost and accuracy always matter in a fixed proportion & Considered only if the error is out of bounds and the refinement level is not correct\\
\hline
\textbf{Reward: Computational Cost Contribution } & Measured based on the polynomial order (DOFs) & Can be measured in various ways (RAM, CPU usage, number of elements, etc.) & None\\
\hline
\textbf{Actions} & Refine, coarsen, do nothing & Refine, coarsen, do nothing & Refine, coarsen\\
\hline
\hline
\textbf{RL-based Error Estimation} & Yes & No & No\\
\hline
\textbf{Validation Cases} & Complex 3D cases, e.g. Navier-Stokes with turbulence & 1D and 2D equations with analytical solutions & Linear advection and compressible Euler equations\\
\hline
\textbf{Validation Equations} & Hyperbolic & Hyperbolic and elliptic & Hyperbolic\\
\hline
\textbf{Validation solver} & DG & DG and HDG & DG\\
\hline
\textbf{Comparison with other AMR Sensors} & Yes & Yes & Yes, with an extremely simple AMR sensor\\
\hline
\end{tabular}
\caption{Comparison of state-of-the-art RL-based mesh adaptation methods.}
\label{tab:AMR_comparison}
\end{table}

\FloatBarrier

\section{State of the art in p-adaptation
}\label{sec:state_of_the_art}
Adaptation strategies in computational fluid dynamics are typically categorized based on the type of error measurement that is used. These categories include feature-based adaptation, adjoint-based adaptation, and local error-based adaptation. Comparative studies of these approaches have been conducted by Fraysse et al. \cite{Fraysse2012} for finite-volume approximations and by Kompenhans et al. \cite{Kompenhans2016a} and Naddei et al. \cite{Naddei2018} for high-order discontinuous Galerkin (DG) methods.

The feature-based adaptation, a classical approach, utilizes easily computable error measures that depend on flow features. This approach operates under the assumption that areas with complex flow are likely to have high errors, thus predicting refinement in regions with high velocity, density, or pressure gradients \cite{Aftosmis1994,Persson2006}. For DG discretizations, an accessible feature-based adaptation criterion is the evaluation of jumps across element interfaces \cite{Krivodonova2003,krivodonova2004shock,Remacle2003}. However, these methods have the disadvantage of lacking a direct correlation between the adaptation criterion and numerical errors, making accuracy prediction challenging. Furthermore, steady-state problems can only be solved by iterative adaptation. Recently, the feature-based approach has been enhanced with machine learning techniques to provide a highly automated adaptation based on a clustering approach \cite{tlales2024machine}.

The adjoint-based adaptation defines a functional target (e.g., drag or lift in external flow aerodynamics). The adjoint problem is then solved to obtain a spatial distribution of the functional error, which guides the mesh adaptation. This technique, originally developed for structural analysis using the Finite Element Method (FEM) by Babuška and Miller \cite{babuvska1984,babuvska1984a}, has been applied to adaptation strategies in DG methods \cite{Hartmann2006,Hartmann2002,Wang2009}. Despite its sophistication, this approach has the drawback of high computational cost and storage requirements to solve the adjoint problem and store error estimators, especially in unsteady flows. Moreover, it only guarantees the reduction of the analyzed functional error, which could deteriorate the error of other functionals.

A computationally efficient alternative is the adaptation based on local errors, which assesses any measurable local error (not feature-based) in all cells in the domain \cite{Hartmann2002}. Unlike feature-based methods, local error-based adaptation methods offer a way to predict and control overall accuracy and are computationally less expensive than adjoint-based schemes \cite{Kompenhans2016,Kompenhans2016a,laskowski2022functional}. Significant effort has been put into developing reliable local error-based adaptation methods. Mavriplis \cite{Mavriplis1989,Mavriplis1994} used estimations of the local discretization error to develop h/p-adaptation techniques for the spectral element method. Residual-based p-adaptation, another local error-based adaptation method, uses residual to measure the accuracy of the local approximation. This method was originally developed for finite elements (FE) and has been successfully applied to DG methods \cite{Hartmann2006,Rubio2015,Naddei2018}. For modal (hierarchical) DG methods, low-cost error estimates utilizing the modal approximation can drive p-adaptation procedures, such as the Variational Multiscale (VMS) indicator by Kuru and De la Llave Plata \cite{Kuru2016}, or the spectral decay indicator by Persson and Peraire \cite{Persson2006}.

Other techniques used to dynamically adapt the dissipation properties of high-order methods include troubled cell indicators, multidimensional optimal order detection (MOOD), and adaptive dissipation control. Troubled cell indicators were initially introduced in \cite{cockburn1989tvb} and are commonly used in conjunction with TVD-TVB limiters \cite{zhu2009hermite,krivodonova2004shock,fu2017new}. MOOD methods assess the maximum achievable accuracy order in each cell that ensures the fulfillment of physical and numerical admissibility conditions \cite{diot2013multidimensional, boscheri2022high, dumbser2014posteriori, vilar2019posteriori, maltsev2023hybrid, gassner2021novel}. Adaptive dissipation control, on the other hand, aims at adapting the dissipation properties of the scheme by means of modal filters (which are closely related to p-adaptation) \cite{flad2016simulation,hamedi2022optimized,winters2018comparative,dzanic2023anti}.

In this work, we apply reinforcement learning as an alternative to other local approaches. RL provides an efficient, robust, and automated p-adaptation strategy that can be applied in real time with minimal computational overhead.

\section{Methodology}\label{sec:Methodology}

\subsection{High-order discontinuous Galerkin solvers}\label{subsec:DGSEM}
\begin{figure}[h]
  \centering
  \includegraphics[width=\textwidth]{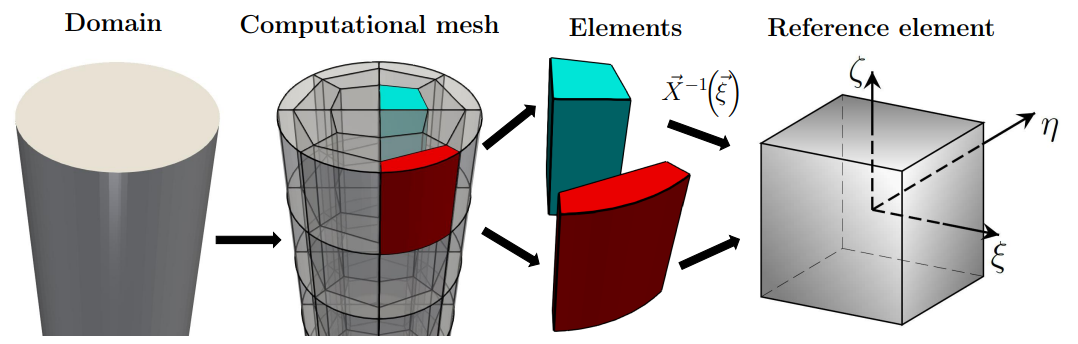}
  \caption{Geometrical transformations in the DGSEM method \cite{thesis_Juan_DG}.}
  \label{fig:DG_geometry}
\end{figure}

In this work we use the nodal discontinuous Galerkin variant called DGSEM (Discontinuous Galerkin Spectral Element Method) \cite{kopriva2009implementing}.
DG methods are characterized by their ability to use mesh refinement through polynomial enrichment (p-adaptation) to achieve highly accurate solutions; that is, low numerical errors. Such high-order polynomial methods produce an exponential decay of the error for sufficiently smooth solutions instead of the algebraic decay characteristic of low-order techniques \cite{horses3d_paper}. The DGSEM method allows us to compute the approximate solution in each element of the computational mesh, which is a tessellation of non-overlapping hexahedral elements for a 3D case, through Lagrange polynomials. Within this framework, the elements of the mesh have to be transformed from their original shape to a cube, which is used as a reference element $E=[-1,1]^3$, as shown graphically in Figure \ref{fig:DG_geometry}. Once the equations are solved in computational space, the inverse transform is applied to recover the original geometry of the problem. The geometric transformation from the physical space to the computational space provides a mechanism to use curvilinear meshes up to an arbitrary order. We provide more details on the formulation in \ref{sec:cNS} and \ref{sec:dg}.

The DGSEM approach is based on the weak formulation of the equations, where the integrals are approximated with Gauss quadratures, which are solved using Legendre-Gauss nodes 
to interpolate the Lagrange polynomials. Gauss points provide an exact integration for polynomials of order $2p+1$ or lower
\cite{kopriva2009implementing}, 
with $p$ as the selected order for the Lagrange polynomials. Therefore, the solution is computed on a set of nodes, which defines the DGSEM procedure as a nodal approach, in contrast to other DG solvers where a modal approach is used instead. A major advantage of the nodal approach is that the final three-dimensional solution inside one element is reconstructed by computing the tensor product of one-dimensional solutions. Later, we will exploit this fact to develop a reinforcement learning agent that will be trained with 1D solutions in computational space, drastically reducing the training time, but that can be applied as it is to a 3D problem, by taking advantage of the way the solution is computed with the tensor product.

\begin{figure}[h]
  \centering
  \begin{tikzpicture}[scale=2.0]
		
    \draw[thick] (-0.25, 0) -- (3.25, 0);
    
    \foreach \x in {0,1,2,3} {
        \draw[dashed] (\x, -0.5) -- (\x, 1);}
    \node[name] at (0.5,-0.5) {$\Omega^{e-1}$};
    \node[name] at (1.5,-0.5) {$\Omega^{e}$};
    \node[name] at (2.5,-0.5) {$\Omega^{e+1}$};
    
    \draw[thick, domain=0:1, samples=100] plot (\x, {0.25*(\x)*(\x-2)*(\x-3)});
    \draw[thick, domain=1:2, samples=100] plot (\x, {0.25+0.5*(\x)*(\x-1.2)*(\x-2.15)});
    \draw[thick, domain=2:3, samples=100] plot (\x, {0.7-0.5*(\x-2.5)*(\x-3.5)});
    
    \node at (0.5, 0.9) {${\mathbf{q}^{e-1}}^{N}$};
    \node at (1.5, 0.9) {${\mathbf{q}^{e}}^{N}$};
    \node at (2.5, 0.9) {${\mathbf{q}^{e+1}}^{N}$};
		
\end{tikzpicture}
  \caption{Example of a 1D subdivision into finite elements with piece-wise solutions.}
  \label{fig:DG_discontinuities}
\end{figure}
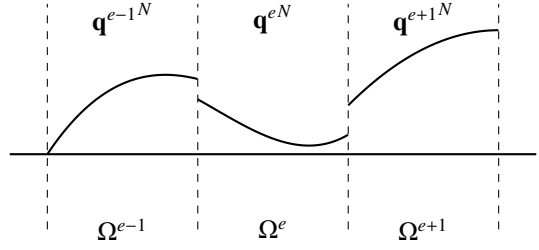

The DGSEM methodology leads to a discontinuous solution along the interfaces between elements, which is a characteristic feature of all DG methods, as shown schematically in Figure \ref{fig:DG_discontinuities}. These discontinuities generate a Riemann problem at the interfaces, which has to be solved to ensure a correct coupling between elements; see also \ref{sec:dg}.
Furthermore, when a p-adaptation algorithm is used, the coupling between the faces of adjacent elements with different polynomial orders is performed using the mortar method \cite{Kopriva2002}.

\FloatBarrier

\subsection{Reinforcement learning framework}\label{subsec:RL_framework}
\begin{figure}[h]
    \centering
    \includegraphics[width=0.7\textwidth]{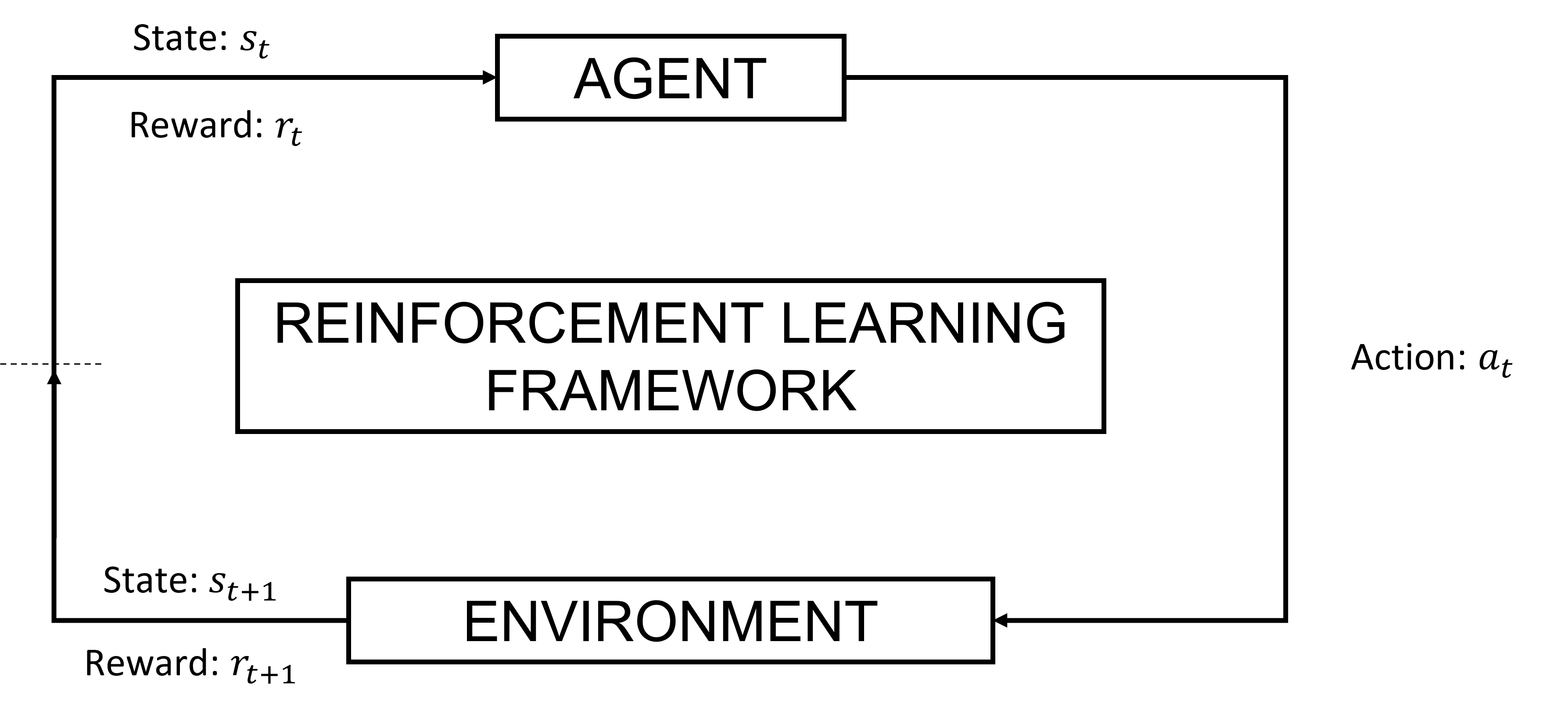}
    \caption{Schematic definition of the RL framework.}
    \label{fig:RL_scheme}
\end{figure}

Reinforcement learning is generally considered a semi-supervised approach, as the agent learns by itself the optimal policy through the interaction with an environment. However, the user must define an objective function to reward the agent when it shows a positive behavior. Most RL approaches are based on the scheme shown in Figure~\ref{fig:RL_scheme}.

The environment is considered to be a finite Markov Decision Process (MDP). That is, the state space $\mathcal S$, the action space $\mathcal A$ and the reward space $\mathcal R$ are finite and the MDP dynamics are given by the set of probabilities $\mathbb P(s',r \mid s, a)$ for all $s\in\mathcal S, a \in \mathcal A, r\in\mathcal R$ and $s'\in\mathcal S^+$ (where \( \mathcal S^+ \) is \( \mathcal S \) plus a terminal state if the problem is episodic and $s'$ is a reachable state from $s$ in a single step by applying an action $a$). This probability reads as the probability to reach the state $s'$ and receive the reward $r$ after taking action $a$ on the state $s$, and it is called the probability transition function \cite{Sutton1998}.

This reinforcement learning framework highlights several concepts that must be defined for a correct understanding of the proposed p-adaptation methodology:
\begin{enumerate}
    \item \textbf{Agent}: The brain behind reinforcement learning, which decides the best action to take. The agent is modeled with the policy function $\pi(s)$, which is obtained through the training process and determines the action $a$ to take from the state $s$. 
    \item \textbf{Environment}: The high-order numerical solver and each element of the computational mesh in our problem. It allows to compute a solution in one element given the polynomial order, $p$.
    \item \textbf{Action $(a)$}: It is an output for the agent and an input for the environment. Three different actions will be considered: to increase the polynomial order by one unit, to decrease the polynomial order by one unit, and to keep the polynomial order constant.
    \item \textbf{State $(s)$}: It is an output for the environment and an input for the agent. It must contain enough information to define the current scenario, and hence it should provide the solution in each element of the computational mesh. 
    \item \textbf{Reward $(r)$}: The user-defined objective function: $r(s, a, s')$. It should have a higher value when the agent is performing better. It provides a trade-off between accuracy and computational cost. 
\end{enumerate}

While the definitions of the environment and the actions are straightforward, the state and the reward must be carefully designed to obtain good performance. These elements of the RL framework are explained in detail in Sections \ref{subsubsec:State}, \ref{subsubsec:State_extrapolation_3D}, and \ref{subsubsec:Reward}. 

\subsection{The value iteration algorithm}\label{subsec:Value_iteration}
In this research, we use a family of model-based algorithms known as Dynamic Programming (DP) to solve the RL problem. These algorithms allow to compute optimal policies given a known model of the environment as a finite MDP \cite{Sutton1998}. Therefore, the probability transition function $\mathbb P(s',r\mid s,a)$ for the environment is explicitly required. 
%
DP does not rely on neural networks (NN), but uses tabular representation of the variables of interest instead, which limits the size of the state space. This limitation has driven the development of Deep Reinforcement Learning methods. However, classic reinforcement learning methods such as DP excel in robustness, reproducibility, and reliability, along with fast convergence, all of which are highly desirable for numerical simulations. Therefore, whenever feasible, these methods should be considered.


Dynamic programming, as well as many other RL approaches, is based on value functions $v_\pi(s)$, also called \textit{v-values}. The value function of a state $s$ under the policy $\pi$ is the expected return when starting the interaction with the environment from $s$ and following $\pi$ thereafter,
\begin{equation}
v_{\pi}(s)\ = \mathbb E_{\pi}\left[\sum_{k=0}^{\infty}\gamma^{k}R_{t+k+1}\ \mid S_t=s\right]\quad \text{for all }s\in\mathcal S,
\label{eq:v-value-definition}
\end{equation}
where $\gamma$ is the discount factor, which measures the importance of future rewards on the present \textit{v-value}. An action taken $k$ steps in the future is worth $\gamma^{k-1}$ times what it would be worth if it were received immediately. For convergence reasons, the discount factor is bounded in $0\leq\gamma<1$, and its value must be selected depending on the specific RL problem to be solved.

The main difference between algorithms within the DP framework is the method used to compute these \textit{v-values}. However, the ultimate goal of all these algorithms is the same: to find an optimal policy, $\pi^*$. By following that policy, it is possible to obtain the highest cumulative reward from any state. The \textit{v-values} associated with this policy are the optimal \textit{v-values}, $v^*(s)$. As shown in \cite{Sutton1998}, the optimal \textit{v-values} follow the \textit{Bellman optimality equation}: 

\begin{equation}
v^*(s) = \max_{a} \mathbb E_{\pi^*}[R_{t+1} + \gamma v^*(S_{t+1}) \mid S_t = s, A_t = a] 
       = \max_{a} \sum_{s', r} \mathbb P(s', r \mid s, a) \left[ r + \gamma v^*(s') \right].
       \label{eq:Bellman_equation}
\end{equation}

The exact solution to the previous problem can be obtained by solving a system of $N$ nonlinear equations (due to the nonlinearity introduced by the $\max$ operator), with $N$ as the total number of states in \( \mathcal S \). In addition, there are different approaches to solving the problem approximately using an iterative algorithm. Among the available options, the \emph{value iteration} algorithm provides an update rule for $v(s)$, based on the Bellman equation, that converges to the optimum $v^*$ for an arbitrary initialization $v_0$ \cite{Sutton1998}:
\begin{equation}
v_{k+1}(s) = \max_{a} \mathbb{E}[R_{t+1} + \gamma v_k(S_{t+1}) \mid S_t = s, A_t = a] 
       = \max_{a} \sum_{s', r} \mathbb P(s',r \mid s, a) \left[ r + \gamma v_k(s') \right].
       \label{eq:Value_iteration}
\end{equation}

This iterative algorithm would require an infinite number of iterations to exactly converge. Hence, it is stopped once the change in the value function is below a small threshold. The pseudocode for this algorithm can be found in \ref{app:Value_iteration}.
Once the optimal value function, $v^*$, is known for every state, $s$, the optimal policy consists in selecting the action that leads to the state with the highest value among the available options.

\subsection{Reinforcement learning strategy for p-adaptation}\label{subsec:RL_strategy}
The main objective of this research is to provide a flexible p-adaptation RL agent that can be used to solve a wide variety of problems (or PDE) and that is not dependent on the computational mesh. Furthermore, the resulting agent, once trained, should be used as it is in any high-order solver, and it should provide accurate results for every possible scenario.
This approach would require training the agent in a way that is common for every problem, which is not a simple task. In this work, we have defined the state and the reward to provide enough information to the agent, while being general enough. In this way, the same agent can be used for different problems.

\subsubsection{State definition}\label{subsubsec:State}
The state can be encoded as a set of variables that must provide enough information about the environment to the agent. Based on this state, the trained agent can select the best course of action for each scenario. An incomplete state may lead to a poor performing agent.

As our goal is to create an agent as general as possible, the state must be as general as possible as well. Here, we consider a nodal-based discontinuous Galerkin solver, as was explained in Section \ref{subsec:DGSEM}. Therefore, we have to focus on the only thing that is common for every DGSEM solver: the solution is computed at the Gauss nodes 
inside each element. Hence, the state is simply defined as the values of the variable of interest, $q$, at the Gauss nodes, as represented in Figure~\ref{fig:Gauss_nodes_p3_training}. The size of the resulting state is $p+1$ (the same as the number of Gauss nodes) and is different depending on the polynomial order. This approach considers a different state in each element of the computational mesh; that is, the RL agent will handle each element independently, leading to a mesh-independent strategy.
However, the main issue behind this definition of the state is that, in a general problem, $q$ may take any value, since $q \in \mathbb R$. 
This would lead to an unbounded and continuous state space, which does not match neither the MDP framework nor the \textit{value iteration} algorithm, as they both deal with a finite state space. Hence, the possible values of the variable of interest $q$ must be discretized, so we can define an appropriate finite space state $\mathcal S$. 

The proposed methodology takes the values of $q$ at every node within an element and normalizes the values from $-1$ to $1$. In this way, for each element there will always be one node with a value of $-1$ and another with a value of $1$. If it is the case that the value of $q$ is exactly the same at every node (below a very small tolerance), then the value at each node is set to $0$ (see Section \ref{subsec:Implementation_p_adaptation} for additional implementation details).
Then, the state is discretized in $N_l$ levels. The higher the value of $N_l$, the more accurate the final agent will be, but the slower the training phase. After some trial and error, we discovered that a value of $N_l=11$ provides a good trade-off.


Finally, the agent must select the optimal action for each element individually, but different elements may have different shapes. To create an agent as general as possible, the solution inside each element is mapped to the computational space (see Section \ref{subsec:DGSEM} for details) before the state is computed.
This step is used to facilitate the integration in finite element and discontinuous Galerkin solvers, and hence the transformation does not create an additional overhead during the p-adaptation process.

\subsubsection{State extrapolation for 3D problems}\label{subsubsec:State_extrapolation_3D}
The previous definition of the state should be applicable regardless of the dimension of the problem (the mesh can be 1D, 2D or 3D). Furthermore, we aim to design a reinforcement learning agent that is able to show anisotropic behavior when adapting the mesh; that is, each element may have a different polynomial in each local axis $x$, $y$ and $z$. With this objective in mind, we go one step further: the agent will be trained for 1D cases only, but the same agent will be able to handle 3D simulations without any modification. This approach saves computational time during the training of the RL agent. To accomplish this objective, we take advantage of the fact that the 3D solution in a DGSEM solver can be obtained as a tensor product of 1D solutions (see Section \ref{subsec:DGSEM}).
Therefore, for each local axis inside an element, the agent is provided with each row of values at the Gauss nodes that are aligned following that axis. The agent will handle one row of nodes at a time and generate an action for each row. Then, we select the most restrictive action, and this operation is repeated for each local axis.

To provide an example of this methodology, let us consider that we have a 3D element with a uniform polynomial $p=3$ for every axis. In this scenario, there are $(p+1)^2 = 16$ rows of nodes aligned with the local $x$ axis, with $p+1=4$ nodes per row. First, we feed the agent with each row of nodes and the agent will provide as an output 16 actions, one per row. Then, the most restrictive action among the 16 options will be selected. We consider that increasing the polynomial order is more restrictive than keeping the polynomial constant, which is, in turn, more restrictive than decreasing the polynomial.
The same procedure has to be applied for each local axis in each element: twice per element for 2D problems and three times per element for 3D problems. The final output is a set of 3 actions, one for each local axis of the element, that are not necessarily the same. Therefore, the p-adaptation approach can show an anisotropic behavior.
It is very important to note that the agent only handles rows of values at the nodes, which are one-dimensional arrays of size $p+1$. The tensor product expansion of 1D cases ensures that the agent trained for 1D can also be effective in 2D and 3D problems.

\subsubsection{Reward definition}\label{subsubsec:Reward}
The reward can be defined as the objective function that the agent should
learn to decide the best course of action based on the current state. This function must represent the desired behavior of the agent for every possible scenario. An appropriate definition of the reward, together with a well-designed state, is the key to an effective RL agent.
In general, the reward function is defined as $r(s,a,s')$, which reads: the reward that the agent will obtain if it is in a state $s$, chooses an action $a$ and arrives at the state $s'$. The first conclusion one may extract from this definition is that you might reach different states $s'$ from a state $s$ taking an action $a$, if your environment is stochastic; that is, if the probability $\mathbb P(s',r \mid s, a)$ from the Bellman equation (\ref{eq:Bellman_equation}) is different from 0 or 1 for some combination of $(s, a, s')$. A detailed description of these probabilities is included in the following Section \ref{subsubsec:Training}.
Depending on the problem, the reward may depend on the original state, $s$, the final state, $s'$, the chosen action, $a$, or all of them.
Here, we explore different possibilities to decide which one is better suited to solve the p-adaptation problem.

\subsubsection*{\textbf{General considerations}}
The definition of the reward will be influenced by the framework used to implement the p-adaptation strategy. In general, the p-adaptation process in a high-order solver is highly time-consuming. Therefore, we cannot afford to adapt the mesh every iteration. Furthermore, for numerical stability reasons, each time the mesh is adapted, only one action is taken in each element; that is, the polynomial order can be increased or decreased by one unit (see Section \ref{subsec:Implementation_p_adaptation} for additional details regarding the implementation). These details should be considered when defining an appropriate definition of the reward function.

Additionally, we want the agent to find the optimal polynomial order for each element of the computational mesh, but it is not relevant how it reaches that polynomial, and therefore the reward must not depend on the action.
However, it is important to decide whether the reward should depend on the current state $s$, or on the next state $s'$. Both approaches are possible and valid, but the meaning is slightly different in each case.
The key behind this decision is based on the knowledge that the trained agent will rely on the value function $v(s)$ to follow the optimal policy, which is simply to select the action that leads to a state with a higher $v(s')$ (see the previous Section \ref{subsec:Value_iteration}). Furthermore, as it is stated in eq.~(\ref{eq:Bellman_equation}), each value function is directly related to the reward $r(s,a,s')$, and hence there is a direct correlation between the selected reward and the final policy.

\subsubsection*{\textbf{Reward options: $r(s)$ vs $r(s')$}}
Let us consider, for the sake of argument, that we are facing a deterministic environment and that the discount factor is $\gamma = 0$. In this hypothetical case, $v(s) = \max\limits_{a} r(s,a,s')$ and hence we do not care about the long term.
On the one hand, if $r=r(s)$, then $v(s)$ will be higher when the current state is beneficial. On the other hand, if $r=r(s')$, then $v(s)$ will be higher when, from the current state, it is possible to reach a future state $s'$ that is beneficial. In other words, if the reward is $r=r(s)$, then the present is more important, while if $r=r(s')$ then the future (even if it is a short-term future) is more important.

Within this framework, if we choose the second reward option, $r=r(s')$, the agent will try to find a state from which it is possible to reach a highly beneficial future state. However, if we can only perform one action at a time, the agent will not be able to reach the desired state until we can adapt a second time, at least. By the time we adapt again (several iterations later), the flow field might have changed, and that desirable future state might not be beneficial anymore.
In contrast, if we define the reward as $r=r(s)$, the agent will take the action that leads to a highly beneficial state right now, even if the future is uncertain. In this way, at least, we can ensure that the chosen action is the best possible one right now. This is the reason why we have selected this approach when defining the reward function.

Finally, it is important to note that for large values of the discount factor, $\gamma = 1 - \varepsilon$, with $\varepsilon <<1$, both approaches are very similar and should lead to equivalent solutions. However, when the number of actions that can be performed is limited and the time between actions is large (many iterations), in presence of a constantly changing environment, a very high value of $\gamma$ might not be desired.

\subsubsection*{\textbf{Reward function}}
Once we have selected the variables that should influence the reward, we must define the objective function for the agent. This function should have a higher value as the error decreases. In addition, it should have a higher value as the polynomial order decreases, because the objective is to use the minimum polynomial order possible to reduce the computational cost while preserving the accuracy.
%
%
The error should provide a measurement of the difference between the DGSEM polynomial approximation and the real solution within one element of the computational mesh. Therefore, the error can only be computed during the training as an analytical reference solution is required, which is unknown during a real simulation. We will propose a solution to this problem in the following Section \ref{subsubsec:Training}. 

Consider the polynomial approximation $y^*(x)$ and the analytical reference solution $y(x)$, where $x$ belongs to the computational space, and both functions have been normalized to values between -1 and 1. To evaluate the error between both functions, the Chebyshev weighted $L^2$ norm is used. This metric is chosen to avoid error propagation through the interfaces of the elements, as it penalizes more the boundaries of the integration interval. The Chebyshev weighted $L^2$ norm is defined as:
\begin{equation}
    e_r^2 = \lVert y-y^* \rVert^2_{L^2,\text{Chebyshev}} = \int_{-1}^{1}\lvert y(x)-y^*(x) \rvert ^2 \frac1{\sqrt{1-x^2}} dx.
    \label{eq:Chebyshev_L2_norm}
\end{equation}

In general, this integral is computed numerically. In particular, it can be approximated by a quadrature rule sampling in $N_e$ Chebyshev nodes: 
\begin{equation}
\label{eq:error_L2}
e_r^2 = \lVert y-y^* \rVert^2_{L^2,\text{Chebyshev}} \approx \frac\pi {N_e}\sum_{i=1}^{N_e} (y(x_i)-y^*(x_i))^2,
\end{equation}
where $x_i=\cos\left({\frac{i\pi}{N_e-1}}\right)$ are the Chebyshev node coordinates and $N_e=2 \, (p_{\max}+1)$, with $p_{\max}$ as the maximum allowed polynomial order. To incorporate the error in the reward, we introduce the \textit{root mean squared error} ($rmse$), which is related to the error in equation (\ref{eq:error_L2}) but omits the constant $\pi$. Since this term is constant, it does not influence the RL algorithm, so it can be excluded. The $rmse$ is defined as follows:
\begin{equation}
    rmse = \sqrt{\frac{\sum_{i=1}^{N_e} (y(x_i) - y^*(x_i))^2}{N_e}}.
    \label{eq:rmse_reward}
\end{equation}

Finally, we define the reward function, $r$, as a smooth Gaussian distribution, which provides a trade-off between cost (low polynomial order) and accuracy (low error):
\begin{equation}
    r = \overbrace{\left(\frac{p_{\max}}{p}\right)^\alpha}^\text{COST} \overbrace{\exp\left({-\frac{rmse^2}{2 \sigma^2}}\right)}^\text{ERROR},
    \label{eq:reward}
\end{equation}
where $\sigma$ is the standard deviation and $\alpha$ is a control parameter.
The most important contribution to the reward function is the exponential part, which measures the accuracy (or the error). This part will drop the reward to zero if the $rmse$ is large compared to the standard deviation, $\sigma$. The second contribution is the cost, and it will only be relevant when the $rmse$ is small enough. This part will decrease the reward when a high polynomial order is used, as it will require a high computational cost. Therefore, the reward function prioritizes the reduction of the error; but if the error is acceptable (based on the value of $\sigma$), then the polynomial degree is reduced as much as possible (without significantly increasing the error).

On the one hand, the value of the parameter $\alpha$ provides a way to measure the trade-off between accuracy and computational cost. When $\alpha$ is small, the accuracy is more important than cost and vice versa. 
On the other hand, the value of the parameter $\sigma$ provides an estimate of the error that must be achieved in one element before the cost is taken into account. Therefore, the smaller the value of $\sigma$, the more important is the precision over the cost.
Based on preliminary tests (not shown here), we have selected $\alpha = 0.9$ and $\sigma = 0.05$, which provides an appropriate balance between efficiency and accuracy of the p-adaptation strategy.

\subsubsection{Training}\label{subsubsec:Training}

The reinforcement learning agent is trained to select the optimal polynomial order for each element of the computational mesh. The proposed strategy for the training process does not depend on a CFD solver. This decision results in a faster and general training that does not depend on the problem to be solved.
%
As explained in Section \ref{subsec:Value_iteration}, we select a \emph{value iteration} algorithm, which is a classic model-based reinforcement learning method. The training process consists in executing an iterative algorithm until convergence is reached (details of the algorithm are shown in \ref{app:Value_iteration}). The \textit{v-values} are initialized to $v_0=0$ and the value function for each state is calculated using the update rule from eq.~(\ref{eq:Value_iteration}). 
%
%
%
In this equation, the discount factor $\gamma = 0.5$ is known and fixed. This low value for $\gamma$ has been selected to increase the importance of the short term, as explained in the previous Section \ref{subsubsec:Reward}. 
The remaining variables on the right-hand side of the update rule are the probabilities $\mathbb P(s',r \mid s, a)$, the reward function $r$ and the future states $s'$. All of them require additional considerations to be computed. In particular, a reference solution, which plays the role of the analytical solution of the problem during the training, has to be defined, together with the polynomial approximation for each state within the DGSEM framework.

\subsubsection*{\textbf{Definition of the approximate solution from the state vector}}
For each state $s$, we only know the values of the approximate solution at the Gauss nodes and the polynomial order $p=size(s)-1$. This information allows to reconstruct, for each state, the approximate solution $y^*$ within the DGSEM framework, which is generated through Lagrange polynomials based on the Gauss nodes:
\begin{equation}
    y^*(x) = \sum_{j=0}^{k} s_{j} \ell_{j}(x),
    \label{eq:Lagrange_polynomials}
\end{equation}
\begin{equation}
    \ell_{j}(x) = \prod_{\substack{i=0 \\ i \neq j}}^{k} \frac{x - x_{i}}{x_{j} - x_{i}} 
    \label{eq:Lagrange_basis}
\end{equation}
where $s_j$ are the components of the state vector and $x_i$ are the coordinates of the Gauss nodes in the computational space.

\subsubsection*{\textbf{Definition of the reference solution}}
The reference solution, $y$, represents possible outcomes that can be found within an element in a real simulation. In particular, we define the reference solution as a polynomial function. This hypothesis is reasonable within the discontinuous Galerkin framework, as the final solution is approximated using Lagrange polynomials (see Section \ref{subsec:DGSEM}). 
Therefore, we limit the space of possible reference solutions to polynomials of different orders.
For each state, $s$, we consider three possible scenarios, where the reference solution, $y$, is a polynomial function:
\begin{enumerate}
    \item One order higher than the approximate solution $y^*$.
    \item Of the same order as the approximate solution $y^*$.
    \item One order lower than the approximate solution $y^*$.
\end{enumerate}

\begin{figure}[h]
    \centering
    \includegraphics[width=0.75\textwidth]{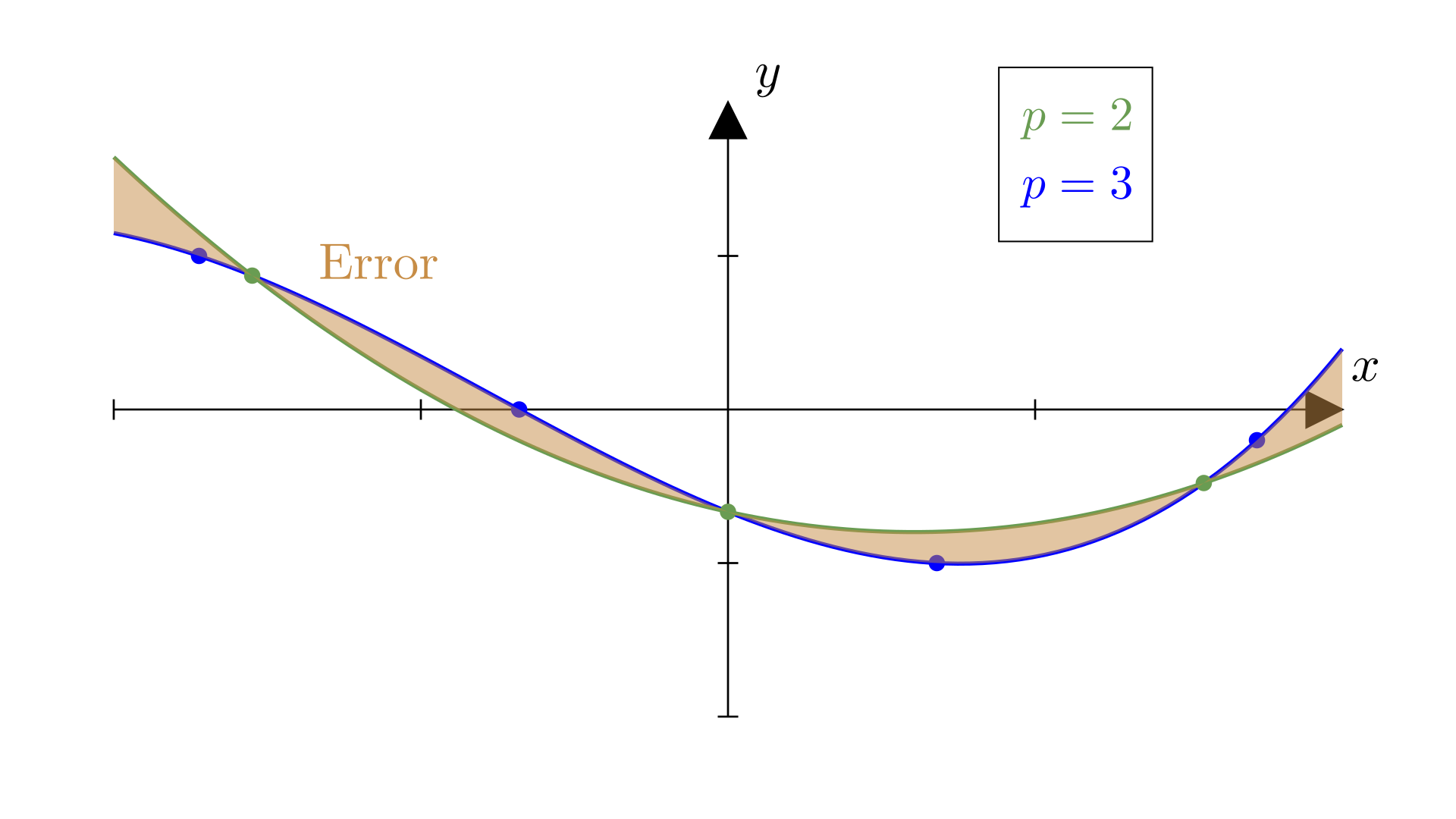}
    \caption{Example of a comparison between the approximate solution $y^*$ ($p=3$ in this example) and the low-order reference solution ($p=2$ in this example). The error between both functions is estimated through the \emph{root mean squared error} defined in eq. (\ref{eq:rmse_reward}).}
    \label{fig:Gauss_Lagrange_Extended}
\end{figure}

With these scenarios in mind, for each state $s$, the three reference solutions must be defined from the information available at the Gauss nodes. Our objective is to design functions that are capable of representing the current state with high fidelity. Therefore, the reference solutions will be constructed using Lagrange polynomials that match all the values at the Gauss nodes. The definition of these functions depends on each specific scenario, among the three aforementioned possibilities:
\begin{enumerate}
\item In the first case, there is an infinite number of high-order polynomial functions that match the current values at the Gauss nodes. Therefore, we select the high-order polynomial with minimum $L^2$ norm that goes through all the nodes for the current state. 
\item In the second case, the reference solution, $y$, is exactly the approximate solution, $y^*$.
\item Finally, in the third case it is necessary to check whether it is possible to approximate the current state with a low-order polynomial. Hence, the approximate solution for the present state, $y^*$, is evaluated at the low-order Gauss nodes; and then, these values are used to define a low-order polynomial using eq.~(\ref{eq:Lagrange_polynomials}) and (\ref{eq:Lagrange_basis}). Next, the low-order (possible reference solution) and the current polynomial (approximate solution) are compared using the same \emph{root mean squared error} defined for the reward in eq.~(\ref{eq:rmse_reward}). This step is sketched in Figure~\ref{fig:Gauss_Lagrange_Extended}. If the $rmse$ is below a threshold $rmse_{th}=0.1$, then we consider that the low-order reference solution is possible for the current state. In other case, we reject this third solution. 
\end{enumerate}

\subsubsection*{\textbf{Definition of the probability transition function}}
The probabilities $\mathbb P(s',r \mid s, a)$ define the likelihood of being in one or another scenario; that is, the likelihood of using one reference solution over the other ones. In this case, since the real analytical solution will strongly depend on each specific problem, we consider that all reference solutions are equally likely. Therefore, when all three scenarios are possible, $\mathbb P(s',r \mid s, a)=1/3$ . However, when the third scenario (i.e., the reference solution is a polynomial
function one order lower than the approximate solution) is not possible (the current state cannot be well captured using a low-order polynomial function), then $\mathbb P(s',r \mid s, a)=1/2$ for the first two scenarios, and $\mathbb P(s',r \mid s, a)=0$ for the third scenario.

Once the specific scenario is selected using $\mathbb P(s',r \mid s, a)$, 
the transition from a state $s$ to the next state $s'$ can be performed employing the reference solution for the current scenario. The steps to compute $s'$ are as follows:

\begin{enumerate}
    \item The polynomial order of the current state is obtained as $p=size(s)-1$.
    \item For each action $a = -1, \, 0 \, \textrm{and} \, 1$ (decrease, keep constant and increase the polynomial order, respectively), the polynomial order of the following state, $p'$, can be calculated as $p'=p+a$. Note that $p'$ must be bounded in $[p_{\min}, p_{\max}]$.  We have selected $p_{\min}=2$ and $p_{\max}=6$ as it provides a reasonable range for several cases and the cost of the training phase is acceptable.
    \item The reference solution for the current scenario is evaluated at the Gauss nodes of the new polynomial order $p'$.
    \item The values at the Gauss nodes are normalized between $[-1, 1]$, resulting in the new state $s'$.
\end{enumerate}

\subsubsection*{\textbf{Example of the iterative training process}}
\begin{figure}[h]
\centering
  \begin{subfigure}[b]{0.48\textwidth}
    \includegraphics[width=\textwidth]{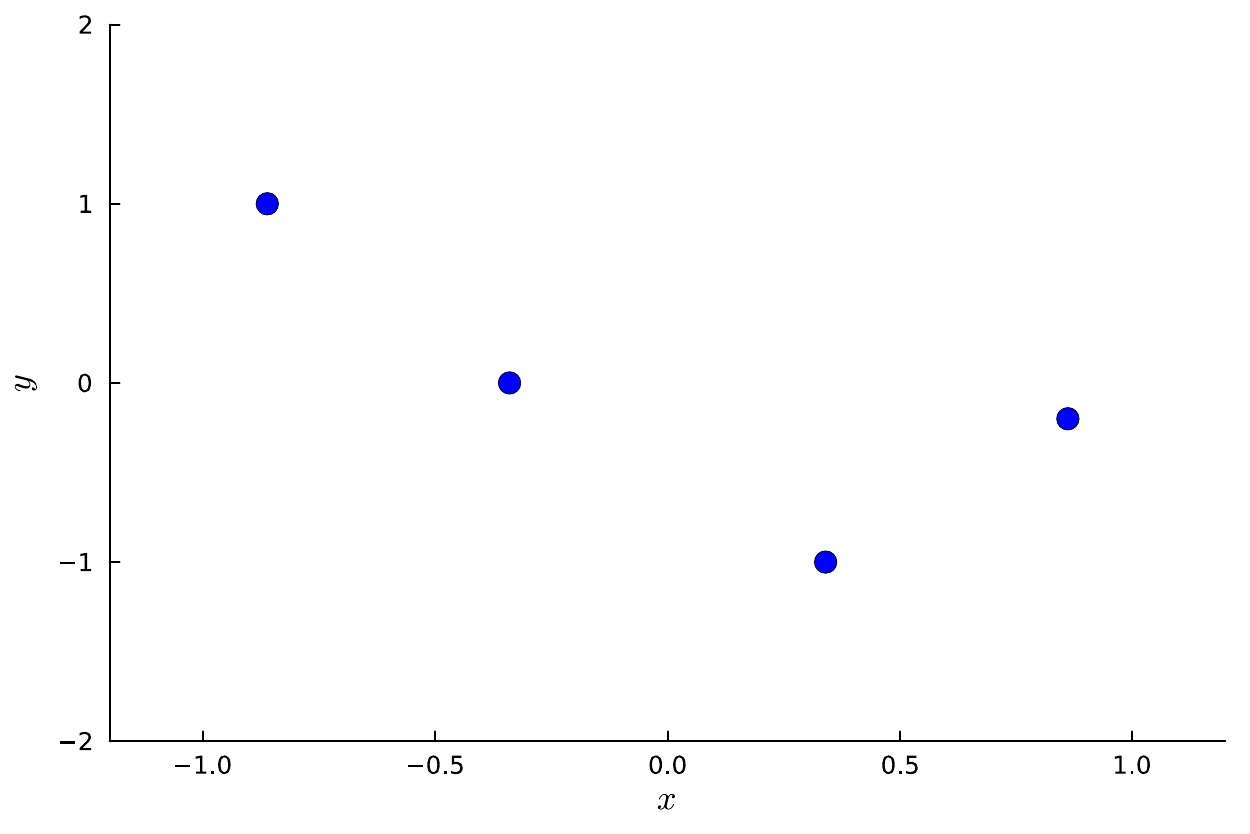}
    \caption{Example of values at the Gauss nodes for a $p=3$ approximate solution.}
    \label{fig:Gauss_nodes_p3_training}
  \end{subfigure}
  \quad
  \begin{subfigure}[b]{0.48\textwidth}
    \includegraphics[width=\textwidth]{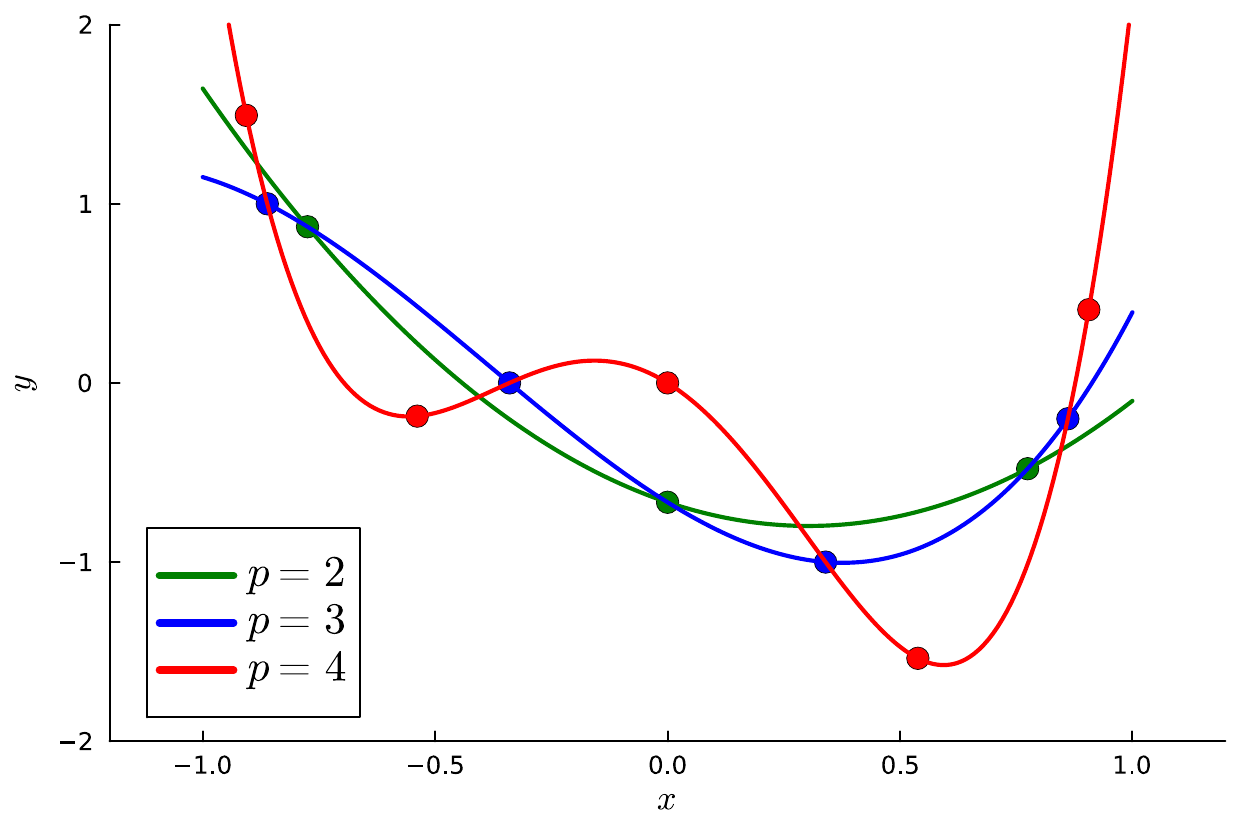}
    \caption{Possible reference solutions generated from the Gauss nodes defined on Fig \ref{fig:Gauss_nodes_p3_training}. The polynomial order of each possible reference solution is included.}
    \label{fig:Gauss_nodes_p234_training}
  \end{subfigure}
  
\caption{a) Values of the state vector for an approximate solution $y^*(x)$ and b) reference solutions $y(x)$ calculated during the training process.}
\label{fig:Analytical_solution_training}
\end{figure}

To provide a detailed overview of the resolution of the problem, let us consider an example of the iterative process. For each iteration of the \textit{value iteration }algorithm, $k$ (see eq.~(\ref{eq:Value_iteration})), we loop through every state, $s \in \mathcal S$. We focus on one of those states, that will be called $s$ for simplicity. That state, $s$, represents a set of values at the Gauss nodes, as shown schematically in Figure~\ref{fig:Gauss_nodes_p3_training}. Given these values, we first define the reference solution for the three possible scenarios that have been defined before, as is represented in Figure~\ref{fig:Gauss_nodes_p234_training}.

For each scenario, $j=1, \, 2 \, \textrm{and} \, 3$, we calculate the reward $r_j (s)$ using eq.~(\ref{eq:reward}). Then, for each action $a$
, we compute the next state $s'_{j, a}$ by evaluating the reference solution of the current scenario, $j$, into the Gauss nodes for the new polynomial order $p'=p+a$. It is important to note that the new polynomial order, $p'$, must be bounded between $p_{\min}$ and $p_{\max}$. 
Finally, once we have computed the reference solutions, the rewards, and the next states for each scenario, we have to introduce all the values in the \textit{value iteration} algorithm. Using the new notation for each scenario:

\begin{equation}
v_{k+1}(s) = \max_{a} \sum_{j} \mathbb P_j(s', r \mid s, a) \left[ r_j(s) + \gamma v_k(s'_{j,a}) \right].
       \label{eq:Example_value_iteration}
\end{equation}

\vspace{2cm}
The final solution to the RL problem is obtained by solving the previous equation for each state $s \in\mathcal S$, a number of iterations $k$, until the stop criterion defined in eq.~(\ref{eq:Stop_criterion_value_iteration}) is fulfilled:
\begin{equation}
    \frac{\sum_{p=p_{\min}}^{p_{\max}} \Delta v_p(s)}{p_{\max} - p_{\min} + 1} < 10^{-3},
    \label{eq:Stop_criterion_value_iteration}
\end{equation}
where $\Delta v_p(s)$ is the maximum difference between two consecutive \textit{v-values}, among every state $s$ that defines a polynomial of order $p$; or in other words, among every state $s$ whose size is: $size(s)=p+1$.

\subsubsection*{\textbf{Problem Complexity}} 
The \textit{value iteration} algorithm is simple and reproducible since it does not rely on neural networks. However, each iteration involves evaluating the equation (\ref{eq:Value_iteration}) for every state, which can be expensive, since the computational complexity relates to the growth of the state space. The number of operations required per iteration is $\mathcal O( \lvert\mathcal S\rvert\times \lvert\mathcal A\rvert)$, where $|\mathcal S|$ denotes the number of states and $|\mathcal A|$ the number of actions. Note that the classical computational cost of \textit{value iteration} typically has a quadratic dependence on the number of states.
However, in our case, this is not applicable, as the summation over future states, $\sum_{s'}$, does not encompass the entire state space. Instead, only three future states $s'$, one for each defined scenario, are considered from a state $s$ and for each action $a$.

The state space $\mathcal S$ has been discretized to make it suitable for an MDP framework. Therefore, the total number of states can be explicitly calculated:
\begin{equation}
    \lvert\mathcal S\rvert = \sum_{p=p_{\min}}^{p_{\max}} 1 + \frac{p(1+p)}{2}N_{l}^{(p-1)},
\end{equation}
with $N_l=11$ as the number of discretization levels for the state (see Section \ref{subsubsec:State} for details). The previous equation highlights the rapid increase in the number of states with the polynomial order $p$. Furthermore, in this expression additional information has been taken into account to reduce the effective number of states:
\begin{enumerate}
    \item Each state must have at least one element whose value is 1 and another element whose value is $-1$, with the exception of the zero state, where all values are 0. 
    \item Symmetries are applied; that is, if two states are symmetric, only one of them is computed.
\end{enumerate}
Both rules drastically reduce the number of possible states during the training phase, and hence the effective training time.
Overall, the computational complexity of the \textit{value iteration} algorithm grows exponentially with the maximum polynomial order considered. Furthermore, since the number of states shows the same trend as the computational complexity, there is a corresponding increase in the memory requirements needed to store all the \textit{v-values}. Despite these challenges, the algorithm is highly parallelizable, which enhances its scalability even with high polynomial orders. More importantly, training is only done once and offline. The same trained agent is then used for multiple cases (all the results provided in Section \ref{sec:Results} are simulated with the same trained agent). 

\subsubsection{Error Estimation}\label{subsubsec:Error_estimation}
The optimal policy obtained during the training process determines the best action to take for each state. This policy is based on the value function $v(s)$, which provides highly valuable additional information that can be used to calculate an error estimate. This estimation is the error between the numerical approximation and the analytical solution of the problem (even if it is unknown) that the RL agent considers as real for the current state.

First, let us start with the definition of the optimal value function, $v^*(s)$, given by the \emph{Bellman optimallity equation} in eq.~(\ref{eq:Bellman_equation}). Considering that we can reach an optimal state $s^*$ in $n$ steps, from that point onward the best action will always be to keep the current polynomial order as it is, and the next state, $s^{n+1)}$, will be equal to the optimal state: $s^{n)}=s^{n+1)}=s^*$. Following these considerations, the value function definition in eq.~(\ref{eq:v-value-definition}) can be explicitly written as: 

\begin{equation}
    v^*(s) = \bar{r} + \gamma \bar{r}' + \dots + \gamma^{n-1} \bar{r}^{n-1)} + \gamma^n v^*(s^*) = \bar{r} + \gamma \bar{r}' + \dots + \gamma^{n-1} \bar{r}^{n-1)} + \frac{\gamma^n \hat{r}}{1 - \gamma},
    \label{eq:V_value_error_estimation}
\end{equation}
where $\bar{r}$, $\bar{r}'$, ..., are the average reward (considering the three possible scenarios described in the previous Section \ref{subsubsec:Training}) for each state $s$, $s'$, ..., respectively.

During the training process, we know each state $s$ and the corresponding future states $s'$, but no information is available beyond that. Hence, the final reward in the optimal state, $\hat{r}$, can be obtained from the previous equation for a value of $n=2$ steps:

\begin{equation}
    \hat{r} = \frac{(1-\gamma)}{\gamma^2} \, \left( v^*(s) - \bar{r} - \gamma \bar{r}' \right).
    \label{eq:reward_error_estimation}
\end{equation}

The estimated reward will be the real reward if an optimal state can be reached in 2 steps, and it will provide an estimation of the reward in other case. However, given that $p_{\min}=2$ and $p_{\max}=6$ (see the previous Section \ref{subsubsec:Training} for details), if the polynomial order is frequently adapted, it is likely to be close to an optimal state, and hence the estimated reward will be an appropriate approximation of the real one.

The value of $\hat{r}$ compresses all the information on future probabilities stored within the value function. The estimated reward takes into account not only the current state $s$, but also the underlying information of $v^*(s)$, which defines how beneficial a state is based on future states $s'$ that can be reached from $s$.
Given that $\hat{r}$ is an estimated reward, it can be replaced with the definition of the reward provided in eq.~(\ref{eq:reward}). The resulting expression has one unknown, the $\widehat{rmse}$ estimated error:

\begin{equation}
     \hat r = \left(\frac{p_{\max}}{p}\right)^\alpha  \exp\left({-\frac{\widehat{rmse}^2}{2 \sigma^2}}\right) = \frac{(1-\gamma)}{\gamma^2} \, \left( v^*(s) - \bar{r} - \gamma \bar{r}' \right),
     \label{eq:Reward_equivalence_error_estimation}
\end{equation}
and the estimation of the error becomes:

\begin{equation}
    \widehat{rmse} = \sqrt{-2 \sigma^2 \log{\left( \frac{v^*(s) - \bar{r} - \gamma \bar{r}'}{\gamma^2 v_{\max, p}} \right)}} \, ,
    \label{eq:rmse_error_estimation}
\end{equation}
where $v_{\max, p}$ is equivalent to the maximum \textit{v-value} that can be obtained given a polynomial order $p$:

\begin{equation}
    v_{\max, p} = \frac{1}{1 - \gamma} \left(\frac{p_{\max}}{p}\right)^\alpha.
    \label{eq:Vmax_error_estimation}
\end{equation}

One may notice that, although we estimate the reward $\hat{r}$ in two steps in the future, the polynomial order used in eq.~(\ref{eq:Vmax_error_estimation}) is the polynomial order $p$ of the present state $s$. The reason behind this decision is that we want to estimate the error that we potentially have based on the measurements from the environment that can be collected in real time. Therefore, the use of the same polynomial order as is used in the environment when computing the state leads to more accurate estimations.

It is important to note that this error estimation is the error that the RL agent believes to be real, but the only feedback received by the agent comes from the reward function. 
In addition, the average rewards $\bar{r}$ and $\bar{r}'$ are calculated based on the probabilities defined during training. Hence, it may be impossible to obtain an estimate for some specific situations. 
However, even if an estimate of the error cannot be obtained, the solution has a true meaning behind it.
\begin{enumerate}
    \item $v^*(s) - \bar{r} - \gamma \bar{r}' < 0$: The agent believes that the error is much higher than expected. This situation is unusual, but it may happen if the polynomial order used to solve the problem is very small compared to the optimal polynomial order (e.g., $p=2$ and $p^*>4$).
    \item $v^*(s) - \bar{r} - \gamma \bar{r}' > 1$: The agent believes that the error is very small compared to the standard deviation of the reward $\sigma$ (see eq.~(\ref{eq:reward}) for details). This situation may occur if the polynomial order is high in relation to the optimum polynomial order (e.g., $p=4$ and $p^*\leq2$).
\end{enumerate}

In both cases, even if a numerical value for the error cannot be obtained, it is possible to extract a valuable information. Furthermore, we could assign a numerical value for each case: a very high error could be used for the first case, $\widehat{rmse}=10 \sigma$, and a very small error could be used for the second case, $\widehat{rmse}=10^{-3} \sigma$.

Finally, we have to link this error, defined for dimensionless variables, with an actual error estimate for a variable $q$ with dimensions. 
The dependency between the normalized solution, $y$, and the real solution, $q$, at the Gauss nodes is given by the following transformation:
\begin{equation}
    y_i = 2 \frac{q_i - q_{\min}}{q_{\max} - q_{\min}} - 1,
\end{equation}
where $q_{\min} = \min (q_i)$ and $q_{\max} = \max (q_i)$ for $i$ in each Gauss node. Applying this transformation to the original definition of the normalized $rmse$ from eq.~(\ref{eq:rmse_reward}), we obtain:
\begin{equation}
    rmse = \sqrt{\frac{\sum_{i=1}^{N_e} (y_i - y^*_i)^2}{N_e}} = 
\frac{2}{q_{\max} - q_{\min}} \sqrt{\frac{\sum_{i=1}^{N_e} (q_i - q^*_i)^2}{N_e}} = \frac{2}{q_{\max} - q_{\min}} rmse_{q},
    \label{eq:rmse_scaled_error_estimation}
\end{equation}
with $rmse_q$ as the \textit{root mean squared error} scaled for the specific variable $q$. Combining eq.~(\ref{eq:rmse_error_estimation}) and eq.~(\ref{eq:rmse_scaled_error_estimation}), we obtain the final expression for the estimated error of the variable $q$:

\begin{equation}
    \widehat{rmse_q} = \frac{q_{\max} - q_{\min}}{2} \sqrt{-2 \sigma^2 \log{\left( \frac{v^*(s) - \bar{r} - \gamma \bar{r}'}{\gamma^2 v_{\max, p}} \right)}} \, .
\end{equation}
This error estimation can be computed beforehand during the training process, with the exception of the scaling factor for the specific variable $q$. Therefore, the computational cost of this estimator at run time, in a real simulation, is negligible, resulting in an inexpensive error estimator that can be applied to any problem or variable of interest. The effectiveness of the error estimator will be tested in the results section.

\subsubsection{Error estimation extrapolation for 3D problems}\label{subsubsec:Error_estimation_extrapolation_3D}
As we can use the RL p-adaptation agent for 3D problems, though it was trained with 1D cases (see Section \ref{subsubsec:State_extrapolation_3D}), we also want to apply a similar approach to use the error estimation capability for 3D problems. In a similar way as we did for the state, inside each element we compute the estimated error for each row of Gauss nodes that aligns with each local axis $x$, $y$ and $z$, and we average the results among every row of nodes, resulting in three different values of the error, one for each direction. Then, depending on how these three values are combined, we may obtain different error estimations for 3D problems. In particular, we provide two possible alternatives:

\begin{enumerate}
    \item Average error: $\widehat{rmse_q} =\frac{\widehat{rmse_{q,x}} + \widehat{rmse_{q,y}} + \widehat{rmse_{q,z}}}{3}$. This approach provides, in general, a good estimate of the error, which is of the same order of magnitude as the real error for the selected variable $q$.
    \item Maximum error: $\widehat{rmse_q} =\max{\left( \widehat{rmse_{q,x}}, \, \widehat{rmse_{q,y}}, \, \widehat{rmse_{q,z}}\right)}$. This estimate of the error tends to slightly overestimate the real error. Furthermore, it could be used as a maximum value that should not be reached in the real solution. 
\end{enumerate}

The choice of one alternative or another to accurately estimate the error depends on each specific problem. Both methods will be compared in the results section. 

\subsection{Implementation of p-adaptation in DGSEM solvers}\label{subsec:Implementation_p_adaptation}
The entire methodology that has been explained in previous sections has been implemented in the open-source CFD solver HORSES3D \cite{horses3d_paper}, developed at ETSIAE-UPM (The School of Aeronautics in Madrid). This code provides all the necessary elements to apply the novel p-adaptation strategy to complex 3D problems in the field of fluid mechanics. However, it is important to discriminate between the RL p-adaptation sensor, which is the whole algorithm described so far, and all the minor additional details that must be taken into account in a real CFD solver to make all the components work in a soft and efficient way.

First, we have to consider that the p-adaptation process in a CFD solver is highly time-consuming. Although the time required to evaluate the RL agent is negligible, as it is based on a look-up table, the whole adaptation process requires significant memory reallocation and, in some cases, dynamic load balancing in MPI simulations. Consequently, adapting the mesh at every iteration is impractical and computationally prohibitive. Therefore, we should adapt the mesh every $100-10000$ iterations, or when we have detected that it is really necessary to perform an adaptation (e.g., based on the RL-based error estimation). Furthermore, a sharp increase or decrease in the polynomial order may lead to a numerical divergence. For that reason, each time the mesh is adapted, only one action is taken in each element; that is, the polynomial order can be increased or decreased by 1 unit.

Furthermore, when a new simulation is started, the original mesh is usually not p-adapted. A common practice is to initialize all elements with a uniform low polynomial (e.g., $p=2$) to reduce computational costs during the first stage of the simulation, which is, in general, a transient phase of little interest. However, this is not a fixed rule: the simulation can also be started from an uniform high-order polynomial, or even from another p-adapted mesh.
Later, the p-adaptation algorithm may take control over the simulation and decide which polynomial order is desirable in each region.

Additionally, a common problem that may arise in a real implementation is related to the state (set of scaled values at the Gauss nodes) that is provided to the RL agent to perform the adaptation, which is always scaled between $[-1, 1]$, as previously explained in Section \ref{subsubsec:State}. Although this step is completely necessary to obtain an RL agent as general as possible, it may be problematic if the values at the Gauss nodes that allow us to create the state are extremely small. In those cases, the variable of interest that has been measured at the nodes can be considered to be constant along that row of points. However, after the scaling process, the state is no longer constant and the RL agent may perceive solution gradients that are actually negligible. To solve this issue, when the spatial change inside one element in the non-dimensional variable used to compute the state, $q$, is below a small tolerance, $\Delta q < 5\cdot 10^{-3}$, 
we set the state to 0. By doing so, we can provide this additional information to the RL agent, obtaining a significant improvement in performance. In addition, as the variable of interest $q$ is, in our solver, always non-dimensional, this tolerance is general enough to be used in a wide variety of problems without modifying its default value. 
The default tolerance has been successfully used for all the problems reported in this work and for the different reported variables. 
Finally, note that the tolerance is independent of the training process, and hence the RL agent, previously trained, is still valid for any value of the aforementioned tolerance.

Another implementation decision is that the polynomial adaptation is bounded between $p_{\min}=2$ and $p_{\max}=6$.
This range is appropriate for a wide variety of problems; however, in large simulations, it could be desirable to decrease the polynomial below $p_{\min}=2$ in some regions, reducing the computational cost. To do so, we consider that if nothing relevant is happening inside one element; that is, the maximum change in the variable of interest is below the aforementioned tolerance, $\Delta q$, then we can decrease the polynomial order to $p=1$. Once we detect that the variations of the variable are above the tolerance, we switch to $p=2$ and the RL agent is in charge of the adaptation again.

In addition, it is important to decide which variables of interest (there can be more than one) should be considered to perform the adaptation process. For all the following problems, the state will be computed based on three non-dimensional components of momentum:
\begin{equation}
    q_1 = \rho u \qquad q_2 = \rho v \qquad q_3 = \rho w
    \label{eq:Variables_of_interest_results}
\end{equation}
where $\rho$ is the density and $u$, $v$ and $w$ as the velocity components. When selecting the polynomial adaptation, the RL agent will choose one action for each variable $q_1$, $q_2$ and $q_3$, and we select the action that is the most restrictive among the three.

\section{Results}\label{sec:Results}
To show the applicability and flexibility of the proposed methodology, different problems with increasing complexity are included. For all the cases, we show the effectiveness of the RL-based p-adaptation, leading to accurate solutions and a significant reduction of the computational time, when compared to using homogeneous polynomials and other adaptation criteria. Note that for those problems in which the RL-based error estimation methodology is used, we consider the \emph{average error} as default (see Section \ref{subsubsec:Error_estimation_extrapolation_3D}), unless explicitly stated otherwise.

Finally, for all the following cases, an explicit Runge-Kutta 3 temporal scheme is used. Furthermore, to solve the Riemman problem at the interfaces between elements, a Roe scheme is used for the convective fluxes, and Bassi Rebay 1 for the viscous fluxes.

\subsection{Error estimation for an Euler flow around a cylinder}\label{subsec:Results_cylinder_Euler}
\begin{figure}[htpb]
\centering
  \begin{subfigure}[b]{0.37\textwidth}
    \includegraphics[width=\textwidth]{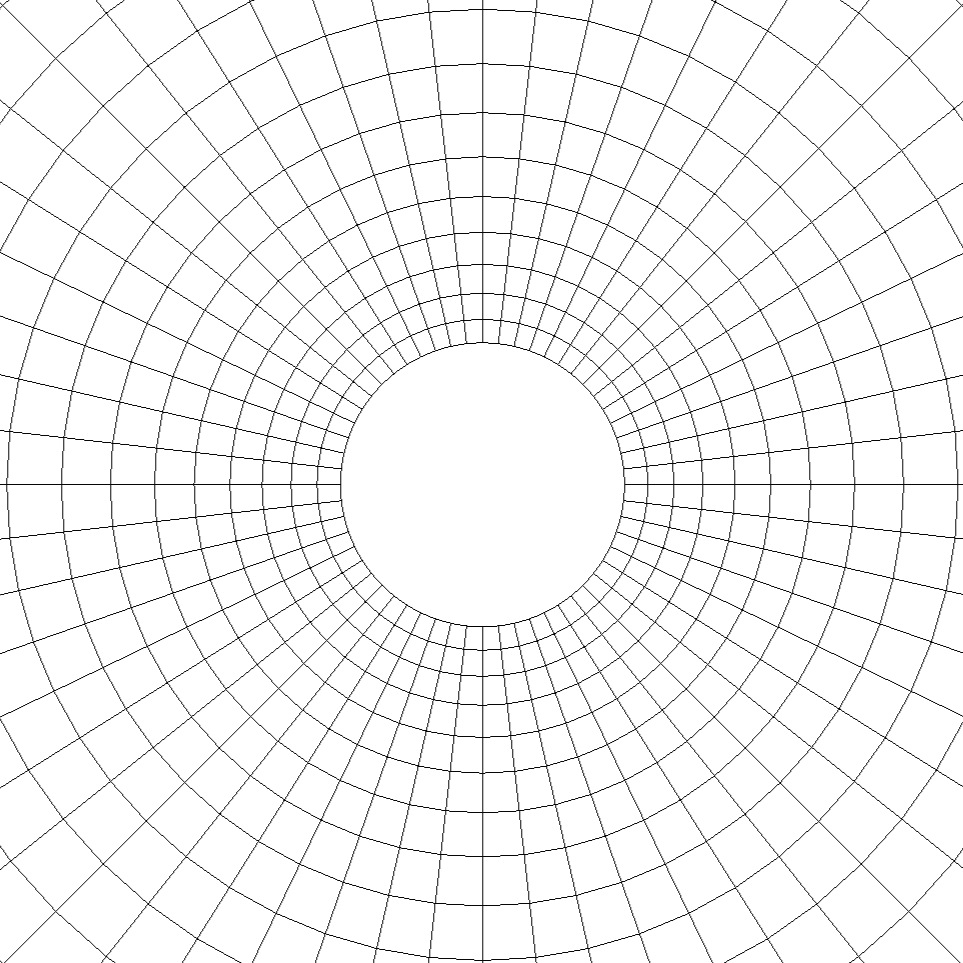}
    \caption{Detail view.}
    \label{fig:Cylinder_mesh_detail_view_results}
  \end{subfigure}
  \quad
  \begin{subfigure}[b]{0.37\textwidth}
    \includegraphics[width=\textwidth]{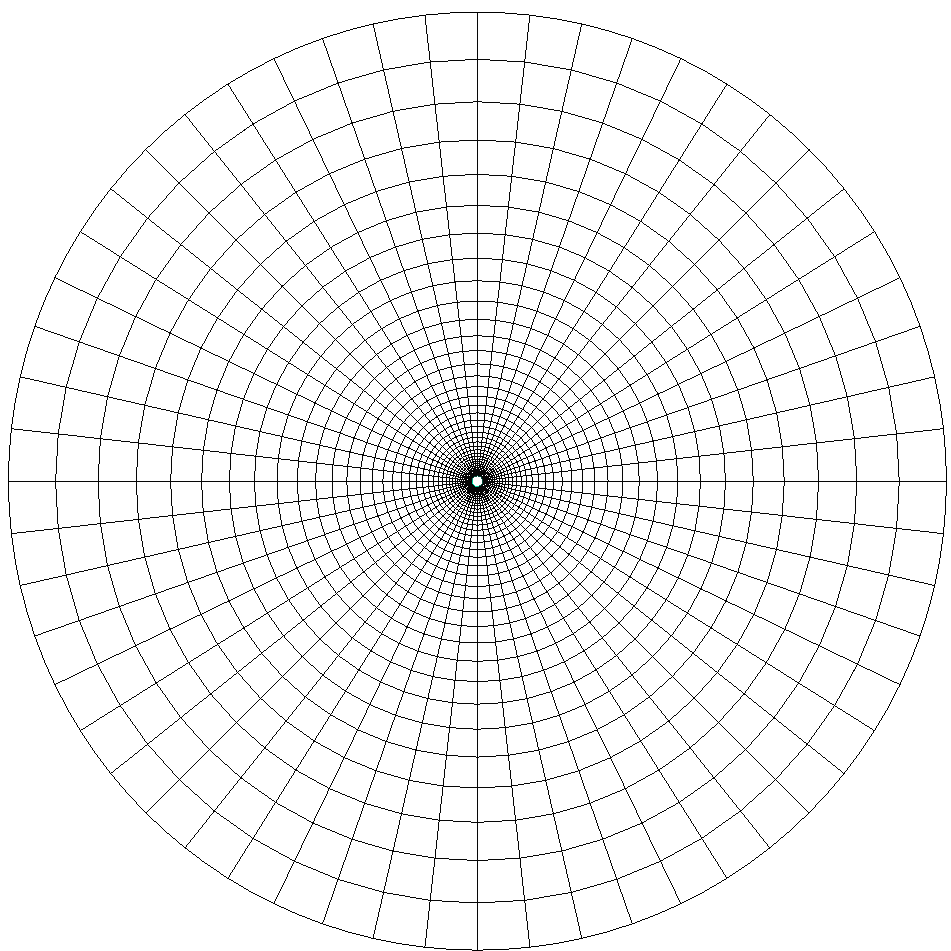}
    \caption{Full view.}
    \label{fig:Cylinder_mesh_full_view_results}
  \end{subfigure}
  
\caption{Representation of the mesh used for the simulation of the Euler flow around a cylinder. 
}
\label{fig:Cylinder_mesh_results}
\end{figure}

The first problem to be tackled is the flow field around a cylinder at $M=0.1$ solved using the Euler equations; that is, the viscosity is not considered in this first problem. This simple solution provides a way to test our approach before addressing the Navier-Stokes equations.
The mesh used for this problem is shown in Figure~\ref{fig:Cylinder_mesh_results}. It only has 6664 elements that have been extruded one unit in the $z$ axis. This mesh has been created slightly coarse on purpose, and it is completely symmetric,  without providing any prior knowledge regarding the flow field. In this way, the p-adaptation algorithm will have to increase the polynomial order to obtain an accurate solution in those regions where strong gradients are present.

\begin{figure}[htpb]
\centering
  \begin{subfigure}[b]{0.36\textwidth}
    \includegraphics[width=\textwidth]{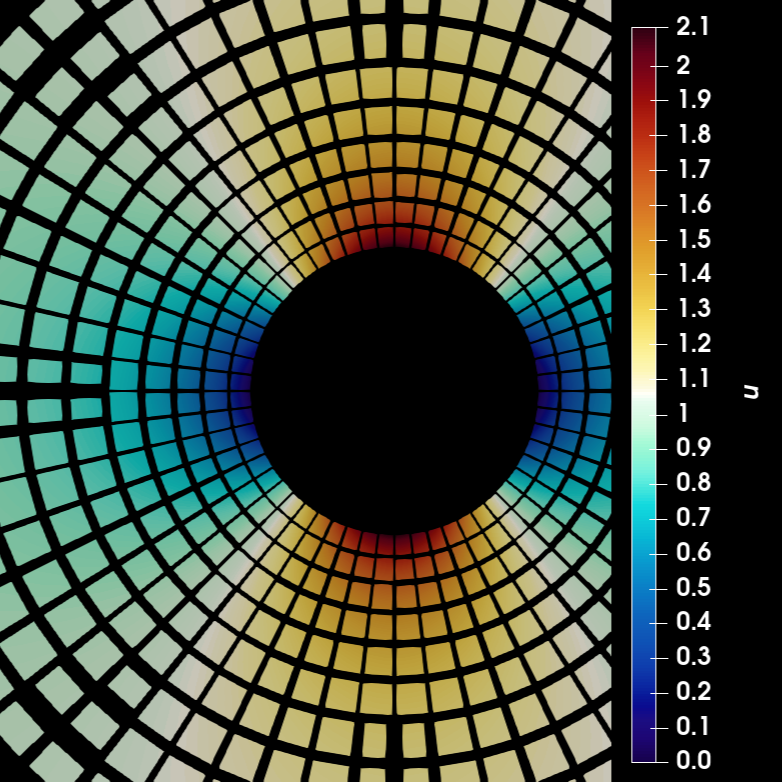}
    \caption{Non-dimensional $u$ velocity.}
    \label{fig:Cylinder_Euler_u_velocity}
  \end{subfigure}
  \quad
  \begin{subfigure}[b]{0.36\textwidth}
    \includegraphics[width=\textwidth]{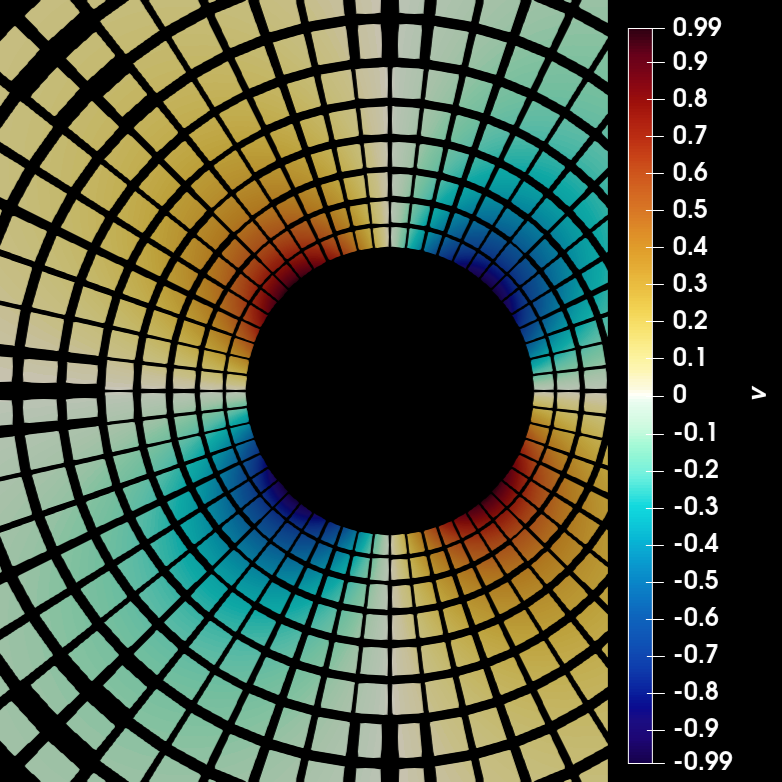}
    \caption{Non-dimensional $v$ velocity.}
    \label{fig:Cylinder_Euler_v_velocity}
  \end{subfigure}
  
\caption{Steady-state velocity field following a) the $x$ axis and b) the $y$ axis, for the Euler flow around a cylinder at $M=0.1$.}
\label{fig:Cylinder_Euler_velocity_results}
\end{figure}

\begin{figure}[h]
    \centering
    \includegraphics[width=0.5\textwidth]{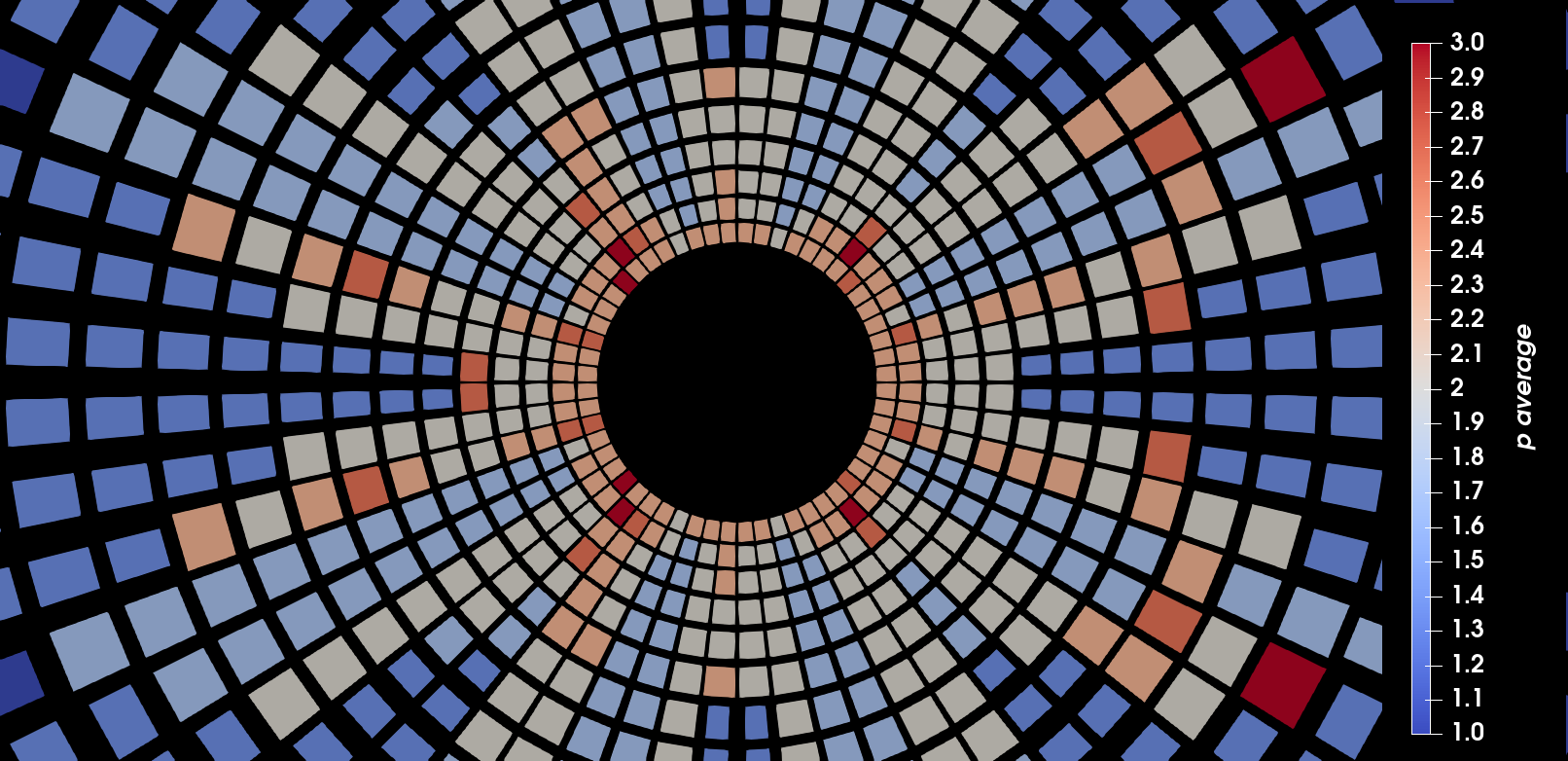}
    \caption{Average polynomial order in each element of the mesh after the p-adaptation process in steady-state conditions for the Euler flow around a cylinder at $M=0.1$.}
    \label{fig:Cylinder_Euler_pavg_results}
\end{figure}

Once steady-state conditions are achieved, the velocity field and the average polynomial order (average among the three anisotropic polynomials) selected in each element by the reinforcement learning agent are represented in Figures~\ref{fig:Cylinder_Euler_velocity_results} and \ref{fig:Cylinder_Euler_pavg_results}, respectively. The high-order polynomials are gathered very close to the cylinder, where the velocity gradients are maximum. In those regions where the flow is far enough from the cylinder, the velocity variations are small and the polynomial order is reduced to save computational costs.

To test the accuracy of the error estimator, a comparison has been performed for different uniform polynomial orders and the RL p-adaptation.
Although the flow around the cylinder is almost a potential flow, the compressible solution for the Euler equations at $M=0.1$ is not completely potential. Therefore, instead of the analytical solution of the potential flow, we consider as a reference solution a uniform polynomial order $p=5$ for the three axes in every element of the mesh. We consider this reference solution to be accurate enough, as the maximum polynomial order selected by the RL agent for this simple test case never exceeds $p=4$ on a single axis. Therefore, the reference solution will always be more accurate than the p-adapted solution.

\begin{figure}[htpb]
\centering
  \begin{subfigure}[b]{0.38\textwidth}
    \includegraphics[width=\textwidth]{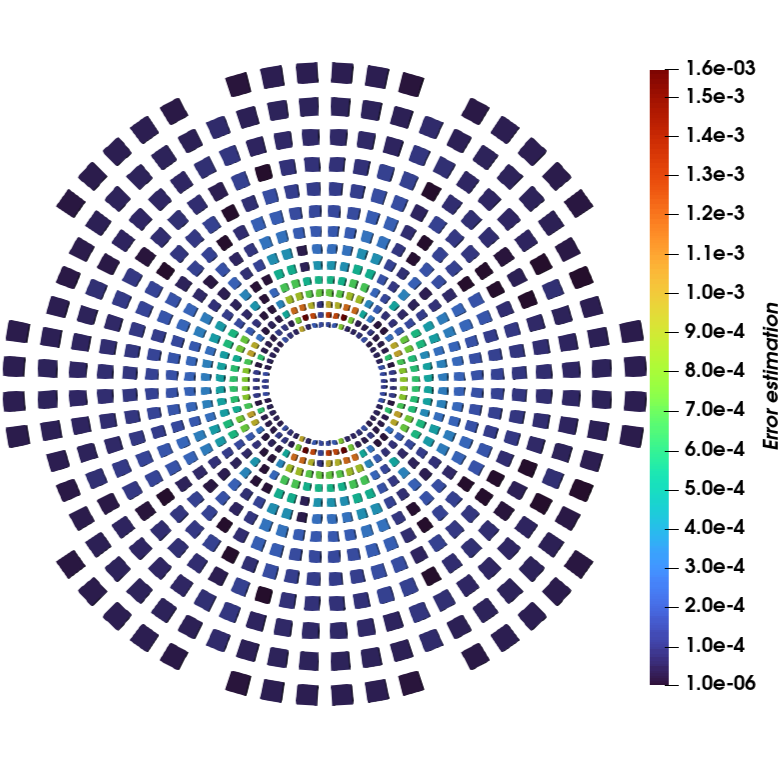}
    \caption{Average error estimation for the RL p-adapted solution.}
    \label{fig:Cylinder_Euler_error_estimation_pAdapted}
  \end{subfigure}
  \quad
  \begin{subfigure}[b]{0.38\textwidth}
    \includegraphics[width=\textwidth]{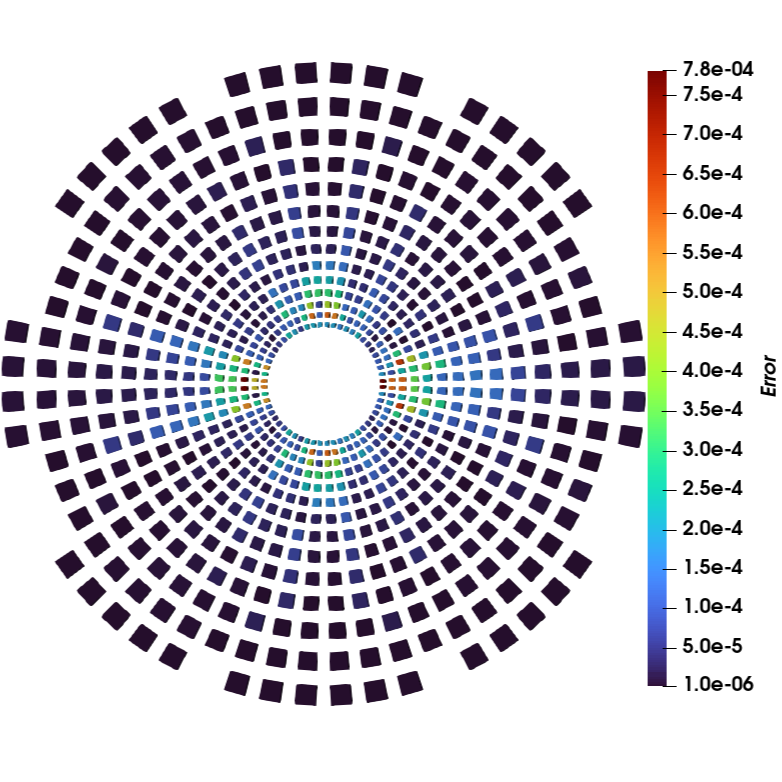}
    \caption{Error between the p-adapted and the reference solution.}
    \label{fig:Cylinder_Euler_real_error_pAdapted}
  \end{subfigure}
  
\caption{Comparison of the error estimated by the RL agent and the real error for the non-dimensional $u$ velocity under steady-state conditions for the Euler flow around a cylinder at $M=0.1$. The error was computed by comparing the p-adapted solution with a uniform $p=5$ reference solution. Only those elements of the mesh with an error higher than $10^{-6}$ are represented.}
\label{fig:Cylinder_Euler_error_pAdapted}
\end{figure}

The average error estimation and the real error between the reference and the RL p-adapted solution, for the non-dimensional $u$ velocity, are shown in Figure~\ref{fig:Cylinder_Euler_error_pAdapted}.  
The error estimator is capable of providing the trend of the error, highlighting those regions where the real error is higher. 
Of course, the exact value of the error cannot be estimated, given that the RL agent is exactly the same for every problem and PDE and no information regarding the equations to be solved, the mesh or the boundary conditions is provided. Even with this little knowledge, the RL agent shows an excellent behavior and versatility, and a fairly good estimation can be obtained. Furthermore, the computational cost of the estimation is negligible, as the error estimator is (almost) completely defined during the training of the RL agent.
A summary of the degrees of freedom, the computational cost, and the maximum error (real and estimated) in the whole domain for each solution is represented in Table~\ref{tab:Cylinder_Euler_summary_results}. First, we can see how the real error decreases as the polynomial order increases, which is the expected behavior. In addition, the order of magnitude of the estimation matches the order of magnitude of the real error in all cases. Although some estimations have slightly underestimated the error, that can be attributed to the fact that we are computing the average error estimation (see Section \ref{subsubsec:Error_estimation_extrapolation_3D}).
Furthermore, the solution obtained with the p-adaptation algorithm has a maximum error close to $10^{-3}$, which is small given that the maximum non-dimensional velocity is $u_{\max}=2.1$. In addition, the computational cost is even lower than the uniform $p=2$ solution, while being more accurate. The p-adapted solution is the fastest and the accuracy is similar to a uniform $p=3$ solution. Of course, by changing the parameters of the reward function (see eq.~(\ref{eq:reward})) one can obtain a different agent whose priority is to reduce the error even more, even if the cost is increased. 
In the following sections, we show more complex problems in which we can take advantage of the maximum potential of the p-adaptation strategy.

\begin{table}[htpb]
    \centering
    \begin{tabular}{|C{3.0cm} C{1.5cm} C{3.5cm} C{2.25cm} C{3.0cm}|}
    \hline
    \textbf{Polynomial order} & \textbf{DOFs} & \textbf{Computational cost (s)} & \textbf{Real error} & \textbf{Error estimation} \\ \hline \hline 
    $p=2$ & $58968$ & $252$ & $8.1 \cdot 10^{-3}$ & $2.1 \cdot 10^{-3}$ \\ \hline
    $p=3$ & $139776$ & $431$ & $6.7 \cdot 10^{-4}$ & $2.8 \cdot 10^{-4}$  \\ \hline
    $p=4$ & $273000$ & $737$ & $6.5 \cdot 10^{-5}$ & $3.0 \cdot 10^{-5}$  \\ \hline
    $p=5$ & $471744$ & $1181$ & $0.0$ (reference) & $0.0$ (reference)  \\ \hline
    \textbf{p-adapted} & $\mathbf{27708}$ & $\mathbf{197}$ & $\mathbf{7.8 \cdot 10^{-4}}$ & $\mathbf{1.6 \cdot 10^{-3}}$  \\ \hline
    \end{tabular}
    \caption{Comparison between the DOFs, the computational cost, the maximum value of the real error, in relation to the reference solution $p=5$, and the maximum value of the estimated average error for the non-dimensional $u$ velocity, for several solutions obtained with different polynomial orders. All solutions are defined in steady-state conditions for the Euler flow around a cylinder at $M=0.1$. The simulations were run in a node Intel(R) Xeon(R) Gold 6248 CPU @ 2.50GHz, using 40 MPI tasks.}
    \label{tab:Cylinder_Euler_summary_results}
\end{table}

\FloatBarrier

\subsection{Comparison of p-adaptation approaches and accuracy in lift and drag for a cylinder at Reynolds 100}\label{subsec:Results_cylinder_Re100}
Once we have tested that the performance of the p-adaptation strategy is appropriate and that we can trust, in general, the error estimation, we increase the complexity of the problem to validate different aspects of the RL p-adaptation approach applied to the Navier-Stokes equations.
In particular, to improve the quality of the designed test cases and quantitatively measure the performance of the p-adaptation strategy, we compare our novel approach with a p-adaptation sensor based on the original sensor developed by Persson and Peraire \cite{Persson2006}. This sensor compares the solution inside each element using a different number of high-order modes, providing a spectral decay indicator that can be used as an estimation of the smoothness of the solution. Although it was first developed for shock-capturing purposes, based on this sensor some p-adaptation approaches have been successfully used in DG solvers \cite{Naddei2018}. However, the available strategies based on this sensor are very limited, as they have only been used to increase the polynomial order (not to decrease it), and they cannot handle anisotropic adaptation. We have made slight improvements in the sensor-based approaches found in the literature by incorporating these two additional functionalities. In this way, we ensure a fair comparison between our novel method and the state-of-the-art sensor.

\begin{figure}[h]
\centering
  \begin{subfigure}[b]{0.37\textwidth}
    \includegraphics[width=\textwidth]{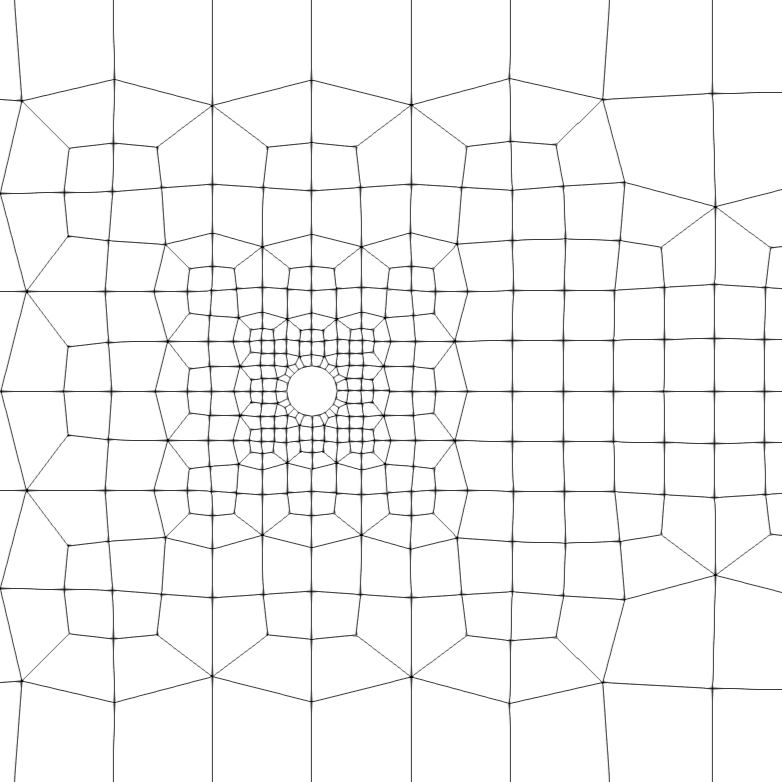}
    \caption{Detail view.}
    \label{fig:Cylinder_Re100_mesh_detail_view_results}
  \end{subfigure}
  \quad
  \begin{subfigure}[b]{0.37\textwidth}
    \includegraphics[width=\textwidth]{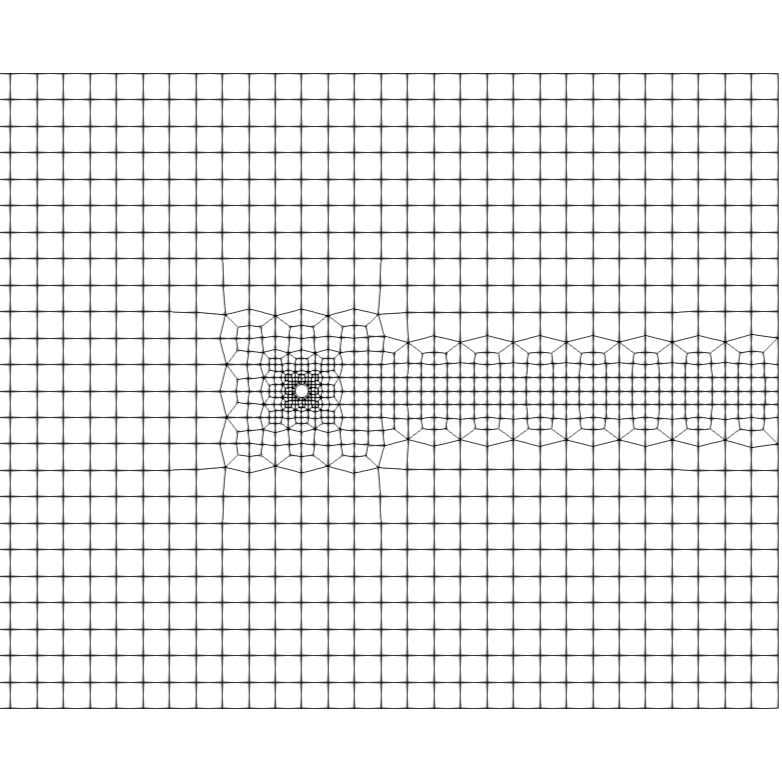}
    \caption{Full view.}
    \label{fig:Cylinder_Re100_mesh_full_view_results}
  \end{subfigure}
  
\caption{Representation of the mesh used for the simulation of the flow around a cylinder at $Re=100$ and $M=0.15$.}
\label{fig:Cylinder_Re100_mesh_results}
\end{figure}

The problem to be solved is the laminar, but unsteady, flow at $Re=100$ around a cylinder at $M=0.15$. The mesh used for this problem is shown in Figure~\ref{fig:Cylinder_Re100_mesh_results}. This coarse mesh has only 1282 elements on purpose, again to highlight the ability of the RL agent to increase the polynomial where required. However, the elements are slightly smaller downstream to capture the vortex shedding.

\begin{figure}[ht]
\centering
\begin{subfigure}[b]{\textwidth}
    \includegraphics[width=\textwidth]{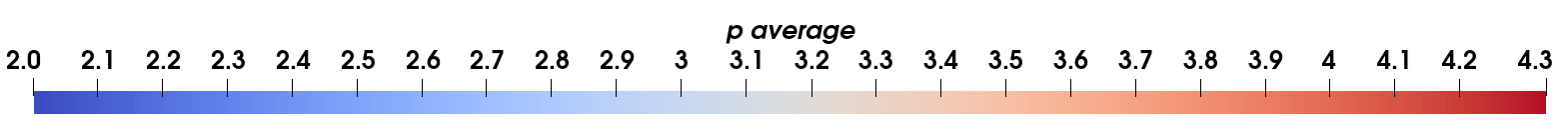}
    \label{fig:Cylinder_Re100_pavg_legend_results}
  \end{subfigure}
  \quad
  \begin{subfigure}[b]{0.48\textwidth}
    \includegraphics[width=\textwidth]{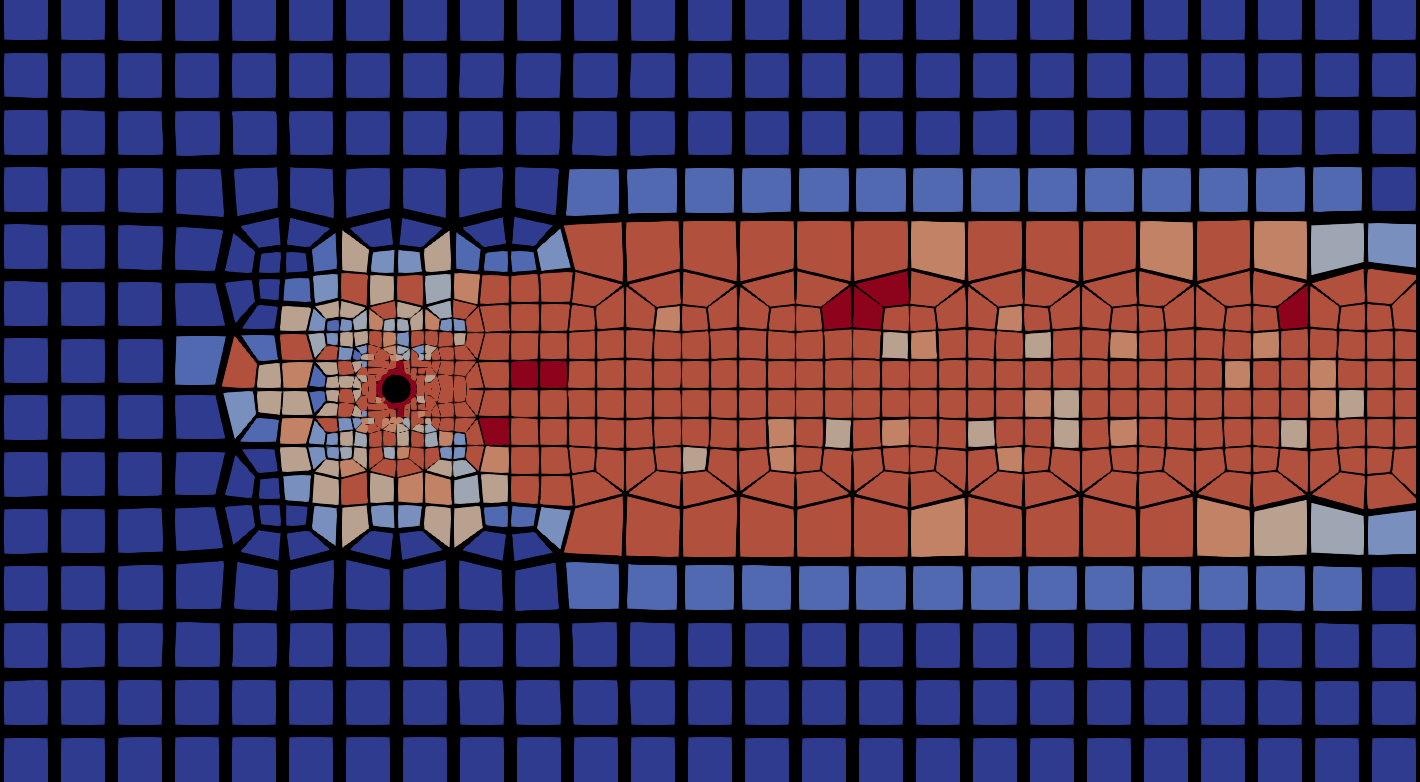}
    \caption{RL p-adaptation.}
    \label{fig:Cylinder_Re100_pavg_RL_results}
  \end{subfigure}
  \quad
  \begin{subfigure}[b]{0.48\textwidth}
    \includegraphics[width=\textwidth]{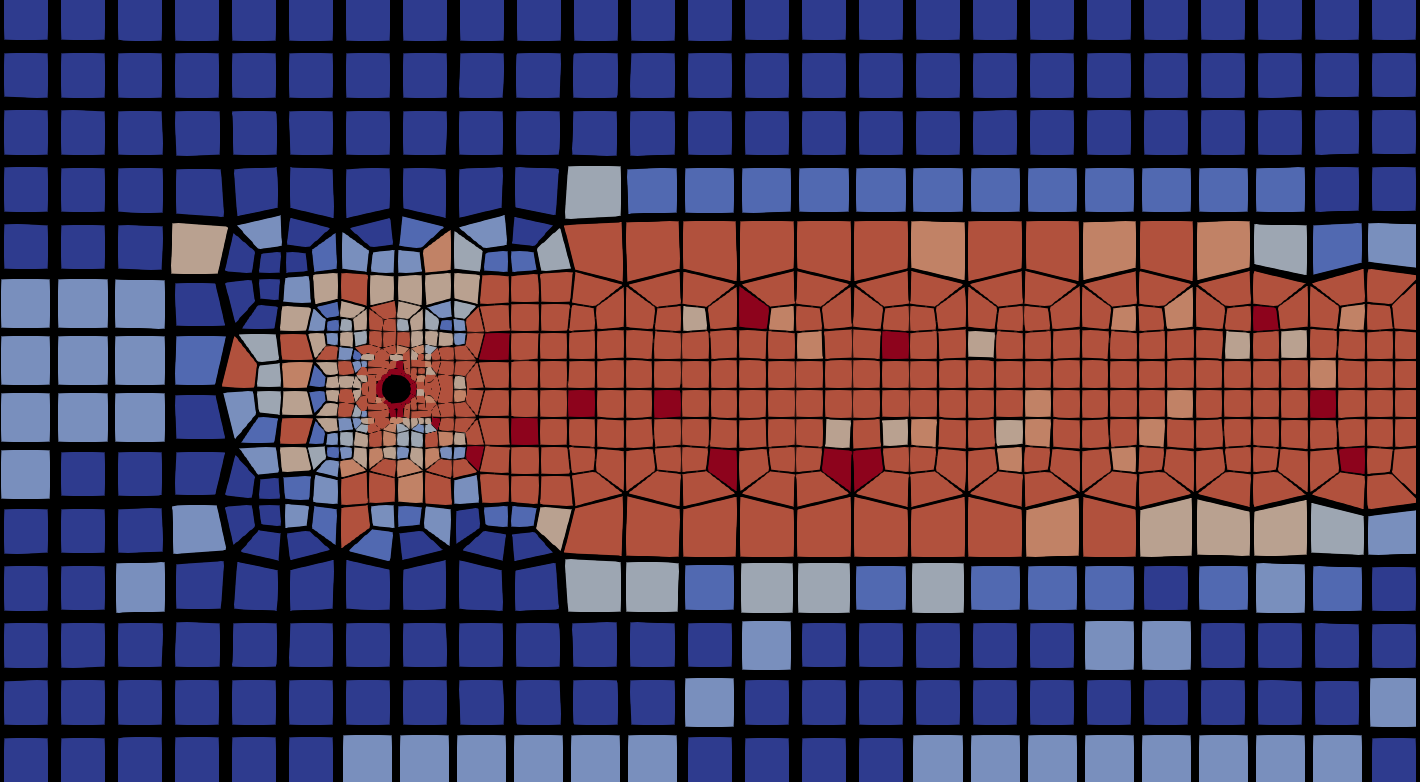}
    \caption{Persson and Peraire sensor (best solution after fine tuning parameters)
    }
    \label{fig:Cylinder_Re100_pavg_modal_results}
  \end{subfigure}
  
\caption{Comparison of p-adaptation between a) the RL agent strategy and b) the improved Persson and Peraire sensor for a simulation of the flow around a cylinder at $Re=100$ and $M=0.15$. A snapshot of the average polynomial order in each element of the mesh is illustrated.}
\label{fig:Cylinder_Re100_pavg_results}
\end{figure}

For this problem, a dynamic p-adaptation is required, as the solution is unsteady. A snapshot of the average polynomial order in each element using both approaches (RL and Persson and Peraire-based sensor), once the flow field is stabilized, is represented in Figure~\ref{fig:Cylinder_Re100_pavg_results}. At first glance, both approaches are similar, as they increase the polynomial near the cylinder and in the wake, where the vortex shedding is present. However, the adaptation performed by our RL agent is more efficient, as it increases the polynomial order only where required, while the Persson and Peraire-based approach increases the polynomial order in some regions upstream and also downstream far from the wake, where the flow field is almost steady.

It is very important to mention that, on the one hand, the Persson and Peraire-based approach relies on a threshold to decide when to increase and decrease the polynomial order, and this threshold is problem-dependent. It took several attempts of trial and error, using different values of the aforementioned threshold, to achieve the polynomial adaptation shown in Figure~\ref{fig:Cylinder_Re100_pavg_modal_results}. 
On the other hand, the parameters of our RL agent are decided beforehand, before the training, and their dependency with each specific problem is small. Therefore, our agent does not require problem-specific tuning,  
which is an important advantage of the proposed methodology, when compared to other state-of-the-art p-adaptation strategies.

\begin{figure}[ht]
\centering
  \begin{subfigure}[b]{0.48\textwidth}
    \includegraphics[width=\textwidth]{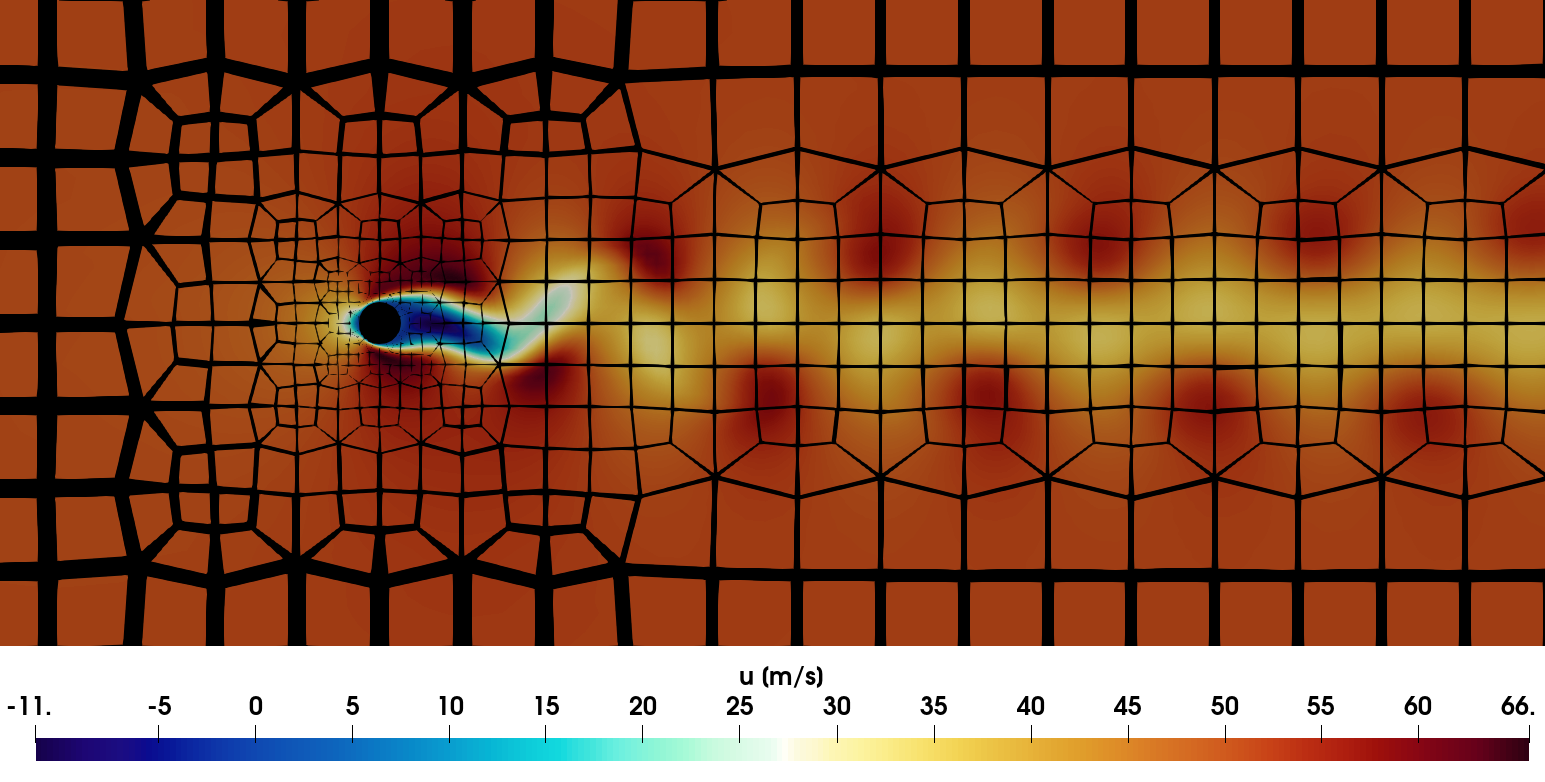}
    \caption{$u$ velocity.}
    \label{fig:Cylinder_Re100_u_velocity_results}
  \end{subfigure}
  \quad
  \begin{subfigure}[b]{0.48\textwidth}
    \includegraphics[width=\textwidth]{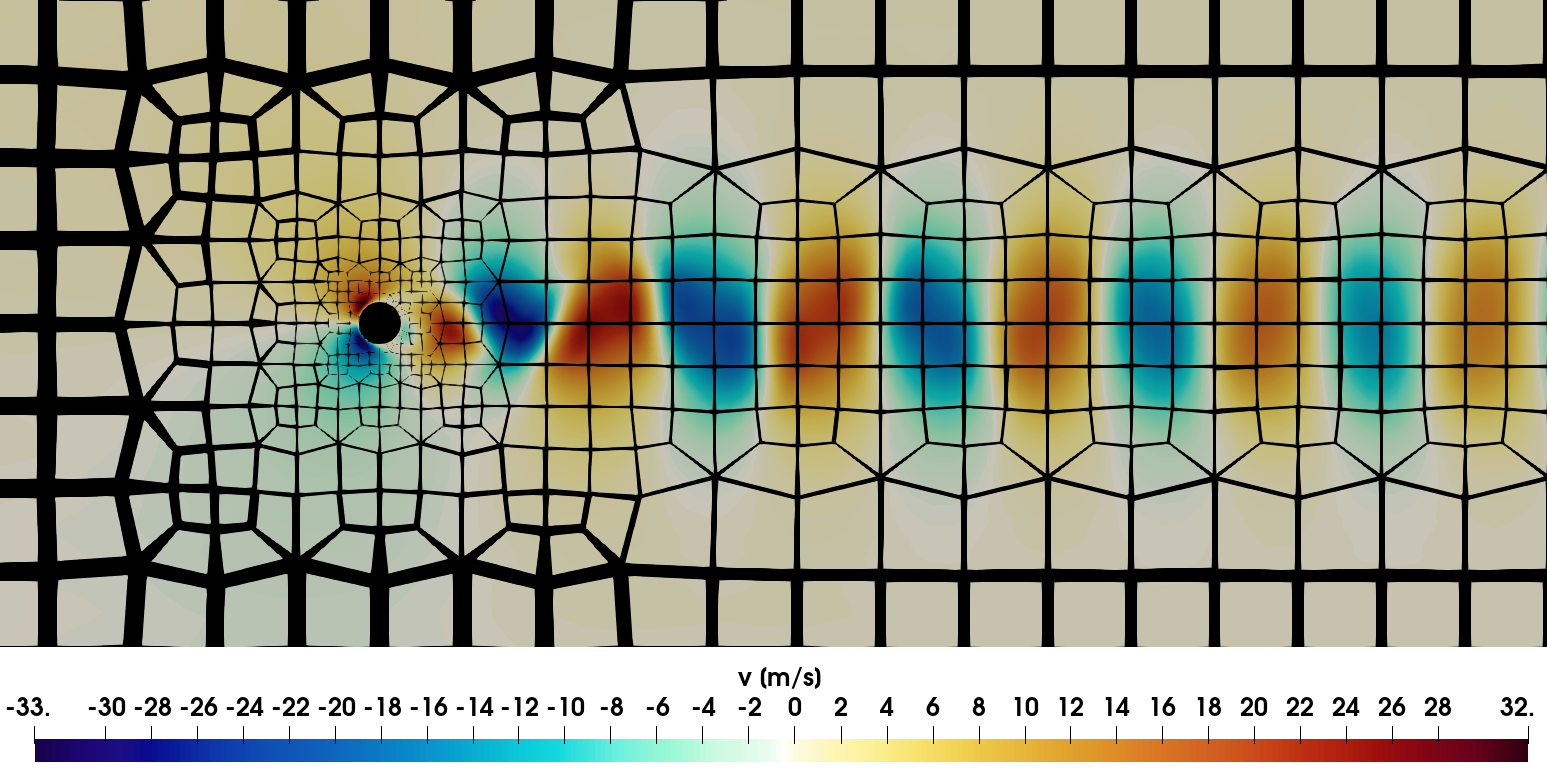}
    \caption{$v$ velocity.}
    \label{fig:Cylinder_Re100_v_velocity_results}
  \end{subfigure}
  
\caption{Snapshot of the velocity field for the simulation of the flow around a cylinder at $Re=100$ and $M=0.15$ using the RL p-adaptation agent. Both, a) the horizontal velocity and b) the vertical velocity are shown.}
\label{fig:Cylinder_Re100_velocity_results}
\end{figure}

For completeness, a snapshot of the velocity field is represented in Figure~\ref{fig:Cylinder_Re100_velocity_results}. The vortex shedding is well captured, as expected. 
Finally, we report the accuracy of the p-adaptation approach in Table~\ref{tab:Cylinder_Re100_lift_drag_results}. A comparison between the average drag, the maximum lift and the Strouhal number is considered. The results obtained with the proposed methodology are very close to those of other sources from the scientific literature.

\begin{table}[htpb]
    \centering
    \begin{tabular}{|C{4.0cm} C{2.0cm} C{2.0cm} C{2.0cm} C{2.0cm}|}
    \hline
    \textbf{Case} & $\overline{C_{d}}$ & $C_{l, \max}$ & $St$ & DOFs\\ \hline \hline
    Braza et al. \cite{braza1986numerical} & $1.28$ & $0.29$ & $0.160$ & $13530$ \\ \hline
    Talley et al. \cite{talley2001experimental} & $1.34$ & $0.33$ & $0.160$ & $42000$  \\ \hline
    Shiels et al. \cite{shiels2001flow} & $1.33$ & $0.30$ & $0.167$  & Unknown\\ \hline
    Gsell et al. \cite{gsell2020multigrid} & $1.41$ & $0.34$ & $0.170$ & Unknown\\ \hline
    Tlales et al. ($p=4$) \cite{tlales2024machine} & $1.35$ & $0.33$ & $0.164$ & $32050$\\ \hline
    Kou et al. ($p=4$) \cite{kou2023combined} & $1.32$ & $0.31$ & $0.165$ & $818800$ \\ \hline
    Persson and Peraire sensor & $1.33$ & $0.32$ & $0.163$ & $24885$ \\ \hline
    \textbf{RL p-adaptation (ours)} & $\mathbf{1.33}$ & $\mathbf{0.32}$ & $\mathbf{0.163}$ & $\mathbf{23675}$ \\ \hline

    \end{tabular}
    \caption{Comparison of mean drag coefficient, amplitude of the
fluctuating lift coefficient, Strouhal number and degrees of fredom (DOFs) for the flow
past a cylinder at Reynolds number 100. The DOFs are computed in the plane (2D) for all cases; additional DOFs in the third dimension in 3D solvers are not considered to show a fair comparison among all methods.}
    \label{tab:Cylinder_Re100_lift_drag_results}
\end{table}

\FloatBarrier

\subsection{Laminar - turbulent transition for the Taylor Green Vortex problem at
Reynolds 1600}\label{subsec:Results_TGV}
The p-adaptation approach showed good behavior in the previous tests, but the flow field was laminar in all cases. Here, we tackle the laminar-turbulent transition by simulating the \emph{Taylor Green Vortex} problem (TGV). The mesh is a perfect cube divided with a Cartesian grid of $32^3$ elements in the domain $x, y, z \in [-\pi, \pi]$ and periodic conditions on all sides. The initial conditions for the variables of the problem are as follows:

\begin{equation}
\left\{ 
    \begin{array}{l} 
\rho = \rho_0,\\
u = V_0 \sin x\cos y \cos z,\\
v = -V_0\cos x \sin y \cos z,\\
w = 0,\\
p = \frac{\rho_0 V_0^2}{\gamma M_0^2} + \frac{\rho_0 V_0^2}{16}(\cos 2x + \cos 2y)(\cos 2z + 2).\\
\end{array}\right. ,
\label{eq:TGV_initial_conditions_results}
\end{equation}
with $\gamma = 1.4$ and $M_0=0.1$. The reported times $t/t_c$ are non-dimensional and have been scaled with the characteristic velocity $V_0$ and the reference length $L=1$.
In the TGV problem, the initial flow is smooth and highly anisotropic. As the simulation progresses beyond $t/t_c > 6$, the flow begins to transition from the original laminar flow to a turbulent anisotropic state. Finally, beyond $t/t_c > 13$, the flow becomes turbulent with isotropic structures. 

\begin{figure}[ht]
\centering
\begin{subfigure}[b]{\textwidth}
    \includegraphics[width=\textwidth]{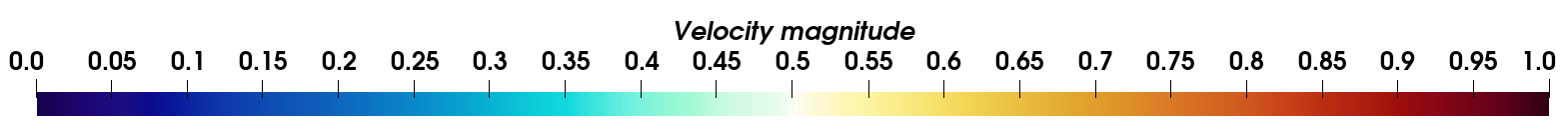}
    \label{fig:TGV_Mach_legend}
  \end{subfigure}
  \quad
  \begin{subfigure}[b]{0.22\textwidth}
    \includegraphics[width=\textwidth]{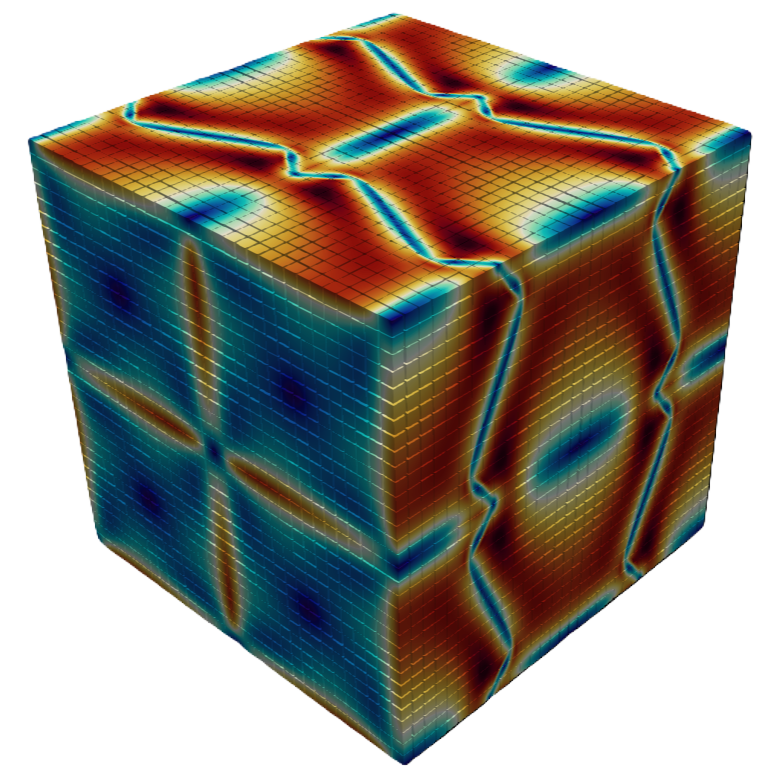}
    \caption{Velocity magnitude at $\displaystyle\frac t{t_c}=4$.}
    \label{fig:TGV_Mach_t4}
  \end{subfigure}
  \quad
  \begin{subfigure}[b]{0.22\textwidth}
    \includegraphics[width=\textwidth]{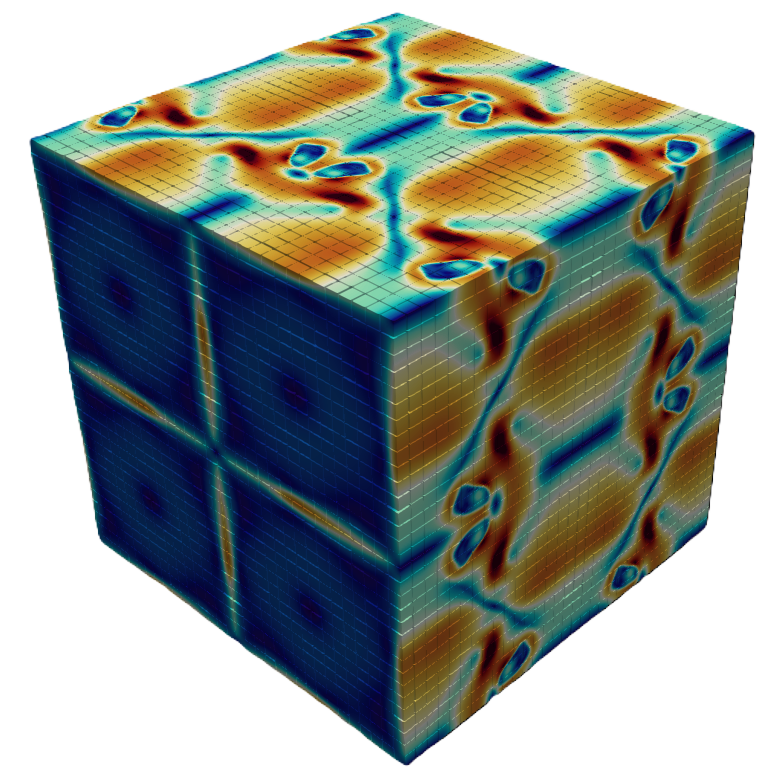}
    \caption{Velocity magnitude at $\displaystyle\frac t{t_c}=8$.}
    \label{fig:TGV_Mach_t8}
  \end{subfigure}
  \quad
  \begin{subfigure}[b]{0.22\textwidth}
    \includegraphics[width=\textwidth]{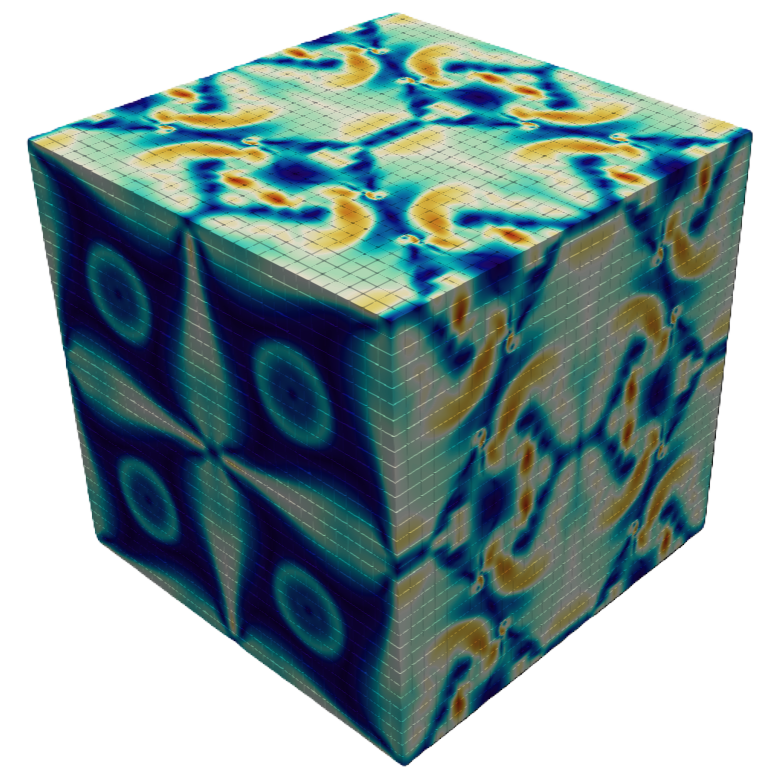}
    \caption{Velocity magnitude at $\displaystyle\frac t{t_c}=12$.}
    \label{fig:TGV_Mach_t12}
  \end{subfigure}
  \quad
  \begin{subfigure}[b]{0.22\textwidth}
    \includegraphics[width=\textwidth]{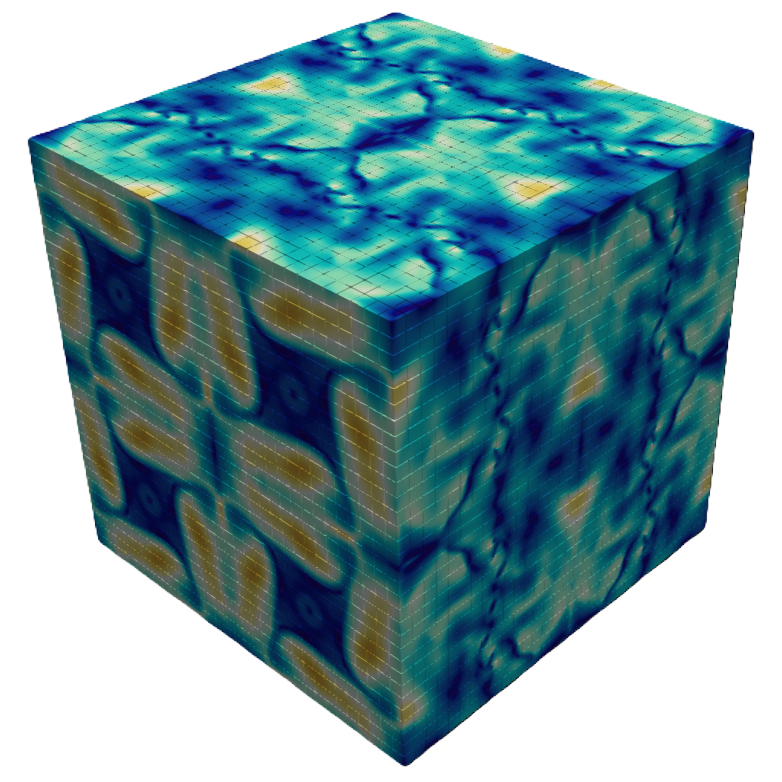}
    \caption{Velocity magnitude at $\displaystyle\frac t{t_c}=16$.}
    \label{fig:TGV_Mach_t16}
  \end{subfigure}
  \quad
  \begin{subfigure}[b]{\textwidth}
    \includegraphics[width=\textwidth]{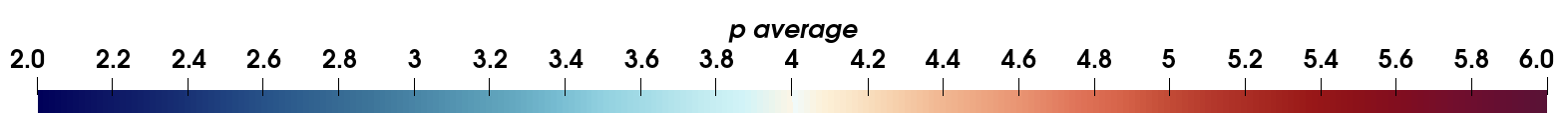}
    \label{fig:TGV_pavg_legend}
  \end{subfigure}
  \quad
  \begin{subfigure}[b]{0.22\textwidth}
    \includegraphics[width=\textwidth]{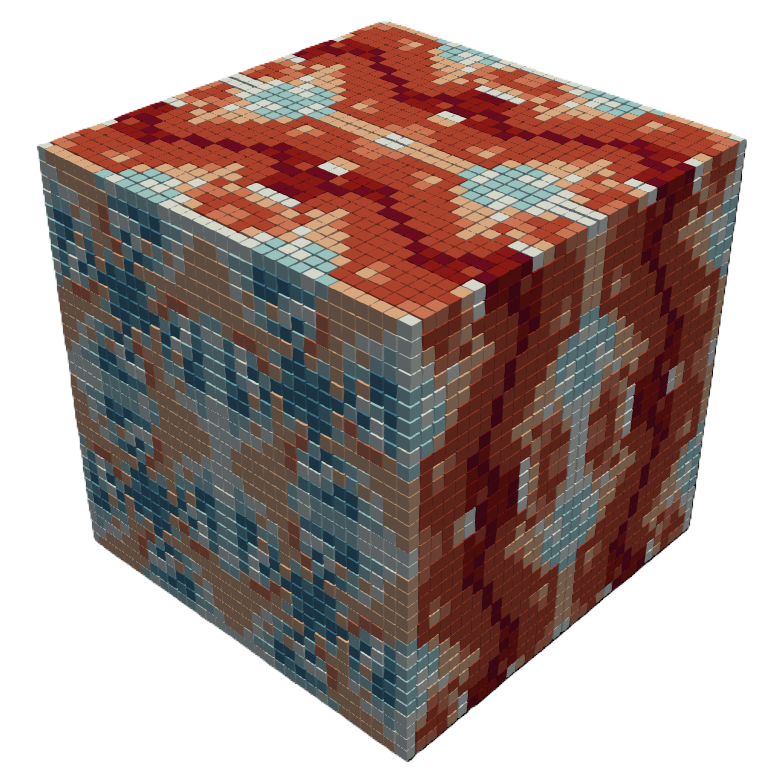}
    \caption{Average $p$ at $\displaystyle\frac t{t_c}=4$.}
    \label{fig:TGV_pavg_t4}
  \end{subfigure}
  \quad
  \begin{subfigure}[b]{0.22\textwidth}
    \includegraphics[width=\textwidth]{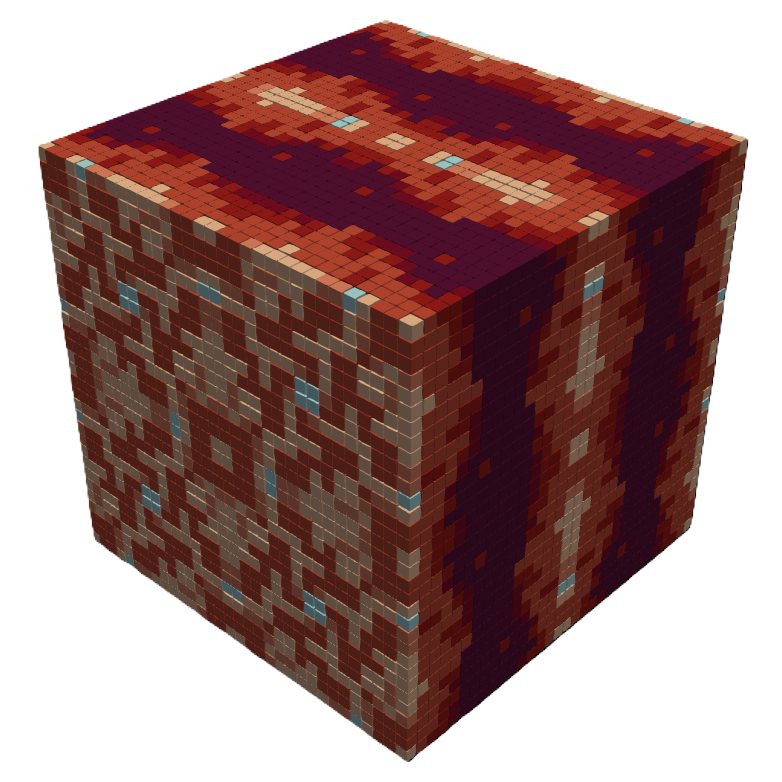}
    \caption{Average $p$ at $\displaystyle\frac t{t_c}=8$.}
    \label{fig:TGV_pavg_t8}
  \end{subfigure}
  \quad
  \begin{subfigure}[b]{0.22\textwidth}
    \includegraphics[width=\textwidth]{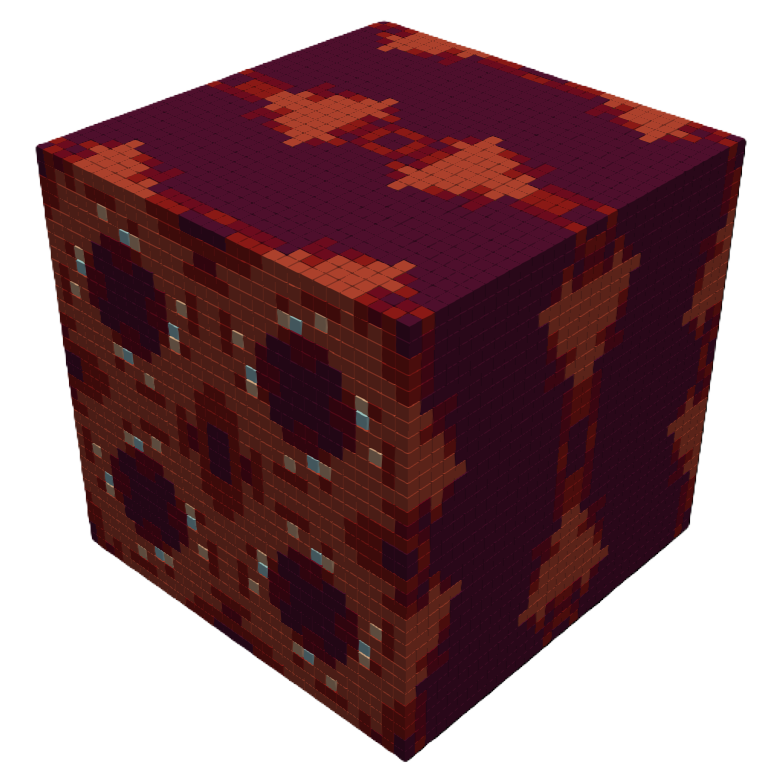}
    \caption{Average $p$ at $\displaystyle\frac t{t_c}=12$.}
    \label{fig:TGV_pavg_t12}
  \end{subfigure}
  \quad
  \begin{subfigure}[b]{0.22\textwidth}
    \includegraphics[width=\textwidth]{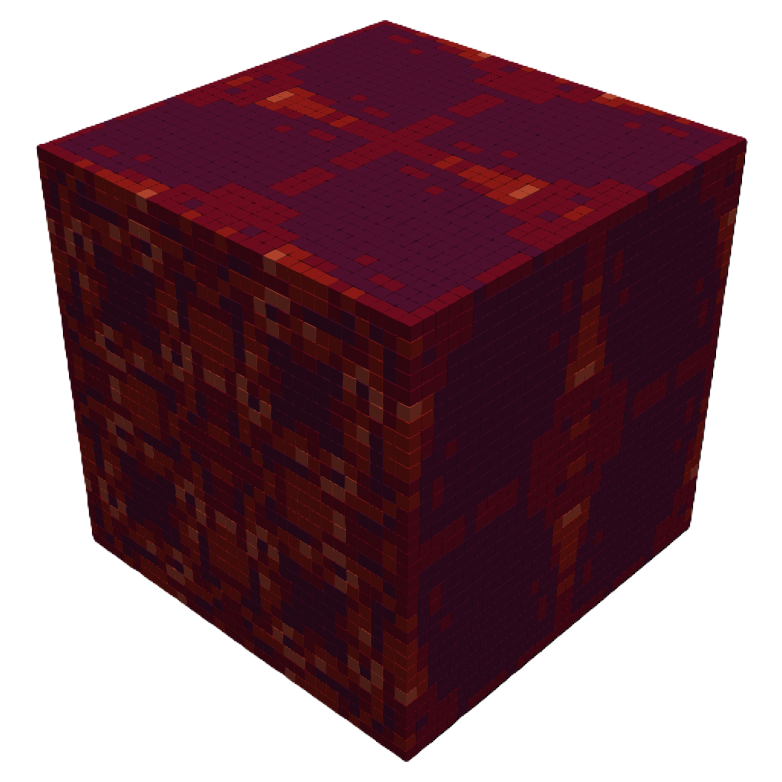}
    \caption{Average $p$ at $\displaystyle\frac t{t_c}=16$.}
    \label{fig:TGV_pavg_t16}
  \end{subfigure}

\caption{(a - d) Non-dimensional velocity magnitude and (e - h) average polynomial order, for the Taylor Green Vortex problem at different timestamps of non-dimensional time.}
\label{fig:TGV_Mach_pavg_results}
\end{figure}
\begin{figure}[h]
    \begin{subfigure}[b]{0.50\textwidth}
        \centering
        \includegraphics[width=\textwidth]{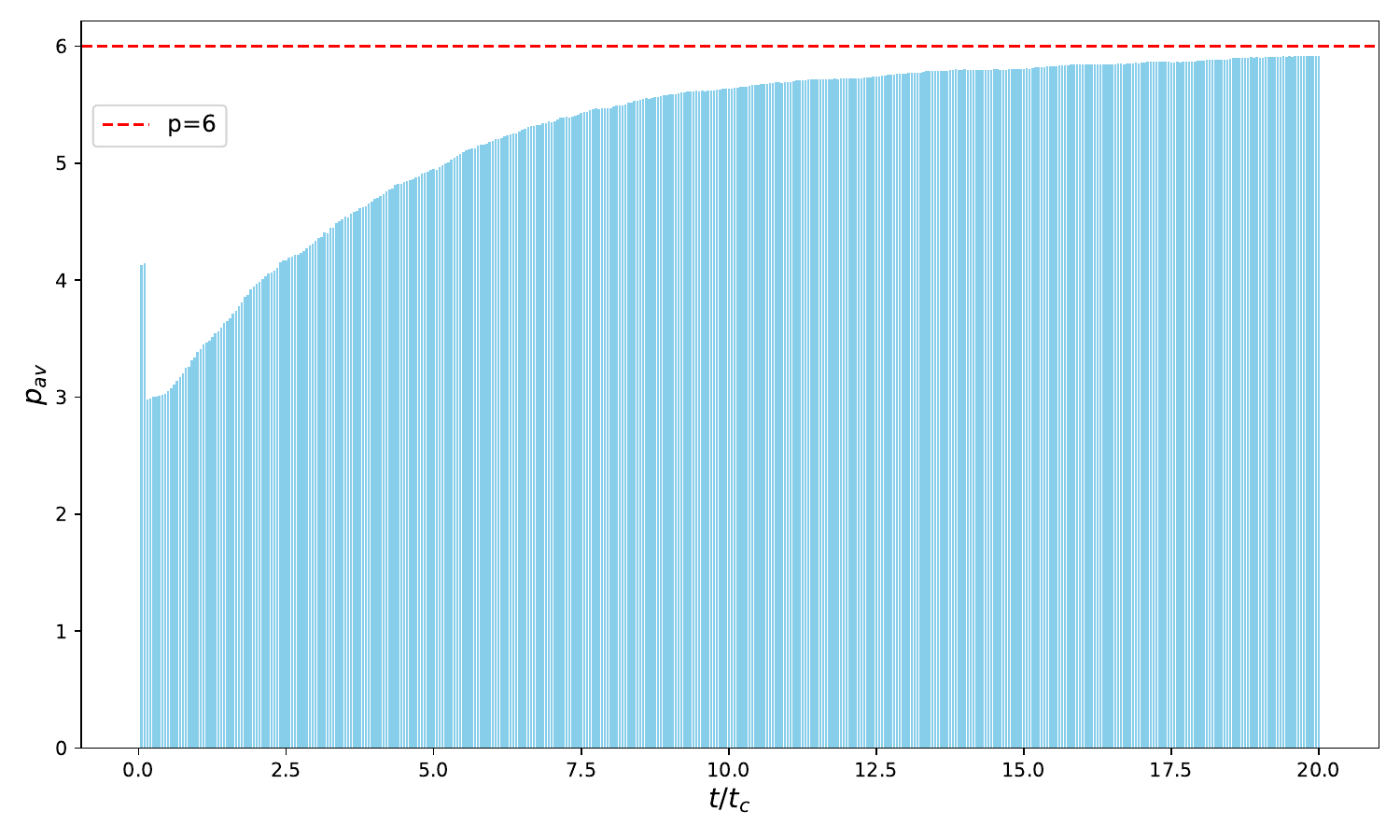}
        \caption{Average polynomial order evolution in the complete domain, using a dynamic p-adaptation every 0.05 non-dimensional seconds.}
        \label{fig:TGV_pavg_evolution_results}
    \end{subfigure}
    \hfill
    \begin{subfigure}[b]{0.42\textwidth}
    \centering
    \includegraphics[width=\textwidth]{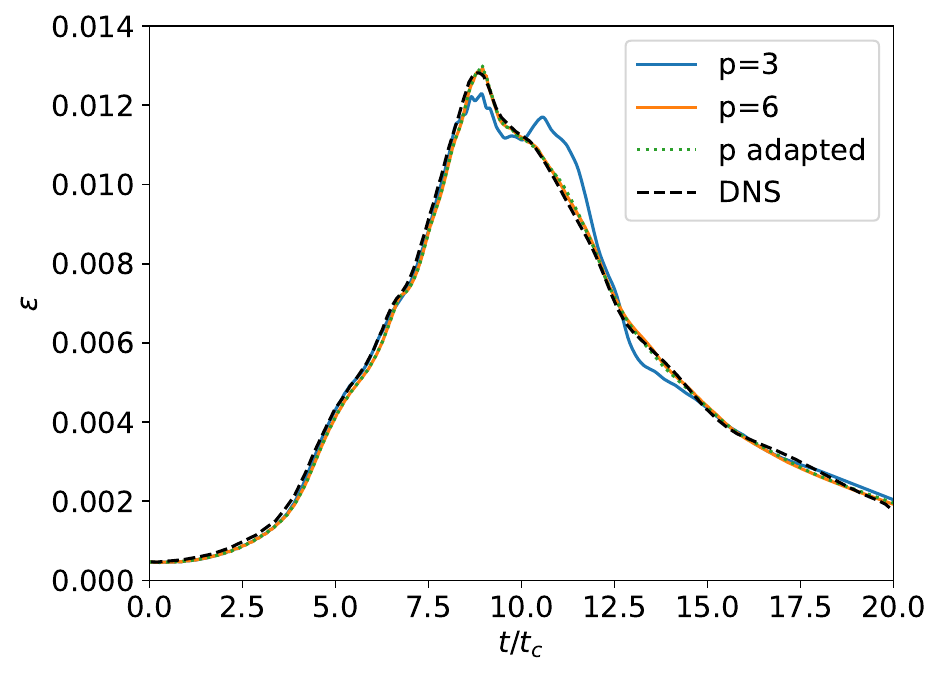}
    \caption{Comparison of the kinetic energy rate in the complete domain for different polynomial orders against DNS results~\cite{Wang2013}.}
    \label{fig:TGV_epsilon_results}
    \end{subfigure}
\caption{Time evolution results of the Taylor Green Vortex problem employing a p-adaptation strategy and homogeneous polynomials. a) Average polynomial and b) kinetic energy rate for different strategies are displayed as a function of the non-dimensional time.}
\end{figure}

The flow field is continuously evolving and it has been widely reported that under-resolution without special numerical treatment creates poor results~\cite{sharma_2019,Gassner_2016,moura2017diffusion,Manzanero_2020}; therefore, a dynamic p-adaptation is an excellent option to accurately solve the problem while maintaining the cost at minimum. A p-adaptation is performed at every $0.05$ non-dimensional time units. The non-dimensional velocity magnitude and the polynomial adaptation for different timestamps are shown in Figure~\ref{fig:TGV_Mach_pavg_results}. Each snapshot shows how the p-adaptation algorithm is able to capture velocity fluctuations by increasing the polynomial order locally. 
As time progresses, the flow transitions to a turbulent state, and hence the average polynomial order in the whole domain increases. This phenomenon is highlighted in Figure~\ref{fig:TGV_pavg_evolution_results}, where it is clear how the polynomial order is progressively increased, almost reaching the maximum polynomial order $p_{\max}=6$ after 20 non-dimensional time units (fully turbulent regime).

Finally, in Figure~\ref{fig:TGV_epsilon_results} we show the evolution of the kinetic energy rate over time for different solutions: two uniform polynomial orders $p=3$ and $p=6$, the RL p-adaptation and a DNS solution from~\cite{Wang2013}. On the one hand, the coarse $p=3$ solution presents large deviations from the reference DNS solution at the transitional times $t/t_c \in[8,14]$. On the other hand, both the p-adapted and the uniform $p=6$ solution are very close to the reference DNS simulation, with both matching almost perfectly. Despite these similar results between the $p=6$ and the p-adapted solutions, the simulation with dynamic p-adaptation was much faster, as described in Table~\ref{tab:TGV_computational_cost_results}. The novel p-adaptation strategy is shown to be accurate in the presence of turbulence while significantly reducing the computational cost by 60\%. 

\begin{table}[ht]
    \centering
    \begin{tabular}{|C{4.0cm} C{2.5cm} C{3.5cm}|}
    \hline

     & $p=6$ & RL p-adapted \\ \hline \hline
    \textbf{Computational cost (h)} & $102$ & $\mathbf{41}$  \\ \hline
    \textbf{Cost reduction (\%)} & $0.0$ & $\mathbf{59.8}$ \\ \hline
    \textbf{DOFs} & $11239424$ & $\mathbf{2066944-10851264}$  \\ \hline
    \textbf{DOFs reduction (\%)} & $0.0$ & $\mathbf{81.6-3.5}$  \\ \hline
    \end{tabular}
    \caption{Comparison of the computational cost, the cost reduction, the DOFs and the DOFs reduction between the uniform $p=6$ and the RL p-adapted solutions for the Taylor Green Vortex problem. The simulations were run on an Intel(R) Xeon(R) Gold 6240R CPU node at 2.40GHz, using 48 OpenMP processes.}
    \label{tab:TGV_computational_cost_results}
\end{table}

\FloatBarrier

\subsection{Simulation of the wind turbine DTU 10MW}\label{subsec:Results_wind_turbine}
Finally, once the reinforcement learning agent has been tested in a wide variety of scenarios and problems, we apply this methodology to solve the flow field around a wind turbine. The objective behind this problem is
to demonstrate that the proposed p-adaptation approach can be used in complex scenarios 
without any previous knowledge regarding the simulation; that is, it is able to adapt to different situations even though it has not been trained for any specific case.

We focus on the 10 MW DTU wind turbine, whose blades are modeled using an actuator line \cite{oscarAL,botero2024low,sorensen2002numerical}, while the nacelle and pylon have been simulated using an immersed boundary approach \cite{kou2022immersed,horses3d_paper}. Furthermore, a no-slip boundary condition has been set at the bottom of the domain, while free-slip boundary conditions have been applied at the top and at the side faces of the domain. Each of these elements adds an additional level of complexity to the p-adaptation algorithm:

\begin{enumerate}
    \item Inside the immersed boundaries (IB) the velocity is zero, and hence the polynomial order should be minimum. At the interface between the IB and the flow, a high polynomial should be used to capture the boundary layer.
    \item Actuator lines add source terms to the Navier-Stokes equations, and we look for an accurate solution near the blades to correctly model the problem.
    \item Near the floor, where there is a wall boundary condition without sliding, a high polynomial order is expected to capture the boundary layer.
    \item Behind the rotor, in the wake of the wind turbine, the flow is fully turbulent, and high polynomials are required.
    An accurate computation of this region is essential to obtain accurate predictions of the power generated by the wind turbine.
    \item An LES model is used to simulate under-resolved scales. In this case, we use the Vreman model \cite{Vreman_2004}, detailed in \ref{sec:cNS}.
\end{enumerate}

The aforementioned elements must be taken into account to obtain an accurate solution. In our case, we use a coarse and almost uniform Cartesian h-mesh with $294492$ hexahedral elements, and the RL agent will select the appropriate polynomial order inside each element. We limit the maximum polynomial order to $p=4$ in the domain and to $p=3$ in the rotor plane, to fix modeling errors associated with the actuator line methodology. Again, it is exactly the same RL agent that was used in all the previous cases and it has no knowledge in relation to this specific problem. Despite this, we expect to obtain an appropriate polynomial adaptation based on the values at the Gauss nodes (the state) that are provided to the agent during the simulation.

The operating conditions of the wind turbine are summarized in Table~\ref{tab:Wind_turbine_conditions_results}. The simulation has been run until the wake has completely developed, reaching the outlet of the domain. In this case, we simulated 30 revolutions of the rotor, around $225 \, \mathrm{s}$.

\begin{table}[htpb]
    \centering
    \begin{tabular}{|C{4.0cm} C{2.5cm}|}
    \hline
    \textbf{Inflow wind speed [$\mathrm{m/s}$]} & $10.0$\\ \hline
    \textbf{Rotor speed [$\mathrm{rpm}$]} & $8.06$\\ \hline
    \textbf{Blade pitch angle [$\mathrm{deg}$]} & $0.0$\\ \hline
    \end{tabular}
    \caption{Wind turbine DTU 10MW operating conditions.}
    \label{tab:Wind_turbine_conditions_results}
\end{table}

\begin{figure}[h]
   \begin{subfigure}[b]{0.48\textwidth}
    \includegraphics[width=\textwidth]{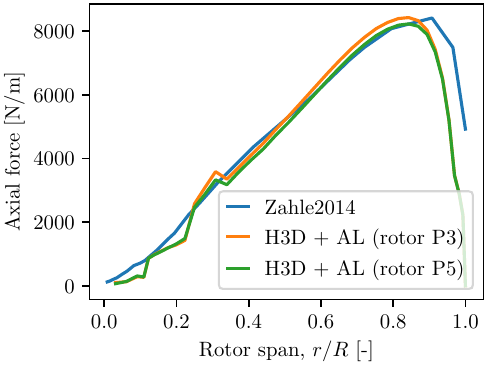}
    \caption{~}
    \label{fig:distributed_forces_h3d_vs_Zahle_normal}
  \end{subfigure}
  \quad
  \begin{subfigure}[b]{0.48\textwidth}
    \includegraphics[width=\textwidth]{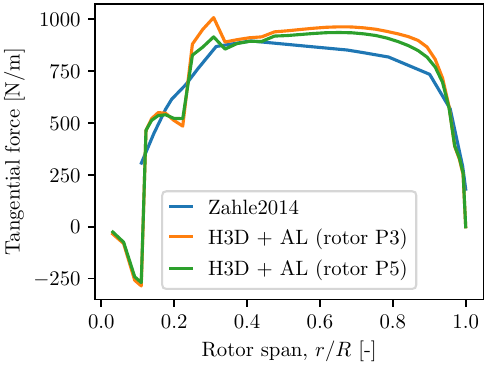}
    \caption{~}
    \label{fig:distributed_forces_h3d_vs_Zahle_tan}
  \end{subfigure}
\caption{Comparison of (a) axial and (b) tangential force distributions along rotor span between blade-resolved RANS simulation in \cite{Zahle2014} (blue) and HORSES3D LES + actuator lines with polynomial orders of $p=3$ and $p=5$ in the rotor zone (orange and green, respectively).}
\label{fig:distributed_forces_h3d_vs_Zahle}
\end{figure}

\begin{figure}[h]
    \begin{subfigure}[b]{0.48\textwidth}
    \includegraphics[width=\textwidth]{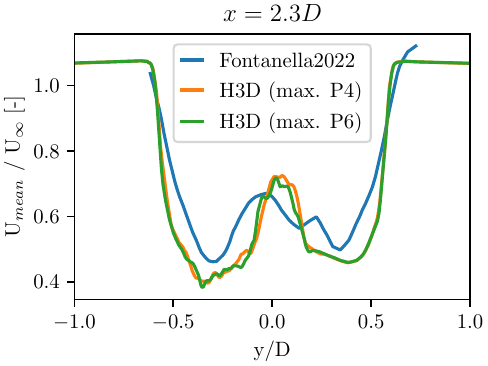}
    \caption{~}
    \label{fig:wake_2p3D_h3d_vs_Fontanella}
    \end{subfigure}
  \quad
  \begin{subfigure}[b]{0.48\textwidth}
    \includegraphics[width=\textwidth]{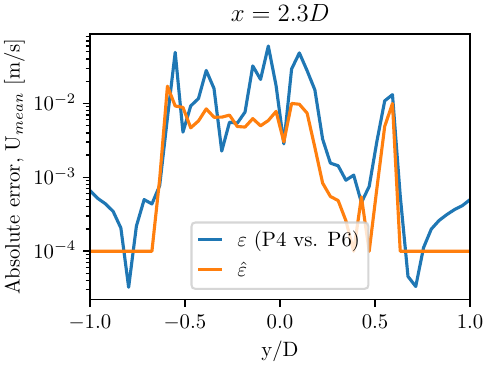}
    \caption{~}
    \label{fig:wake_2p3D_err_vs_est}
  \end{subfigure}
\caption{Axial velocity of the wake averaged in time (over three rotations), at the hub height and a distance of 2.3D downstream of the rotor plane, where D is the diameter of the rotor. Comparison between (a) experimental results presented in \cite{Fontanella2022} (blue) and HORSES3D LES + actuator line simulation with maximum polynomial orders of $p=4$ and $p=6$ (orange and green, respectively) and (b) absolute error in time-averaged axial velocity of the wake for a max. $p=4$ with respect to a max. $p=6$ simulation (blue) and RL-based error estimation (\emph{maximum error}) for a max. $p=4$ simulation (orange).}
\label{fig:wake_2p3D_error_and_estimator}
\end{figure}

The solution is unsteady and turbulent, and a dynamic p-adaptation is performed each time the rotor rotates $1/16 \mathrm{th}$ of revolution. 
First, we validate the results with published data and explore the proposed error estimation. To do so, an additional simulation of the proposed configuration and operational conditions has been carried out, only this time allowing a maximum polynomial order of $p=6$ in the entire domain and setting a constant polynomial order of $p=5$ in the rotor zone. Figure~\ref{fig:distributed_forces_h3d_vs_Zahle} shows a comparison between the axial and tangential force distributions along the blade span, given by both simulations and the blade-resolved simulation presented in \cite{Zahle2014}, which corresponds to the same set of operational conditions. In this case, the increase in the maximum allowed polynomial order results only in a slight improvement of the solution with respect to the blade-resolved simulation, which overall shows good agreement with the exception of the axial components near the tip of the blade. These small differences are expected in actuator line simulations, as the source term used to represent rotating blades is calculated using 2D airfoil data (including a tip correction), but the flow has additional 3D effects at the root and tip of the blades~\cite{Stanly_2022,sorensen2002numerical}.
Figure~\ref{fig:wake_2p3D_h3d_vs_Fontanella} compares the axial velocity of the wake averaged in time (over three rotations), at the hub height and a distance of 2.3D downstream of the rotor plane, where D is the diameter of the rotor, between the numerical simulations and the experimental results presented in \cite{Fontanella2022}, which were obtained in a wind tunnel for a model of scale $1/75 \mathrm{th}$ of the DTU 10MW wind turbine. Simulation results provide a reasonable representation of the characteristic double Gaussian distribution found in the experimental setup, whereas again the increase in the maximum polynomial order does not translate into a significant improvement in results, illustrating the correctness of the initial simulation. The absolute difference between the time-averaged wake velocities of the simulations with maximum polynomial orders of $p=4$ and $p=6$ is presented in Figure~\ref{fig:wake_2p3D_err_vs_est}, as a measure of the error of the first simulation setup. This distribution is then used to assess the performance of the proposed RL-based error estimator (\emph{maximum error}) computed for the simulation in which $p=4$ is used as the maximum polynomial order, which provides a good prediction of the shape, trend, and magnitude of the error, even capturing the drop at approximately $y/D=0.3$ and the peak at $y/D=0.6$. This estimation provides valuable insight into the error, as it can be computed inexpensively for any simulation without prior knowledge of the solution.
\vfill
\vfill

\newpage

\begin{figure}[htpb]
\centering
\begin{subfigure}[b]{\textwidth}
    \includegraphics[width=\textwidth]{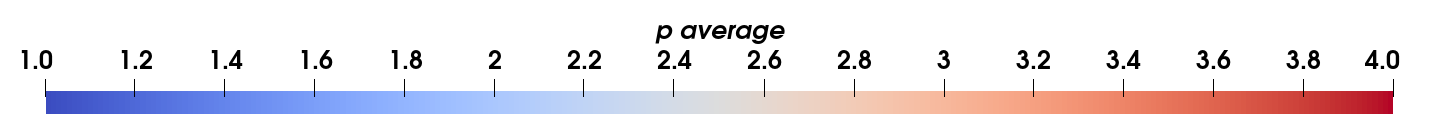}
    \label{fig:Wind_turbine_pavg_legend}
  \end{subfigure}
  \quad
  \begin{subfigure}[b]{0.22\textwidth}
    \includegraphics[width=\textwidth]{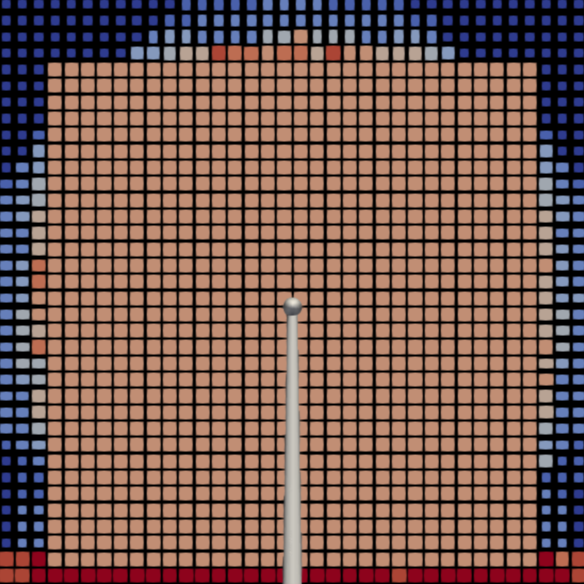}
    \caption{Average $p$ at $1 \mathrm{m}$.}
    \label{fig:Wind_turbine_pavg_1m}
  \end{subfigure}
  \quad
  \begin{subfigure}[b]{0.22\textwidth}
    \includegraphics[width=\textwidth]{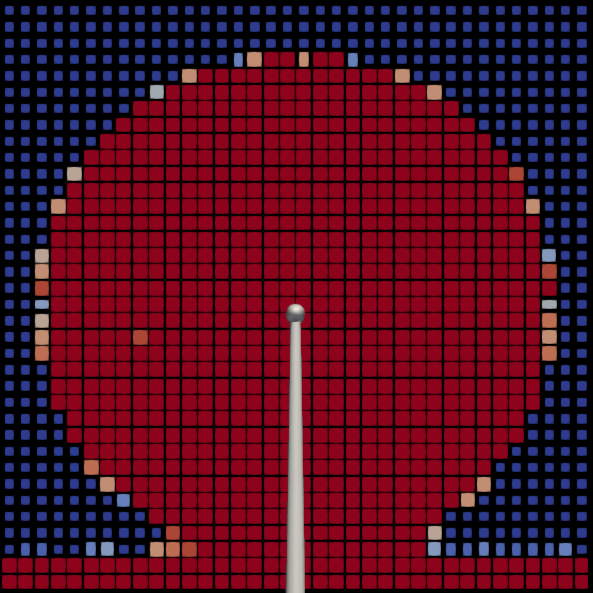}
    \caption{Average $p$ at $1 \mathrm{D}$.}
    \label{fig:Wind_turbine_pavg_1D}
  \end{subfigure}
  \quad
   \begin{subfigure}[b]{0.22\textwidth}
    \includegraphics[width=\textwidth]{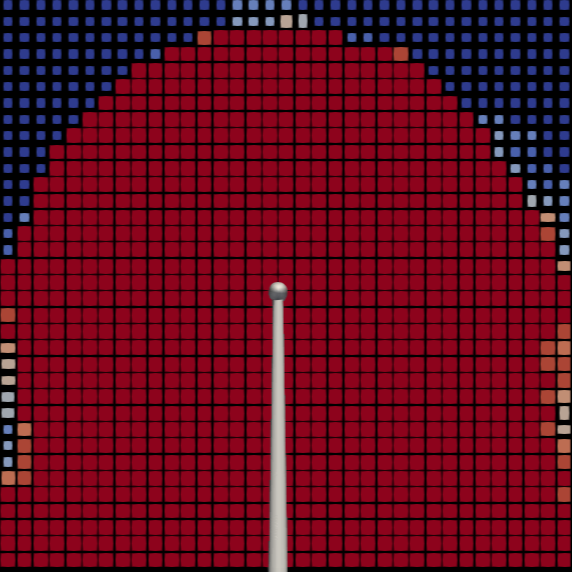}
    \caption{Average $p$ at $3 \mathrm{D}$.}
    \label{fig:Wind_turbine_pavg_3D}
  \end{subfigure}
  \quad
   \begin{subfigure}[b]{0.22\textwidth}
    \includegraphics[width=\textwidth]{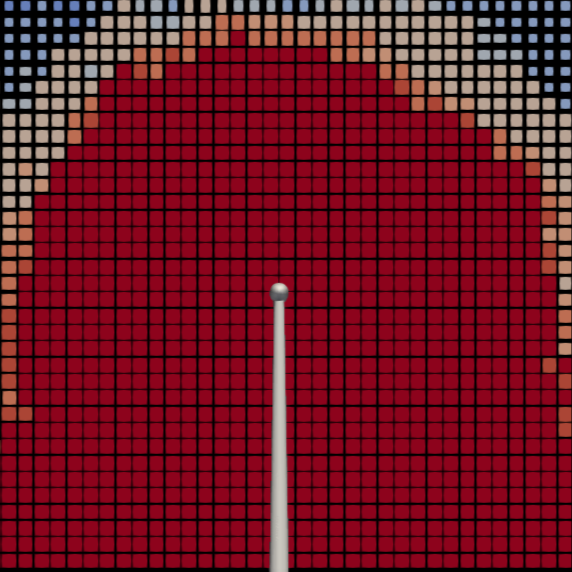}
    \caption{Average $p$ at $5 \mathrm{D}$.}
    \label{fig:Wind_turbine_pavg_5D}
  \end{subfigure}
  \quad
  \begin{subfigure}[b]{\textwidth}
  \includegraphics[width=\textwidth]{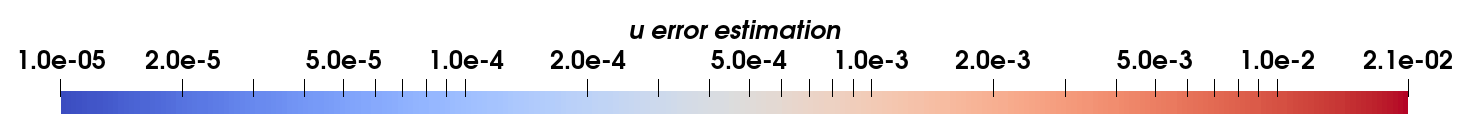}
    \label{fig:Wind_turbine_error_estimation_legend}
  \end{subfigure}
  \quad
   \begin{subfigure}[b]{0.22\textwidth}
    \includegraphics[width=\textwidth]{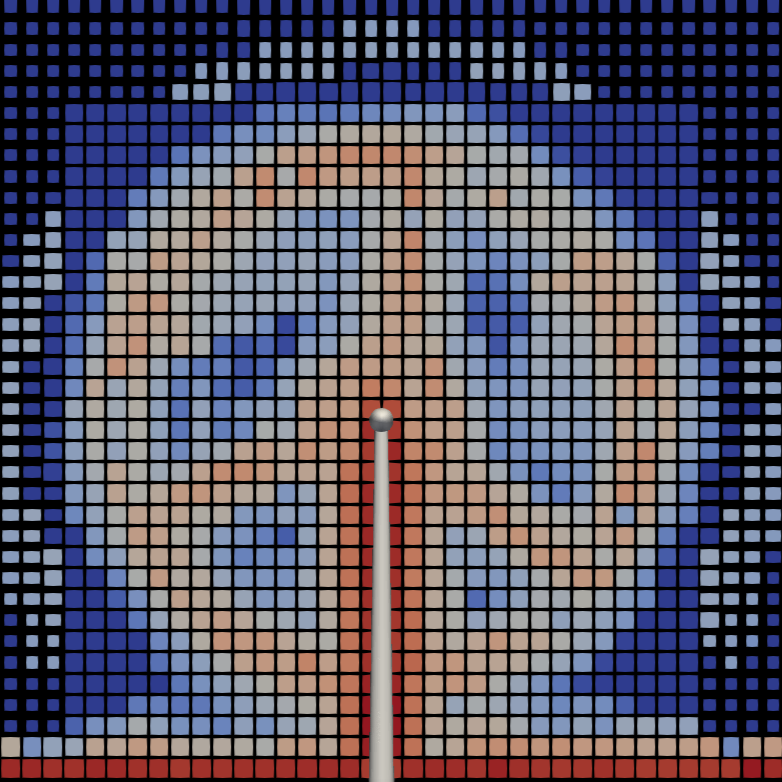}
    \caption{Max error estimation at $1 \mathrm{m}$.}
    \label{fig:Wind_turbine_max_error_1m}
  \end{subfigure}
  \quad
  \begin{subfigure}[b]{0.22\textwidth}
    \includegraphics[width=\textwidth]{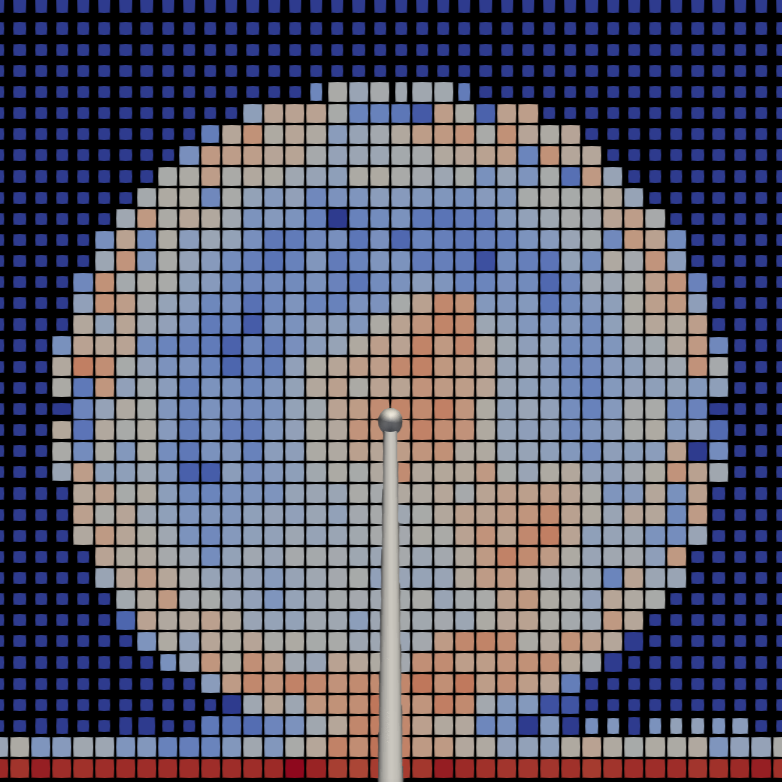}
    \caption{Max error estimation at $1 \mathrm{D}$.}
    \label{fig:Wind_turbine_max_error_1D}
  \end{subfigure}
  \quad
  \begin{subfigure}[b]{0.22\textwidth}
    \includegraphics[width=\textwidth]{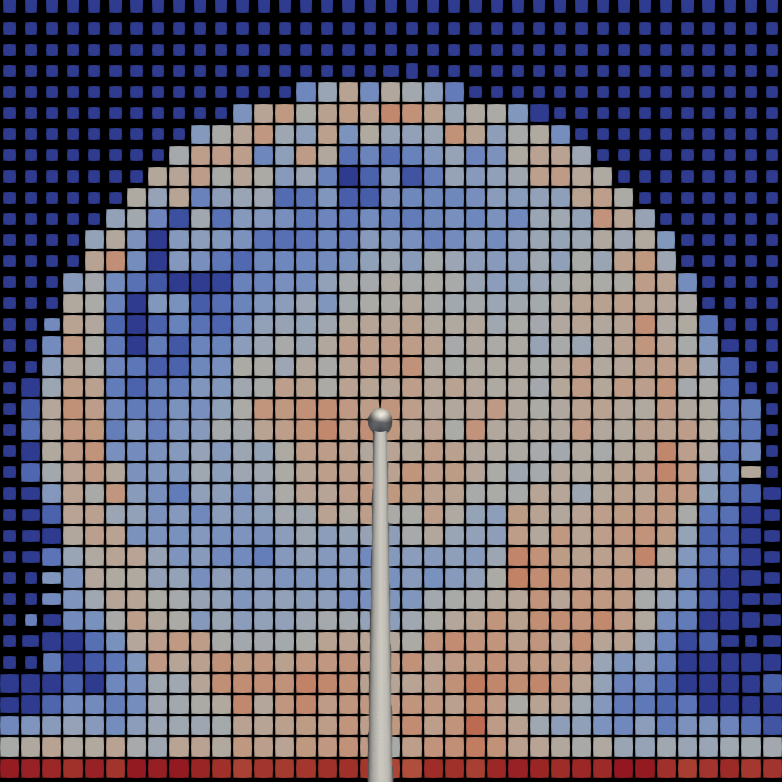}
    \caption{Max error estimation at $3 \mathrm{D}$.}
    \label{fig:Wind_turbine_max_error_3D}
  \end{subfigure}
  \quad
  \begin{subfigure}[b]{0.22\textwidth}
    \includegraphics[width=\textwidth]{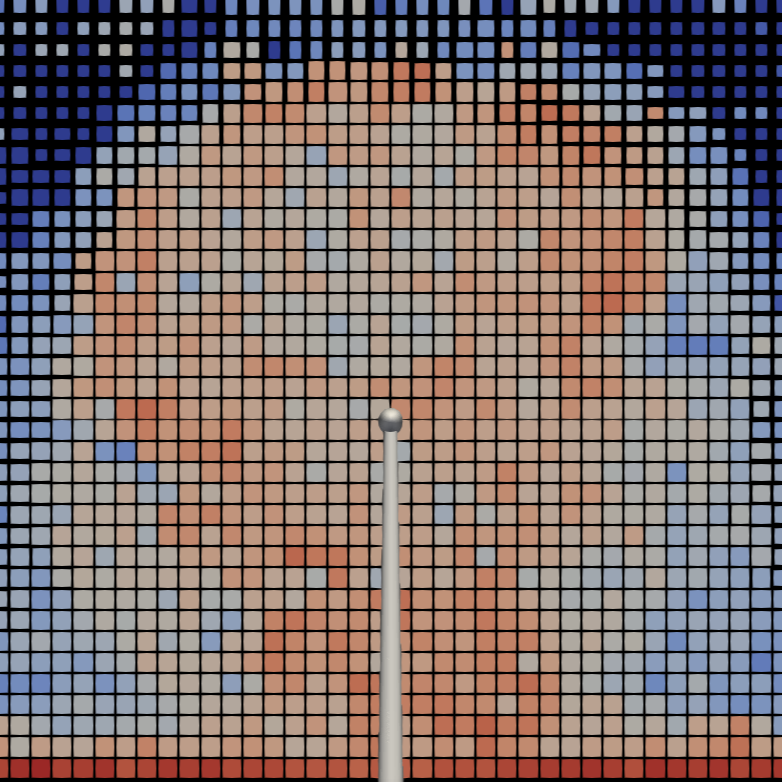}
    \caption{Max error estimation at $5 \mathrm{D}$.}
    \label{fig:Wind_turbine_max_error_5D}
  \end{subfigure}
  \begin{subfigure}[b]{0.22\textwidth}
    \includegraphics[width=\textwidth]{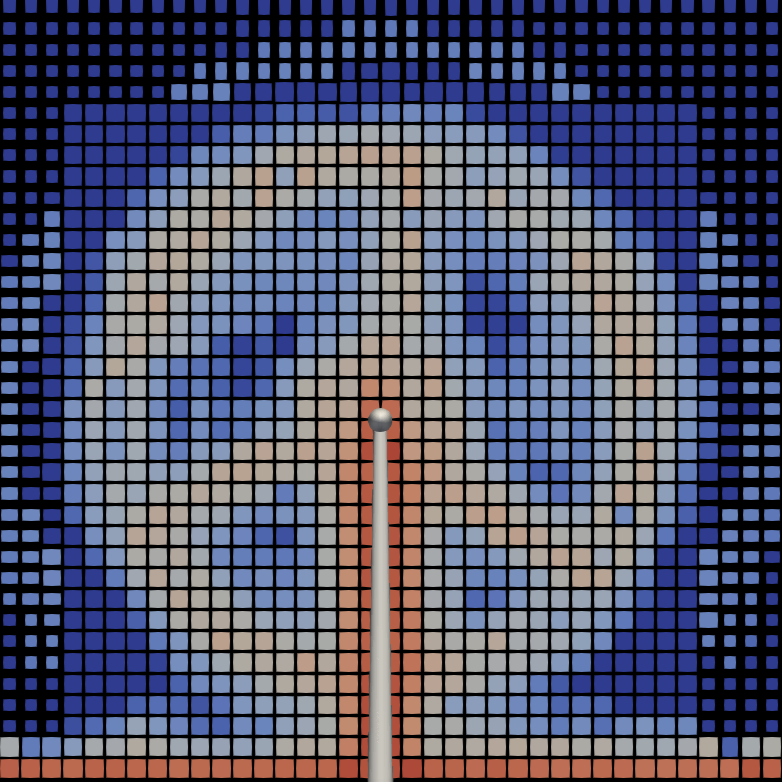}
    \caption{Avg error estimation at $1 \mathrm{m}$.}
    \label{fig:Wind_turbine_avg_error_1m}
  \end{subfigure}
  \quad
  \begin{subfigure}[b]{0.22\textwidth}
    \includegraphics[width=\textwidth]{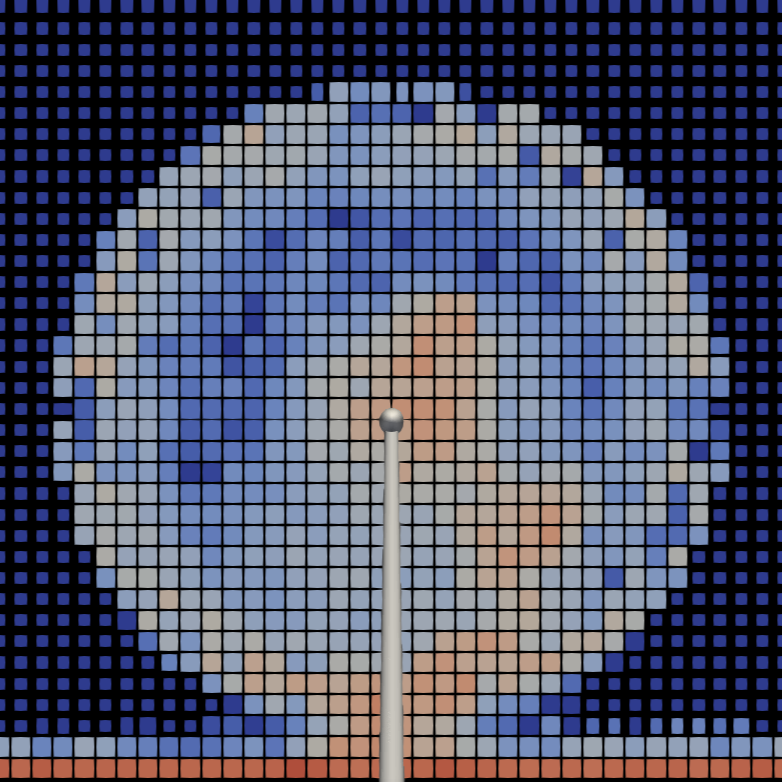}
    \caption{Avg error estimation at $1 \mathrm{D}$.}
    \label{fig:Wind_turbine_avg_error_1D}
  \end{subfigure}
  \quad
  \begin{subfigure}[b]{0.22\textwidth}
    \includegraphics[width=\textwidth]{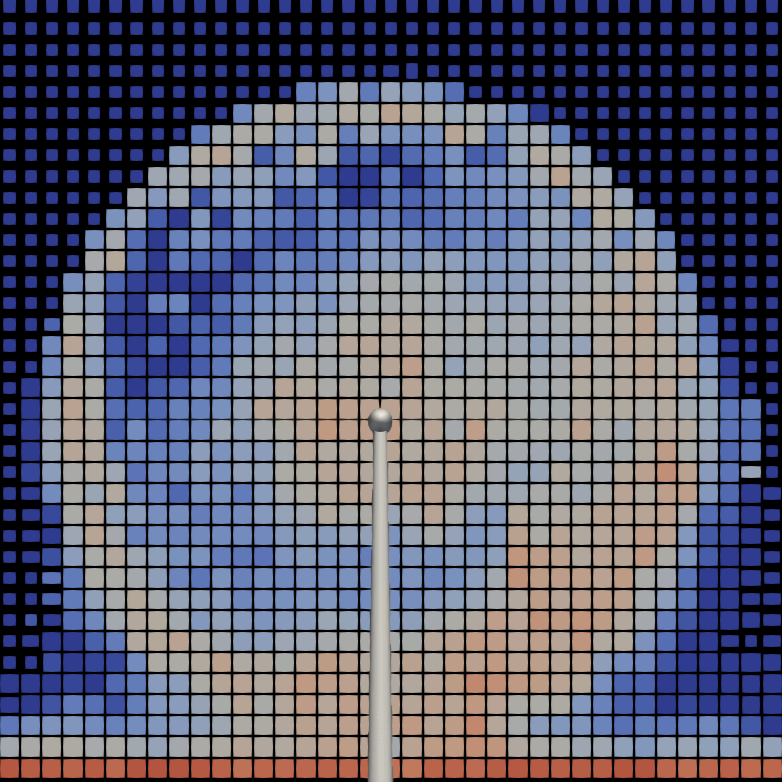}
    \caption{Avg error estimation at $3 \mathrm{D}$.}
    \label{fig:Wind_turbine_avg_error_3D}
  \end{subfigure}
  \quad
  \begin{subfigure}[b]{0.22\textwidth}
    \includegraphics[width=\textwidth]{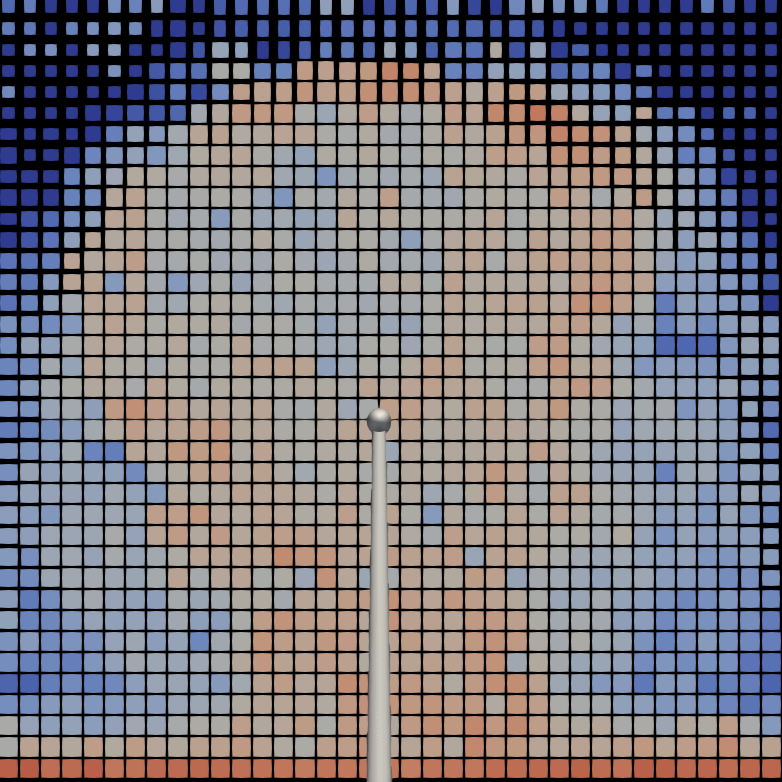}
    \caption{Avg error estimation at $5 \mathrm{D}$.}
    \label{fig:Wind_turbine_avg_error_5D}
  \end{subfigure}

\caption{Snapshot of the (a - d) average polynomial order, (e - h) maximum error estimation of the non-dimensional $u$ velocity and (i - l) average error estimation of the non-dimensional $u$ velocity, for the wind turbine simulation at different planes parallel to the rotor.}
\label{fig:Wind_turbine_rotor_Error_pavg_results}
\end{figure}

\begin{figure}[h]
\centering
\begin{subfigure}[b]{\textwidth}
    \includegraphics[width=\textwidth]{Figures/Wind_Turbine/Nav_Wind_Turbine_legend.png}
    \label{fig:Wind_turbine_pavg_legend}
  \end{subfigure}
  \quad
  \begin{subfigure}[b]{0.9\textwidth}
    \includegraphics[width=\textwidth]{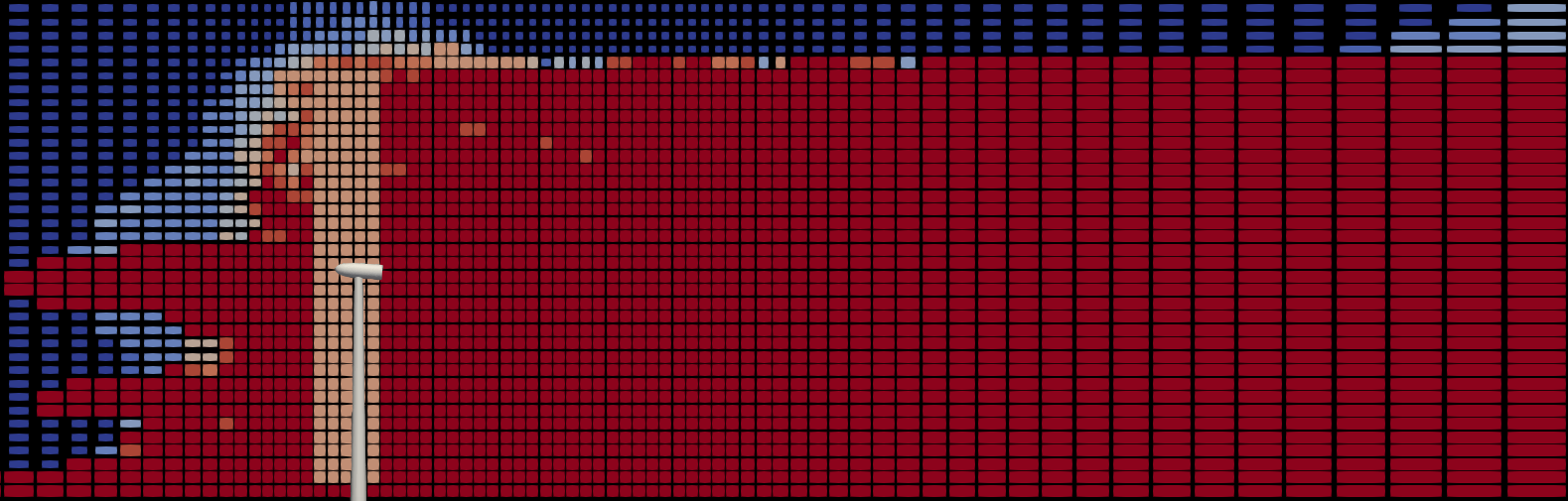}
    \caption{Average polynomial order.}
    \label{fig:Wind_turbine_pavg}
  \end{subfigure}
  \quad
  \begin{subfigure}[b]{\textwidth}
  \includegraphics[width=\textwidth]{Figures/Wind_Turbine/Error_estimation_Wind_Turbine_legend.png}
    \label{fig:Wind_turbine_error_estimation_legend}
  \end{subfigure}
  \quad
  \begin{subfigure}[b]{0.9\textwidth}
    \includegraphics[width=\textwidth]{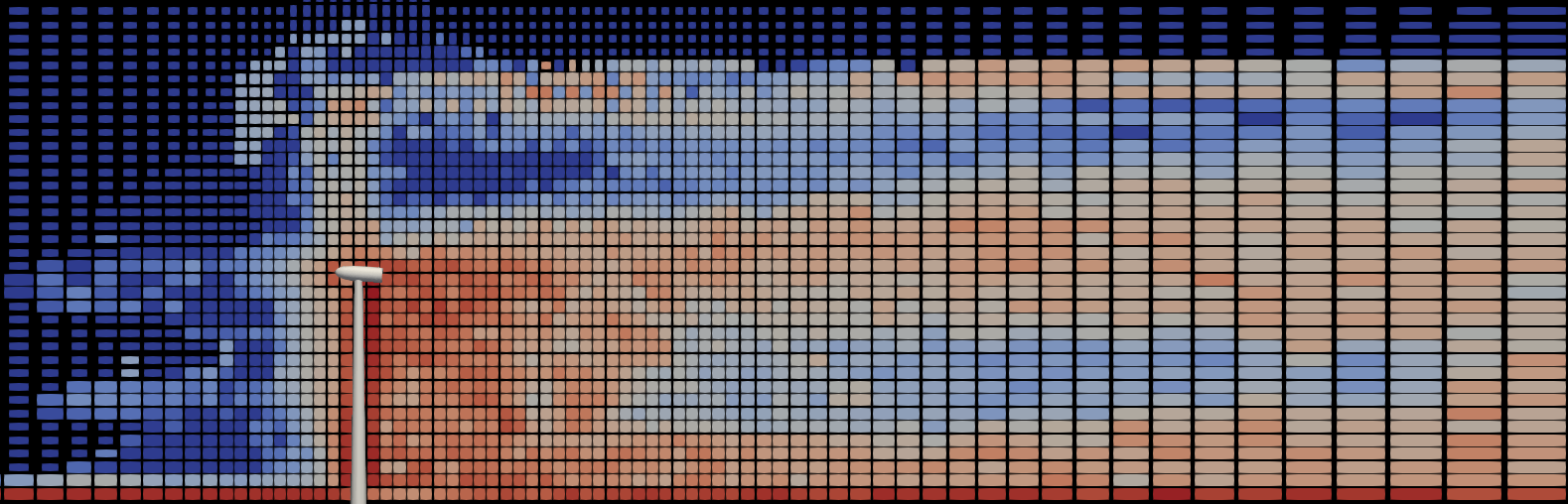}
    \caption{Maximum error estimation.}
    \label{fig:Wind_turbine_max_error}
  \end{subfigure}
  \quad
  \begin{subfigure}[b]{0.9\textwidth}
    \includegraphics[width=\textwidth]{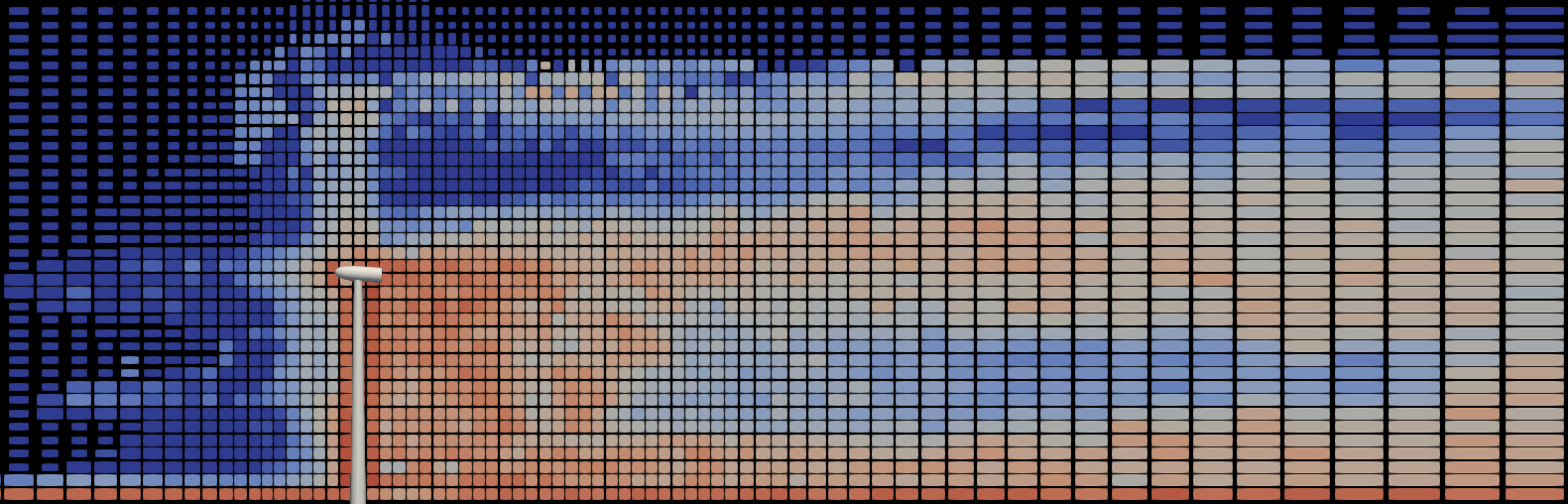}
    \caption{Average error estimation.}
    \label{fig:Wind_turbine_avg_error}
  \end{subfigure}
  
\caption{Snapshot of the a) average polynomial order, b) maximum error estimation of the non-dimensional $u$ velocity and c) average error estimation of the non-dimensional $u$ velocity, for the wind turbine simulation at a plane that divides the wake in two.}
\label{fig:Wind_turbine_Error_pavg_results}
\end{figure}

For completeness, a snapshot of the average polynomial order and the error estimation, once the wake has fully developed, are represented in Figures~\ref{fig:Wind_turbine_rotor_Error_pavg_results} and \ref{fig:Wind_turbine_Error_pavg_results}. The p-adaptation algorithm has effectively increased the polynomial order to the maximum in the wake and at the BC on the floor.  
Also, the agent has selected the minimum polynomial order $p=1$ for the far field region, which means that the change in the non-dimensional variables of interest ($\rho u$, $\rho v$ and $\rho w$) is below the defined tolerance (see Section \ref{subsec:Implementation_p_adaptation} for details). Furthermore, downwind turbulent mixing enlarges the wake size, and consequently the high-order region is increased, as shown in Figures~\ref{fig:Wind_turbine_pavg_1m}, \ref{fig:Wind_turbine_pavg_1D}, \ref{fig:Wind_turbine_pavg_3D} and \ref{fig:Wind_turbine_pavg_5D}.

Finally, these figures also show the error estimation for the non-dimensional axial velocity at different locations. The highest error values, around $10^{-2}$, are located near the immersed boundaries and near the floor (no slip boundary condition). Both results were expected, as the mesh was coarsened on purpose to test the effectiveness of the p-adaptation algorithm and the error estimation. Inside the immersed boundaries (tower and nacelle), there is only one h-element of the computational mesh for each height, so even if the polynomial order is large, the error remains large.
%
The error estimation algorithm is able to identify the two major sources of error, without having any prior knowledge regarding this specific problem and without training a specific agent for this purpose. Furthermore, the computational cost of the estimator is negligible, making it a good choice to capture the trend of spatial errors in numerical simulations.

\FloatBarrier

\section{Conclusions}\label{sec:Conclusions}
In conclusion, this study presents a new approach to automate anisotropic p-adaptation in high-order h/p solvers using reinforcement learning. The RL-based adaptation dynamically adjusts high-order polynomials based on the evolving solution. The proposed methodology is independent of the computational mesh and can be applied to any partial differential equation, illustrating its broad applicability and flexibility.

The approach is based on the \textit{value iteration} algorithm and allows offline training, which is decoupled from the CFD solver, leading to a flexible and efficient approach. Therefore, the actual simulations do not show any significant computational overhead. We showcase the methodology for a variety of flows ranging from laminar 2D to turbulent 3D, including circular cylinders, the Taylor Green Vortex, and a 10MW wind turbine. The method offers a robust, reproducible, and general approach, applicable to complex three-dimensional problems without needing to retrain. In addition, we include an inexpensive error estimation, derived from the RL framework, that offers the possibility to quantify local discretization errors with a negligible cost in numerical simulations. 

These findings open new avenues for further research and potential applications in the solution of complex partial differential equations using RL-based h/p-mesh adaptation. 

\section*{Acknowledgments}
This research has received funding from the European Union (ERC, Off-coustics, project number 101086075). Views and opinions expressed are, however, those of the authors only and do not necessarily reflect those of the European Union or the European Research Council. Neither the European Union nor the granting authority can be held responsible for them.
EF and GR
acknowledge the funding received by the Grant DeepCFD (Project No. PID2022-137899OB-I00) funded by MICIU/AEI/10.13039/501100011033 and by ERDF, EU. OM and EF would like to thank the support of
Agencia Estatal de Investigación (for the grant "Europa Excelencia 2022" Proyecto EUR2022-134041/AEI/10.13039/501100011033) y del Mecanismo de Recuperación y Resiliencia de la Unión Europea. DH and EF acknowledge the funding received by the Comunidad de Madrid according to Orden 5067/2023, of December 27th, issued by the Consejero de Educación, Ciencia y Universidades, which announces grants for the hiring of predoctoral research personnel in training for the year 2023.
Finally, all authors gratefully acknowledge the Universidad Politécnica de Madrid (www.upm.es) for providing computing resources on Magerit Supercomputer and the computer resources at MareNostrum and the technical support provided by Barcelona Supercomputing Center (RES-IM-2024-1-0003).

\appendix

\section{Compressible Navier-Stokes solver}
\label{sec:cNS}
We solve the compressible 3D Navier-Stokes equations for laminar cases and supplement the equations with the Vreman LES model for turbulent flows. The 3D Navier-Stokes equations, when including the  Vreman model, can be written compactly:
%
\begin{equation}
\boldsymbol{q}_t+ \nabla \cdot {\ssvec{F}}_e = \nabla\cdot\ssvec{F}_{v,turb},
\label{eq:compressibleNScompact}
\end{equation}
where $\boldsymbol{q}$ is the state vector of large-scale resolved conservative variables $\boldsymbol{q} = [ \rho , \rho u , \rho v , \rho w , \rho e]^T$, $\ssvec{F}_e$ are the inviscid, or Euler fluxes,
\begin{equation}
\ssvec{F}_e = \left[\begin{array}{ccc} \rho u & \rho v & \rho w \\
                                                                                \rho u^2 + p & \rho uv & \rho uw \\
                                                                                	\rho uv & \rho v^2 + p & \rho vw \\
                                                                                	\rho uw & \rho vw & \rho w^2 + p \\
                                                                                	\rho u H & \rho v H & \rho w H
\end{array}\right],
\end{equation}
where $\rho$, $e$, $H=E+p/\rho$, and $p$ are the large-scale density, total energy, total enthalpy and pressure, respectively, and $\vec{v}=[u,v,w]^T$ is the large-scale resolved velocity components. Additionally, $\ssvec{F}_{v,turb}$ defines the viscous and turbulent fluxes,
\begin{equation}
\ssvec{F}_{v,turb}= \left[\begin{array}{ccc}0 & 0 & 0\\
 \tau_{xx} & \tau_{xy} & \tau_{xz} \\
 \tau_{yx} & \tau_{yy} & \tau_{yz} \\
 \tau_{zx} & \tau_{zy} & \tau_{zz} \\
 \sum_{j=1}^3 v_j\tau_{1j} + \kappa T_x& \sum_{j=1}^3 v_j\tau_{2j} + \kappa T_y& \sum_{j=1}^3 v_j\tau_{3j} + \kappa T_z
\end{array}\right],
\label{eq:viscousfluxes}
\end{equation}
where $\kappa$ is the thermal conductivity, $T_x, T_y$ and $T_z$ are the temperature gradients and the stress tensor $\boldsymbol{\tau}$ is defined as $\boldsymbol{\tau} = (\mu+\mu_t)(\nabla \vec{v} + (\nabla \vec{v})^T) - 2/3(\mu+\mu_t) \boldsymbol{I}\nabla\cdot\vec{v}$, with $\mu$ the dynamic viscosity, $\mu_t$ the turbulent viscosity ( defined through the Vreman 
model) and $\boldsymbol{I}$ the three-dimensional identity matrix. When solving laminar flows, it suffices to set $\mu_t=0$ and reinterpret the large-scale resolved components as the main components (no under-resolved components).
The turbulent dynamic viscosity in the Vreman \cite{Vreman_2004} model is: 
\begin{equation}
\begin{split}
    &\mu_t = C_v \rho\sqrt{\frac{B_\beta}{\alpha_{ij}\alpha_{ij}}},\\
    &\alpha_{ij} = \frac{\partial u_j}{\partial x_i},\\
    &\beta_{ij} = \Delta^2\alpha_{mi}\alpha_{mj},\\
    &B_\beta = \beta_{11}\beta_{22} -\beta_{12}^2 +\beta_{11}\beta_{33} -\beta_{13}^2 +\beta_{22}\beta_{33} -\beta_{23}^2,
\end{split}
\label{eq-iLES:LES_vreman}
\end{equation}

\noindent where $C_v=0.07$ is the constant of the model. 
The Vreman LES model adjusts the model parameters based on the local flow characteristics and automatically reduces the turbulent viscosity in laminar, transitional, and near-wall regions, allowing to capture the correct physics. 

\section{Spatial discretisation: discontinuous Galerkin}
\label{sec:dg}

The discontinuous Galerkin Spectral Element Method (DGSEM), is a particularly efficient nodal version of DG schemes \cite{kopriva2009implementing}. For simplicity, here we only introduce the fundamental concepts of DG discretizations. Further details can be found in \cite{MANZANERO2020109241,FerrerJCP}.

%
The physical domain is tessellated with nonoverlapping curvilinear hexahedral elements, $e$, which are geometrically transformed to a reference element, $el$. This transformation is performed using a polynomial transfinite mapping that relates the physical coordinates $\vec{x}$ and the local reference coordinates $\vec{\xi}$. The transformation is applied to \eqref{eq:compressibleNScompact} resulting in the following:

\begin{equation}
J \boldsymbol{u}_t  + \nabla_\xi\cdot{F}_e = \nabla_\xi\cdot{F}_{v,turb},
\label{eq:compressibleNScompact_transformed}
\end{equation}
where $J$ is the Jacobian of the transfinite mapping, $\nabla_\xi$ is the differential operator in the reference space and ${F}$ are the contravariant fluxes \cite{kopriva2009implementing}. 

To derive DG schemes, we multiply \eqref{eq:compressibleNScompact_transformed} by a locally smooth test function $\phi_j$, for $0\leq j\leq P$, where $P$ is the polynomial degree, and integrate over an element $el$ to obtain the weak form
\begin{equation}\label{eq::NS2}
\int_{el}J \boldsymbol{u}_t\phi_j+\int_{el} \nabla_\xi\cdot{F}_e\phi_j  =\int_{el} \nabla_\xi\cdot{F}_{v,turb}\phi_j.
\end{equation}
We can now integrate by parts the term with the inviscid fluxes, ${F}_e$, to obtain a local weak form of the equations (one per mesh element) with the boundary fluxes separated from the interior
\begin{equation}\label{eq::NS3}
\int_{el}J \boldsymbol{u}_t\phi_j +  \int_{\partial el} {F}_e\cdot\hat{\mathbf{n}}\phi_j-\int_{el} {F}_e\cdot\nabla_\xi\phi_j
=\int_{el} \nabla_\xi\cdot{F}_{v,turb}\phi_j,
\end{equation}
where $\hat{n}$ is the unit outward vector of each face of the reference element ${\partial el}$. 
We replace discontinuous fluxes at inter--element faces by a numerical inviscid flux, ${F}_{e}^{\star}$, to couple the elements, 
\begin{equation}\label{eq::NS4}
\int_{el}J \boldsymbol{u}_t\cdot\phi_j + \int_{\partial el} {{F}_{e}^{\star}}\cdot\hat{\mathbf{n}}\phi_j-\int_{el} {F}_e\cdot\nabla_\xi\phi_j
=\int_{el} \nabla_\xi\cdot{F}_{v,turb}\phi_j.
\end{equation}
This set of equations for each element is coupled through the Riemann fluxes ${F}_{e}^{\star}$, which governs the numerical characteristics, see for example the classic book by Toro \cite{toro2009riemann}. Note that one can proceed similarly and integrate the viscous terms by parts (see, for example, \cite{Unified,ferrer2016a,FERRER2011224,FERRER20127037}). The viscous terms require further manipulations to obtain usable discretisations (Bassi Rebay 1 and 2 or Interior Penalty). Viscous and turbulent terms are discretised following the same spatial discretisation and in this work we retain the Bassi Rebay 1 scheme. For simplicity, here we retain the volume form:
\begin{equation}\label{eq::NS5}
\int_{el}J \boldsymbol{u}_t\cdot\phi_j + \int_{\partial el} \underbrace{{{F}_{e}^{\star}}\cdot\hat{\mathbf{n}}}_\text{Convective fluxes}\phi_j-\int_{el} {F}_e\cdot\nabla_\xi\phi_j
=\int_{el} (\underbrace{\nabla_\xi\cdot{F}_{v,turb}}_\text{Viscous and Turbulent fluxes})\cdot\phi_j.
\end{equation}


The final step, to obtain a usable numerical scheme, is to approximate the numerical solution and fluxes by polynomials (of order $p$) and to use Gaussian quadrature rules to numerically approximate volume and surface integrals. In HORSES3D we allow for Gauss-Legendre or Gauss--Lobatto quadrature points, but we only use Gauss-Legendre in this work.
The connection between non-conforming elements is achieved using the mortar method \cite{Kopriva2002}. For details on the implementation, see \cite{kessasra2024comparison}.

\section{Value Iteration}\label{app:Value_iteration}
The update rule for the \textit{value iteration} algorithm is introduced in eq.~(\ref{eq:Value_iteration}). In Algorithm \ref{alg:Value_Iteration}, we show the pseudo-code that allows to implement this RL approach.

\color{black}
\begin{algorithm}[H]
    \caption{Value Iteration \cite{Sutton1998}}
    \label{alg:Value_Iteration}
    \SetAlgoLined
    \SetKwInOut{Parameter}{Algorithm parameter}
    \Parameter{$\varepsilon > 0$ determining accuracy of estimation}
    Initialize $v(s)$, for all $s \in\mathcal S^+$, arbitrarily except that $v(\text{terminal}) = 0$\;
    \While{$\Delta \geq \varepsilon$}{
        $\Delta \gets 0$\;
        \For{each $s \in\mathcal S$}{
            $V \gets v(s)$\;
            $v(s) \gets \displaystyle\max_a \displaystyle\sum_{s', r} \mathbb P(s',r | s, a) [r + \gamma v(s')]$\;
            $\Delta \gets \max(\Delta, |V - v(s)|)$\;
        }
    }
    \For{each $s \in\mathcal S$}{
        $\pi(s) \gets \displaystyle\arg\max_a \displaystyle\sum_{s', r} \mathbb P(s',r | s, a) [r + \gamma v(s')]$\;
    }
    \KwOut{Deterministic policy $\pi$}
\end{algorithm}

\newpage 
\bibliographystyle{elsarticle-num} 
\bibliography{bibliography}

\end{document}